\documentclass[10pt]{article}
\usepackage{color}
\usepackage{epsfig}
\usepackage{lscape}
\usepackage{amssymb} 
\usepackage{graphicx}
\usepackage{footnote}
\usepackage[utf8]{inputenc}	
\usepackage{mathtools}
\usepackage{amsthm}
\usepackage{wasysym}
\usepackage{amsfonts}
\usepackage[english]{babel}
\usepackage{bigints}
\usepackage{mathrsfs}
\usepackage{lineno}
\usepackage{nicefrac}
\usepackage{enumerate}
\usepackage{appendix}
\usepackage{graphicx}
\usepackage{commath}
\usepackage{multirow}
\usepackage{tabu}

\usepackage{times}
\usepackage[T1]{fontenc}

\usepackage{algorithm} 
\usepackage{algpseudocode} 

\usepackage[most]{tcolorbox}
\usepackage{mdframed}

\usepackage{amsmath}
\usepackage{mathtools} 
\usepackage{mathrsfs}

\usepackage[]{natbib}
\usepackage{csquotes}

\usepackage[colorlinks=true,citecolor=blue,linkcolor=black,urlcolor=blue]{hyperref}
\definecolor{light-gray}{gray}{0.91}

\usepackage[margin=2cm]{geometry}

\allowdisplaybreaks[1]

\makeindex

\usepackage{authblk}

\newcommand{\imag}{\textbf{i}}
\def\tr{\textrm{tr }}
\newcommand{\ImagPart}{\mathfrak{Im}}
\newcommand{\RealPart}{\mathfrak{Re}}
\newcommand{\error}{\mathcal{O}}

\def\dvg{\textrm{Div }}

\def\crl{\text{curl }}

\def\Lm{\mathsf{L}}
\def\Lagr{\mathscr{L}}
\def\Linop{\mathcal{L}}
\def\adLinop{\mathcal{L}^{\dagger}}
\def\adLinopc{\mathcal{L}^{\dagger ^c}}
\def\d{\textbf{d}}
\def\u{\textbf{u}}
\def\n{\hat{\textbf{n}}}
\def\m{\textbf{m}}
\def\v{\textbf{v}}
\def\x{\textbf{x}}
\def\f{\textbf{f}}
\def\e{\textbf{e}}
\def\E{\textbf{E}}
\def\Env{\text{Env}}
\def\dobs{\textbf{d}^{\text{obs}}}
\def\mest{\textbf{m}^{\text{est}}}
\def\G{\textbf{G}}
\def\den{\boldsymbol{\rho}}
\def\stress{\boldsymbol{\sigma}}
\def\strain{\boldsymbol{\varepsilon}}
\def\id{\mathbb{I}}

\usepackage{color}

\begin{document}
	
\title{\vspace*{0cm}Understanding the Adjoint Method in Seismology: Theory and Implementation in the Time Domain}

\date{\vspace{-5ex}}

\author[,1,2]{Rafael Abreu \thanks{email: rabreu@ipgp.fr}}
\affil[1]{Institut de Physique du Globe de Paris, CNRS, Universit\'e de Paris, 75005 Paris, France}
\affil[2]{Institut f\"ur Geophysik, Westf\"alische Wilhelms-Universit\"at M\"unster, 48149 M\"unster, Germany}

\maketitle

\begin{abstract}
    
    The adjoint method is a popular method used for seismic (full-waveform) inversion today. The method is considered to give more realistic and detailed images of the interior of the Earth by the use of more realistic physics. It relies on the definition of an adjoint wavefield (hence its name) that is the time reversed synthetics that satisfy the original equations of motion. The physical justification of the nature of the adjoint wavefield is, however, commonly done by brute force with ad hoc assumptions and/or relying on the existence of Green's functions, the representation theorem and/or the Born approximation. Using variational principles only, and without these mentioned assumptions and/or additional mathematical tools, we show that the time reversed adjoint wavefield should be defined as a premise that leads to the correct adjoint equations. This allows us to clarify mathematical inconsistencies found in previous seminal works when dealing with visco-elastic attenuation and/or odd-order derivative terms in the equation of motion. We then discuss some methodologies for the numerical implementation of the method in the time domain and to present a variational formulation for the construction of different misfit functions. We here define a new misfit travel-time function that allows us to find consensus for the long-standing debate on the zero sensitivity along the ray path that cross-correlation travel-time measurements show. In fact, we prove that the zero sensitivity along the ray-path appears as a consequence of the assumption on the similarity between data and synthetics required to perform cross-correlation travel-time measurements. When no assumption between data and synthetics is preconceived, travel-time Fr\'echet kernels show an extremum along the ray path as one intuitively would expect.

	\medskip
	
	
\end{abstract}

\newpage

\paragraph{Article Highlights}

\begin{itemize}
	\item We present the adjoint method using variational principles only and we clarify mathematical inconsistencies in previous seminal works, related to the treatment of viscoelastic attenuation
	\item We clarify the history and origin of the adjoint method and the adjoint wavefield as it is used within the seismological community, and we solve the well-known and long-standing banana-doughnut (cross-correlation) travel-time sensitivity kernel paradox
	\item We propose a new method for performing adjoint travel-time inversions without relying on waveforms, thereby bridging the gap between ray-theoretical and cross-correlation (travel-time) adjoint inversions
\end{itemize}

\section{Introduction}

Full-waveform modeling (FWM) and full-waveform inversion (FWI) are techniques that, when combined, allow to obtain more detailed images of the interior of the real Earth \citep[e.g.][]{fichtner2010book,liu2012seismic,tromp2005seismic,chen2015full,fichtner2009,tape2010seismic,chen2007full,french2014whole,zhu2015seismic,kawai2010waveform,virieux2009overview,operto2013guided,fichtner2015crust}.

FWM refers to solving the equations of motion that describe the physics of the problem using a (preferred) numerical method, e.g., Finite Difference \citep{moczo2014finite,igel2017computational}, Spectral Element \citep{komatitsch1998spectral,komatitsch1999introduction,chaljub2007spectral,Salvus}, Discontinuous Galerkin \citep{hu1999analysis,chung2006optimal,de2008discontinuous,igel2017computational,duru2022stable}, or any other (see \cite{igel2017computational} for a review), in order to compute synthetic data $(\u)$. This process is also called solving the forward problem, and it depends on the assumed model parameters $(\m)$ such as the distribution in 3D of density and seismic velocities.

FWI refers to finding the model(s) $(\m)$ that fit data (observations) $(\d)$. Thus solving a FWI is equivalent to find model parameters $(\m)$ that minimize the differences $(e)$ between data and synthetics $(e=\d-\u)$ using complete waveforms. This process is also called solving the inverse problem, and it does not have a unique solution. This, however, does not suggest that the forward problem has a unique solution either. The solution of the forward problem depends on the chosen numerical technique, mathematical simplifications and, in general, the degree of accuracy chosen/accepted by the user. Highly heterogeneous models of the Earth built by FWI are not an exact representation of the real earth.

The process of finding the minimum error solution, i.e., the model $(\m)$ that minimizes the differences between data $(\d)$ and synthetics $(\u)$, is expedited by being able to calculate the change of the residual, and/or error as defined by \cite{menke2012geophysical}, $(\delta e)$ due to a perturbation (to an initial assumed model) in the material parameters $(\delta \m)$. This is commonly done using the adjoint method, which allows us to calculate the sensitivity $(K)$ of the total wavefield error $(e)$ to a perturbation/variation of a material property $(\delta \m)$ \citep[e.g.][]{tarantola1984inversion,Tarantola1988,gauthier1986two,tromp2005seismic,fichtner2006adjoint}. This sensitivity $K$ is used to iteratively improve an initial (assumed) model $(\m_0)$. 

\begin{figure}[H]
	\begin{center}
		\includegraphics[width=0.9\textwidth]{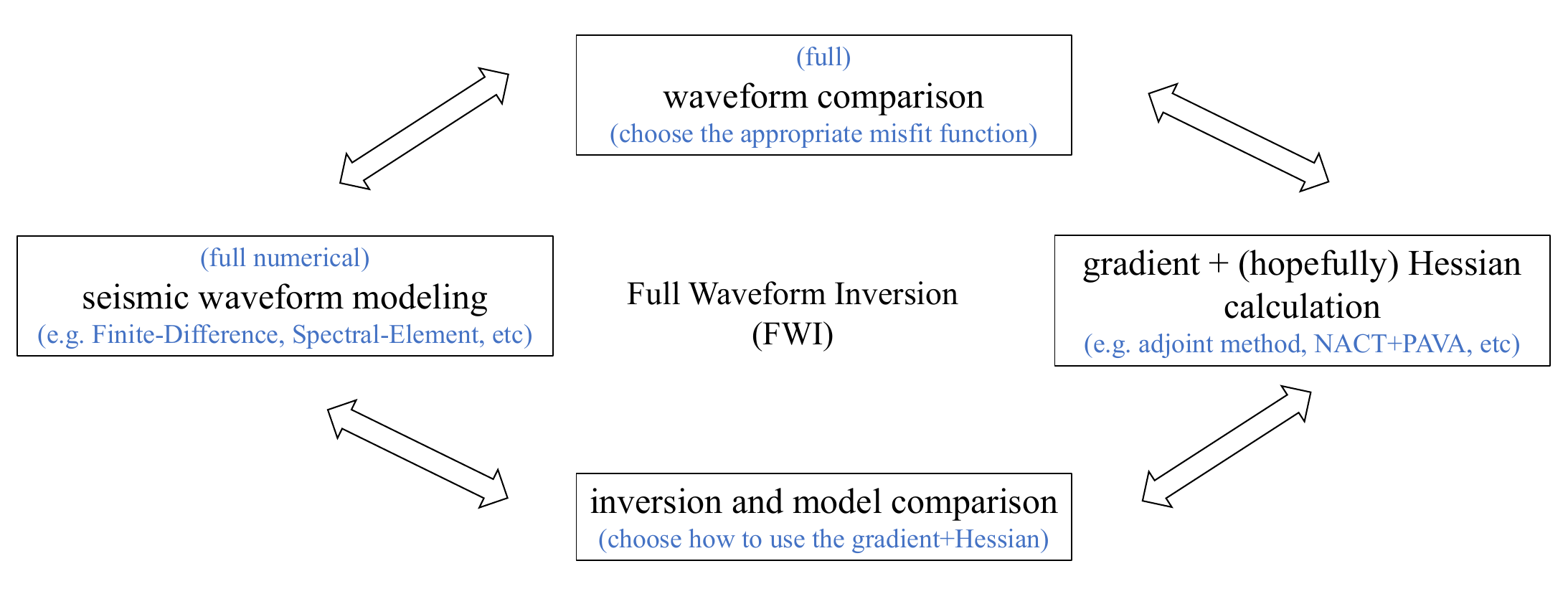}
		\caption{Full-waveform inversion process.}
		\label{Fig.FWI_process}
	\end{center}
\end{figure}

A common misunderstanding and/or misleading statement can be to relate FWI to the adjoint method only. The adjoint method provides way to compute the Fr\'echet derivative (the Hessian and/or any other higher-order functional derivative) that can be used  to minimize the (waveform) differences between data and synthetics, but it is not the only way. A schematic approach to the FWI process is presented in Fig. \ref{Fig.FWI_process}, where we can observe that FWI is an iterative process and the adjoint method helps to compute gradients (and/or Hessians) that are used to solve the minimization/optimization problem. In fact, we will show in this work that we can use the adjoint method for travel-time inversions without having to use data/synthetics waveforms.

A good starting model $(\m_0)$ and our capacity to accurately compute synthetics $(\u)$ are decisive ingredients in a FWI \citep{zhu2016building,asnaashari2013regularized,kohn2012influence,prieux2013building,asnaashari2015time,datta2016estimating,teodor2021challenges}. In seismology, realistic initial models $(\m_0)$ are obtained using large datasets and simplified theories of wave propagation like ray theory and/or normal modes \citep[e.g.][]{kennett2009seismic,cerveny2001seismic,nolet_2008,chapman2004fundamentals,Tromp1998,aki2002quantitative}. While the use of these simplified theories gives an average long wavelength earth model \citep[e.g.][]{kustowski2008anisotropic,kennett1995constraints,kennett1991traveltimes}, it is believed that only FWI is able to produce more refined images. 

This is a point of debate that still exists today: Can FWM and FWI really improve tomographic images (at all scales) to a degree not achievable by any other (simplified) technique? For instance, the first global FWI model called SAW12D \citep{li1996global}, used normal mode perturbation theory and it included across branch mode coupling which provided finite frequency sensitivity of body SH waveforms \citep{li1995comparison}. This model is distinguished from other tomographic models published (to that date) in that the rms profile with depth has more power than other models in the upper mantle/lower mantle transition region. The model provides a particularly good fit to the non-hydrostatic geoid through harmonic degree 12 (79$\%$ variance reduction), as well as good fits to observed splitting functions of S velocity sensitive mantle modes, indicating that both large-scale and small-scale features are very well constrained \citep{li1996global}. It has been shown that FWI can resolve velocity perturbations up to 10--30$\%$, far beyond those achievable by classical (ray theory) tomographic techniques \citep{liu2012seismic}. 
	
\cite{lekic2011inferring} introduced a hybrid approach that computed full 3D forward simulations combining the accuracy of the spectral element method (SEM) to solve wave propagation in a (highly) heterogeneous mantle and the speed of normal mode solutions to solve wave propagation in the relatively homogeneous core \citep{capdeville2003coupling,capdeville2003coupling2}. This combination makes wave propagation in the whole Earth much more efficient while keeping the accuracy of SEM simulation in the whole Earth since, for practical purposes, the Earth's core is assumed to have a 1D structure. \cite{french2014whole} combined this technique with an inversion using 2D sensitivity kernels \citep{li1995comparison}, that allowed them to obtain detailed images of numerous mantle plumes \citep{french2014whole,wamba2021multi}. This was the continuation of the work done by \cite{lekic2011inferring,french2013waveform} for the upper mantle, the latter with a more refined homogenized crust. \cite{french2014whole} extended the inversion to the whole mantle by adding body wavepackets down to 32s period.
	
In particular, the global full-waveform tomographic model SEMUCB \citep{french2014whole} has revealed, for the first time. the presence of fat plumes rooted at the core mantle boundary (CMB) beneath major hotspots within the perimeter of \textit{both} large low-shearwave-velocity provinces (LLSWPs). Although the existence of the two superplumes underneath Africa and the Pacific has been well known \citep{montagner1994can,Masters1996,su1994degree}, \cite{tromp2019seismic} claims that the new images obtained by the FWI model GLAD \citep{bozdaug2016global,lei2020global} show larger perturbation in shear wave speed $(>2\%)$ that are too large to be purely thermal in nature and must include a compositional contribution.  \cite{rickers2012imaging} claim that mantle plumes cannot be properly imaged by travel-times only and that FWI is fully required. 

This, however, does not prevent that global earth models are built today with simplified theories and large amounts of datasets \citep[e.g.][]{ritsema2011s40rts,moulik2014anisotropic,chang2015joint,durand2017confirmation}. The reason is because, despite the benefits, the application of FWM and FWI carries several important challenges: mainly the large amounts of (i) computational power needed and (ii) human time required to process data. Models built with simplified theories are good long wavelength models, since the approximate theory allows the retrieval of the long wavelength structure well. Nevertheless, on one hand, using large amounts of data does not guaranty that a simplified theory will better represent reality, since earthquakes and stations are not distributed uniformly around the globe but almost always in the same places, and on the other hand, using the full physics of the problem and not exploiting all data available does not guaranty a better model either. Solutions to the limitations of both approaches are still needed today.

Due to the large computational costs required, FWI was initially applied in 2D only in exploration seismics \citep[e.g.][]{gauthier1986two,igel1996waveform}. With the advancement of numerical modeling techniques and computational resources, simulations of seismic wave propagation for complicated and realistic 3D heterogeneous media became feasible at all scales \citep[e.g.][]{komatitsch1998spectral,komatitsch2002spectralA,komatitsch2002spectral,peter2011forward}, opening new avenues of applications in general sciences.

Today, the applications of FWI techniques span to several scientific fields like helioseismology \citep[e.g.][]{hanasoge2014full,hanasoge2011adjoint,mandal2017finite,hanasoge2014full2}, medical imaging \citep[e.g.][]{guasch2020full,bernard2017ultrasonic,lucka2021high,bernard2017ultrasonic,Marty2021,martiartu20203,korta2019optimal}, seismic microzonation \citep[e.g.][]{kubina2018adjoint}, tsunamis \citep[e.g.][]{zhou2019adjoint,pires2001tsunami} and non-destructive testing \citep[e.g.][]{van2021electrochemical,kordjazi2021nondestructive}. 

In the last decades, significant efforts have been made by different research groups to refine (and extend) FWI techniques. To mention some of the current areas of ongoing research are: (i) the quantification of model resolution, regularization, the computational cost reduction \citep{fichtner2011hessian,fichtner2011resolution,tape2010seismic,fichtner2015resolution,thrastarson2020accelerating,van2020accelerated,krischer2018automated,menke2015relationship,menke2020connection,guitton2012constrained}, (ii) the combination with statistical learning techniques \citep[e.g.][]{xie2021fly,van2021evolutionary,fichtner2021autotuning,gebraad2020bayesian,fichtner2019hamiltonian,fichtner2019hamiltonian2,keating2021null,fichtner2018collaborative}, (iii) the application of alternative forward solvers and (iv) the development of efficient numerical
methods to compute full-wave sensitivity kernels \citep[e.g.][]{geller1993two,kawai2010waveform,fuji2010methodology,fuji2012finite,zhao2011efficient,zhao2011efficient2,yuan2017localized,borisov2015efficient}, (v) the development of source encoding techniques that allow to compute sensitivities of several sources in a master simulation, thus considerably reducing the computational time \citep[e.g.][]{krebs2009fast,castellanos2015fast,capdeville2005towards,moghaddam2013new,ben2011efficient,schuster2011theory,tromp2019source}. It thus seems highly provable that FWM and FWI have a prominent future and the design of new techniques that can handle present limitations are needed. 

In the opinion of the author, despite the significant success and the prominent future, the mathematics of adjoint techniques , used to perform FWI, remains obscure to a large part of the scientific community. This is in part due to the complexity of the mathematics that is often used to present the technique, e.g., the use of Green's functions \citep{tarantola1984inversion,Tarantola1988,tromp2005seismic,chen2011full,chen2015full,luo1991wave,zhao2011efficient,geller1993two,zhao2006structural,chen2007full1,menke2012geophysical}, modern and elegant (but complex) operator theory \citep{Tarantola1988,fichtner2006adjoint,fichtner2010book,kennett2021exploiting}, and the Born approximation \citep{operto2013guided,geller1993two,monteiller2015three}, when an intuitive physical interpretation is not given \citep{plessix2006review,Tarantola1988,tromp2005seismic,virieux2017introduction} and/or when it is given but relying on complex mathematics like the Born approximation, Green's functions and the representation theorem \citep{tarantola1984inversion,Tarantola1988,luo1991wave,kennett2021exploiting}.

In this contribution, we show that it is not necessary to use such large and complicated mathematical machinery (Green's functions, general operator theory, the Born approximation, the representation theorem) to present and to gain physical understanding of FWI, as it is used in the geophysical community. To do so, we heavily rely on the variational formulation of the adjoint problem already presented by \cite{Liu01122006,liu2008finite}, with some key differences: 
\begin{enumerate}[(i)]
	\item We show that the nature of the adjoint wavefield must be defined at the beginning of the mathematical development. This allows us to show and clarify some mathematical inconsistencies found in several seminal works of \cite{Tarantola1988,fichtner2006adjoint}, where viscoelastic adjoint equations, and general adjoint equations \cite[e.g.][]{tromp2005seismic}, are not properly treated and/or interpreted.
	\item The relax definition of the adjoint wavefield as the time-reversed displacement. We claim that the adjoint wavefield can be defined as a linear functional applied over the displacement field, while at the same time, satisfying the original equations of motion.
\end{enumerate}
Our claims allow us to show mathematical equivalences between waveform, cross-correlation travel-time and amplitude Fr\'echet kernels, to point out mathematical inconsistencies found in previous work when dealing with viscoelastic effects and to find an important consensus to the well known and long standing banana-doughnut paradox.

In general, we aim to clarify and to introduce full waveform (adjoint) inversion to a wider part of the scientific community to whom the knowledge and use of advanced mathematical tools are not an everyday part of their scientific routine. While, however, we do understand that it can be claimed that the variational formulation is by no means easier to understand compared to the other mentioned techniques and, it is simply a matter of personal taste which one is the most appropriate one to use, we can mention that the variational formulation is the most general technique that can be used to find mathematical equations that describe (almost) all processes in nature like quantum mechanics, continuum (and enriched continuum) mechanics \citep[e.g.][]{madeo2015generalized}, relativity theory \citep[e.g.][]{lanczos2012variational}, and in particular the standard model \citep{schwartz2014quantum,burgess2007standard} in which all fundamental particles are described and three fundamental forces: Electromagnetism, strong and weak interactions are unified. Nevertheless, we do understand that these claims do not directly relate to Earth sciences; they simply emphasize the importance of the variational technique in a large number of fields outside the Earth sciences. We do hope that this motivates the reader to appreciate the importance of it. We do agree that, in the end, it is simply a matter of personal decision which one  is the better suited technique to properly understand FWI and the adjoint method, but we emphasize that it is imperative having present the mathematical inconsistencies found in previous seminal works, clarified here for the first time.

This paper is organized as follows: First we set the notation that will be used throughout the work. Since we will use techniques of the calculus of variations we continue by giving a historical introduction to the variational formulation of general initial value problems. The reader who is already familiar can directly jump to the next section where we introduce the formal adjoint (equation) method used to solve initial value problems in a tutorial way. We emphasize and insist on the tutorial way of writing because it creates an easy preamble to understand and to develop mathematical intuition for those unfamiliar with (adjoint) FWI which serves as basis to clarify mathematical inconsistencies found in seminal work. We then continue to formally present the adjoint method used by the geophysical community and its implementation using any numerical solution of the wave equation. We then present the general procedure for the construction of adjoint sources and the construction of new ones, which help us to find a solution, for the first time, of the (well known and long standing) banana-doughnut paradox. We finish with a general discussion and conclusions of this work.

\section{Historical Background}

To present the adjoint method used for the geophysical community we use a general variational formulation. The calculus of variation, developed by Euler (1707--1783) and Lagrange (1736--1813), uses the principle of least action as a general procedure for solving dynamical problems in mechanics \citep{dittrich2021development}. All modern theories of classical (and extended) continuum (and quantum) mechanics can be formulated using the Euler-Lagrange action principle. We can find equations of motion appearing in (almost) all branches of sciences by defining an action functional $I$ as an integral over the Lagrangian $\Lagr$ as follows
\begin{align}
I =  \int_{\Sigma} \int_{\Omega} \Lagr \dif \x \dif t, \qquad \text{with} \qquad \Lagr=K-U,
\label{eq.Euler_Lagrange_Action}
\end{align}
where $K$ is the kinetic energy and $U$ the potential energy over a spatial domain $\Omega$ and time interval $\Sigma$. 

Emmy Noether (1882--1935) stated a theorem \citep{noether1971invariant,Noether1918} manifesting a connection between symmetries and conservation laws for variational problems, asserting that in order to find conservation laws (equations of motion) using the Euler-Lagrange functional eq. \eqref{eq.Euler_Lagrange_Action}, it is necessary that its Lagrangian operator $\Lagr$ be symmetric (if linear) or its Gateaux derivative be symmetric (if nonlinear). This raised and immediate problem: How do we define the Lagrangian $\Lagr$ in order to find non-conservative equations of motion, e.g., the heat/diffusion equation?

\citet[p. 298]{Morse1953} attempted to answer this question by giving birth to what we today call the Adjoint Equation Method. They, however, first described their contribution as a mathematical trick with no physical justification. Their initial development attempted to find the Lagrangian of the one-dimensional oscillator with friction given by the following expression
\begin{align}
M \partial_t^2 u + R \partial_t u + k u = 0 ,
\end{align} 
were $u$ is displacement, $M$ the mass, $R$ the damping coefficient and $k$ the spring constant. Their aim was to obtain this equation from some Lagrangian functional $\Lagr$ by minimizing the action eq. \eqref{eq.Euler_Lagrange_Action}. They found that the following Lagrangian was able to do the trick
\begin{align}
\Lagr = M(\partial_t u \partial_t u^*) - \left[ \frac{1}{2} R \left(u^* \partial_t u-u\partial_t u^*\right) + k u u^*\right] ,
\end{align}
which was the same as considering the Lagrangian functional of two coordinates $(u,u^*)$. Solving for $\delta I = 0$, the resulting equations are
\begin{align}
M \partial_t^2 u + R \partial_t u + k u = 0, \qquad M \partial_t^2 u^* + R \partial_t u^* - k u^* = 0 ,
\end{align}
were the equation involving $u^*$ shows a negative frictional term. Consequently, \cite{Morse1953} were able to find equations of motion of dissipative systems using a Lagrangian formalism, meaning that they were able to treat a non-conservative system like if they were conservative. \citet[p. 299]{Morse1953} state: ``by this arbitrary trick we are able to handle dissipative systems as though they were conservative. This is not very satisfactory if an alternate method of solution is known, but it will be necessary, in order to make any progress when we come to study dissipative fields, as in the diffusion equation.''  

Early important efforts were made in order to find physical meaning to this new variable $u^*$ in the area of reactor physics for transport phenomena \citep[e.g.][]{lewins1965importance,lewins1960derivation,lewins1960approximate}. It was \cite{gurtin1963variational,gurtin1963variational1,tonti1973variational} who formally introduced the method of the adjoint equation, which is the basis of the adjoint method introduced by \cite{bamberger1979stability,bamberger1982inversion,tarantola1984inversion,Tarantola1988} in the geophysical community. 

\section{Notational Agreement}
\label{sec.notational_agreement}

We briefly review fundamental concepts of functional analysis used in this work. For a more detailed introduction we refer to \cite{zeidler1984nonlinear,zeidler1990nonlinear,griffel2002applied}.

\subsection{Inner Products}

We denote by $x_i$ the $i$th component (with $i\in\{1,2,3\}$) of the spatial coordinate, i.e., $(x_1,x_2,x_3)\equiv(x,y,z)$ and by $u^{\alpha}_i$ the $i$th component of the displacement at the $\alpha$ location $x_i^{\alpha}$. We denote by $\hat{n}_i^{\alpha}$ the normal unit vector with the $i$th component at the location $x_i^{\alpha}$. We thus can write the following equivalences
\begin{align}
	\begin{aligned}
		u^{\alpha}_i &= u(x^{\beta}_i,t) =\n^{\alpha} \cdot \u^{\alpha}=\n^{\alpha} \cdot \u(\x^{\alpha},t)=\hat{n}^{\alpha}_i \u^{\alpha} , \\
		\u^{\alpha} & = u_i^{\alpha} \hat{n}^{\alpha}_i ,
	\end{aligned}
	\label{eq.normal_vector_displ}
\end{align}
where $(\cdot)$ denotes the dot product. When we do not need it, the simply omit the location $\alpha$ of the displacement vector. 

We denote by the inner product of two scalar $u(x),v(x)$ and two vector $\u(\x),\v(\x)$ functions, as follows
	\begin{align}
		\left\langle u,v \right\rangle_x = \int_x u v \dif x, \qquad \left\langle \u,\v \right\rangle_x =  \int_{\Omega} \v^T \u \dif \x =  \int_{\Omega} \u \cdot \v \dif \x = \int_{\Omega} u_i v_i \dif \x,
	\end{align}
	respectively. Analogously, the inner product over time and space of two scalars $u(x,t),v(x,t)$ and two vectors $\u(\x,t),\v(\x,t)$ functions, as follows
	\begin{align}
		\left\langle u,v \right\rangle_{x,t} = \int_x \int_T u v \dif t \dif x , \qquad \left\langle \u,\v \right\rangle_{x,t} = \int_{\Omega} \int_T \v^T \u \dif t \dif \x =  \int_{\Omega} \int_T \u \cdot \v \dif t \dif \x = \int_{\Omega} \int_T  u_i v_i \dif t \dif \x,
	\end{align}
	respectively. 

A symmetric matrix is defined as a matrix $A$ such that $A^T=A$, where $A^T$ refers to the transposed matrix defined by $A^T_{ij}=A_{ji}$. Then it follows that for any $x,y\in \mathbb{R}^n$,
	\begin{align}
		\left\langle x,Ay \right\rangle = \left\langle A^Tx,y \right\rangle .
		\label{eq.Matrix_adjoint}
	\end{align} 

On occasions we will denote a double dot product $(:)$ of tensors to signify contraction over two adjacent indices as follows
\begin{align}
	\textbf{T} : \textbf{P} = T_{ij} P_{ij} .
\end{align}

\subsection{Bilinear Forms}

Let $X$ be a real Banach space (a complete normed vector space). A bilinear form on $X$ is a map $A:X\times X\to \mathbb{R}$ with the following properties \citep{zeidler1990nonlinear} 
	\begin{align}
	\begin{aligned}
		\begin{cases}
			A(u,\alpha v + \beta w) & = \alpha A(u,v) + \beta A(u,w) , \\
			A(\alpha v + \beta w,u) & = \alpha A(v,u) + \beta A(w,u) ,
		\end{cases}		
	\end{aligned}
	\end{align}
for all $u,v,w\in X$ and all $\alpha,\beta \in \mathbb{R}$.
 
\subsection{Linear Operators and Their Adjoints}

Let $N,M$ be vector spaces. An operator $\Linop:N\to M$ is linear if 
\begin{align}
	\Linop(ax+by) = a \Linop x + b \Linop y ,
\end{align}
for all scalar $a,b$ and all $x,y\in N$. The adjoint\footnote{the etymology of the word adjoint traces to the Latin \textit{adiungo} which means ad- (``to, towards, at'') + iungo (``join, connect, attach''). } of a linear operator comes from a generalization of the matrix inner product given in eq. \eqref{eq.Matrix_adjoint}. Let $\Linop:H\to H$ be a bounded linear operator on a Hilbert space $H$\footnote{A Hilbert space $X$ is a linear space together with a scalar product with the additional property that each Cauchy sequence is convergent. This means that a Hilbert space defines a distance function (induced by the scalar product) for which the space is a complete metric space (a convergent Cauchy sequence of points in $m$ has a limit that is also in $m$) \citep{zeidler1990nonlinear}. }, then there is a unique operator $\Linop^{\dagger}:H\to H$ such that
	\begin{align}
		\left\langle x,\Linop^{\dagger}y \right\rangle = \left\langle \Linop x,y \right\rangle \qquad \text{for all} \quad x,y\in H .
	\end{align}
	The linear and bounded operator $\adLinop$ is called the adjoint operator of $\Linop$. 

A linear bounded operator $\Linop:N\to M$ is invertible if for each $x\in M$ there is one and only one $y\in N$ such that $\Linop y = x$. The mapping $x \mapsto y$ is called the inverse of $\Linop$ and we denote it by $y=\Linop^{-1}x$. The adjoint operator $\adLinop$ satisfies \citep{griffel2002applied}
	\begin{align}
		\left(\Linop^{\dagger} \right)^{\dagger}=\Linop, \qquad  \left(\Linop^{\dagger} \right)^{-1} = \left(\Linop^{-1} \right)^{\dagger} , \qquad \left(\Linop_1 \Linop_2 \right)^{\dagger} = \left(\Linop_1\right)^{\dagger} \left(\Linop_2 \right)^{\dagger} ,
	\end{align}
	where $\Linop_1,\Linop_2$ refer to two different linear bounded operators. The adjoint operator is called self-adjoint (or Hermitian) if $\adLinop=\Linop$.

\subsection{Differential Calculus of Operators: Functional/Fr\'echet Derivative}

A continuous linear operator $\Linop:N\to M$ is said to be the Fr\'echet derivative of $\boldsymbol{f}:N\to M$ at the point $\textbf{x}\in N$ if \citep{frechet1911notion,frechet1912notion,frechet1925notion,zeidler1984nonlinear,griffel2002applied} 
	\begin{align}
		\boldsymbol{f}(\textbf{x}+\delta \boldsymbol{h})=\boldsymbol{f}(\textbf{x}) + \Linop \delta \boldsymbol{h} + o(\delta \boldsymbol{h}) \qquad \text{as} \quad \delta \boldsymbol{h}\to 0,
		\label{eq.General_Frechet_der}
	\end{align}
	where $o(\delta \boldsymbol{h})$ denotes the set of all functions $\boldsymbol{f}(\delta \boldsymbol{h})=o(\delta \boldsymbol{h})$ that satisfy $\norm{\boldsymbol{f}(\delta \boldsymbol{h})}/\norm{\delta \boldsymbol{h}} \to 0$ as $\delta \boldsymbol{h}\to 0$. 
	
	The expression $\boldsymbol{f}(\delta \boldsymbol{h})=o(\delta \boldsymbol{h})$ means that $\boldsymbol{f}$ is of a smaller order of magnitude than $\delta \boldsymbol{h}$. More precisely, a functional $\boldsymbol{f}$ is called Fr\'echet differentiable if there exists a linear continuous operator $\Linop$ such that 
	\begin{align}
		\lim_{\norm{\boldsymbol{h}}\to 0} \frac{\norm{\boldsymbol{f}(\textbf{x}+\delta \boldsymbol{h})-\boldsymbol{f}(\textbf{x})-\Linop \delta \boldsymbol{h}}}{\norm{\boldsymbol{h}}}= 0,
		\label{eq.Frechet}
	\end{align}
	where $\norm{}$ denotes the operator norm. A variation of any functional $\boldsymbol{h}$ by an infinitesimal but arbitrary amount can be represented in the form
	\begin{align}
		\delta \boldsymbol{h} (\textbf{x}) = \epsilon \boldsymbol{\eta}(\textbf{x}),
	\end{align}
	where $\epsilon$ is an infinitesimal number and $\boldsymbol{\eta}$ is an arbitrary function; we thus can write $\boldsymbol{f}(\textbf{x}+\delta \boldsymbol{h})=\boldsymbol{f}(\textbf{x}+\epsilon\boldsymbol{\eta})$. Considering that $\textbf{x}\in\Omega$, we can write the following
	\begin{align}
		\Linop \delta \boldsymbol{h} =  \boldsymbol{f}(\textbf{x}+\delta \boldsymbol{h})-\boldsymbol{f}(\textbf{x}) = \delta \boldsymbol{f} = \int_{\Omega} \left[\frac{\delta \boldsymbol{f}}{\delta \boldsymbol{h}}\right] \delta \boldsymbol{h} \dif \textbf{x} = \epsilon \int_{\Omega} \left[\frac{\delta \boldsymbol{f}}{\delta \boldsymbol{h}}\right] \boldsymbol{\eta} \dif \textbf{x} = \epsilon \int_{\Omega} \Linop \boldsymbol{\eta} \dif \textbf{x}.
		\label{eq.Frechet_int}
	\end{align}
	
	This definition of the Fr\'echet derivative implies the form of an integral linear function with kernel $\delta \boldsymbol{f}/\delta \boldsymbol{h}$ acting on a function $\boldsymbol{\eta}$. It can be understood as a continuous extension of the first total differential of a function of several variables, i.e.,
	\begin{align}
		\dif \boldsymbol{f} (\textbf{x}) = \sum_{n=1}^N \frac{\partial \boldsymbol{f}}{\partial x_n} \dif x_n .
	\end{align} 

	Note, however, that the definition given in eq. \eqref{eq.Frechet_int} is not guaranteed for arbitrary functionals $\boldsymbol{\eta}$ and $\boldsymbol{f}$. For geophysical applications, consider a (nonlinear) functional $\textbf{G}$ that maps the functional of the (earth) model space $\textbf{m}$ to the data space $\textbf{d}$, i.e,
	\begin{align}
		\textbf{d} = \textbf{G}(\textbf{m}) .
		\label{eq.Geophysical_data}
	\end{align} 
	Combining eq. \eqref{eq.Frechet_int} and eq. \eqref{eq.Geophysical_data}, we can write
	\begin{align}
		\delta \textbf{d} = \delta \textbf{G} = \int_{\Omega} \left[\frac{\delta \textbf{G}}{\delta \textbf{m}}\right] \delta \textbf{m} \dif \textbf{x} =  \int_{\Omega} \Linop \delta \textbf{m}\dif \textbf{x} = \Linop \delta \textbf{m} ,
		\label{eq.Frechet_geophysical}
	\end{align}
    where the Fr\'echet derivative kernel in eq. \eqref{eq.Frechet_geophysical} is $\Linop = \partial \textbf{G}/\partial \textbf{m}$.    
    
    In the geophysical community, the linear functional $\Linop$ in eq. \eqref{eq.Frechet_geophysical} (commonly called the sensitivity kernel and/o Fr\'echet derivative kernel) expresses the sensitivity of $\delta \textbf{d}$ to a perturbation in the earth model parameters $\delta \textbf{m}$ for an arbitrary location $\x$ in the volume $\Omega$ \citep{tarantola2005inverse}. 
    
    Note that eq. \eqref{eq.Frechet_geophysical} can be simply understood as an analogous to the discrete version of the first total differential of a function of several variables \citep{menke2012geophysical}
\begin{align}
	\Delta d_i \sum_{i=1}^M G_{ij}^{(0)}\Delta m_j, \qquad \text{with} \quad G_{ij}^{(0)}=\eval{\frac{\partial d_i}{\partial m_j} }_{m_0} .
\end{align}

\subsection{Linear Inversion vs Imaging}

If we assume that the relation between the functional $\G$ and the model space $\m$ is linear, we can write eq. \eqref{eq.Geophysical_data} as follows
\begin{align}
\d = \G \m .
\label{eq.linear_functional}
\end{align}

As a consequence, we can estimate the model parameters $\mest$ that fit observations as
\begin{align}
\mest = \G^{-1}\d .
\end{align}

However, in most cases $\textbf{G}^{-1}$ does not exists and/or it cannot be computed (the matrix $\textbf{G}$ may not be square for example). One can look to minimize the functional eq. \eqref{eq.linear_functional} by defining $\chi$ given by
\begin{align}
\chi (\m) = \frac{1}{2} \left( \d -\G \m \right)^T  \left(\d -\G \m \right) ,
\end{align}
where the superscript $T$ refers to the conjugate transpose (or adjoint transpose). The functional derivative with respect to model parameters $(\m)$ can be written as
\begin{align}
\frac{\partial \chi}{\partial \m} = - \left( \dobs -\G \m \right) \G^T .
\end{align}
Therefore an extremum of $\chi$ is found at $\partial \chi/\partial \m=0$, which leads to
\begin{align}
\mest = \left( \G^T \G \right)^{-1} \G^T \d ,
\label{eq.Linear_inversion}
\end{align}
assuming of course that $\G^T \G$ is invertible. Equation \eqref{eq.Linear_inversion} is referred in the literature as least squares inversion \citep[e.g.][]{lines1984review,tarantola1982generalized}. \cite{rice1962inverse,claerbout1964error,claerbout1985imaging} showed that $\G^T \G$ can be approximated by
\begin{align}
\G^T \G \approx \alpha \id,
\end{align} 
where $\alpha$ is some real constant and $\id$ is the identity matrix. Therefore the least square solution (eq. \eqref{eq.Linear_inversion}) can be written as
\begin{align}
\mest =\alpha^{-1} \G^T \d .
\label{eq.Imaging}
\end{align}
\cite{kawakatsu2008time} define eq. \eqref{eq.Imaging} as imaging. \cite{claerbout2008basic} states that: \textit{the adjoint (transpose of the matrix) is the first step and a part of all subsequent steps in this inversion process. Surprisingly, in practice the adjoint sometimes does a better job than the inverse! This is because the adjoint operator tolerates imperfections in the data and does not demand that the data provide full information.}

\section{Tutorial Introduction to the Adjoint Equation Method}
\label{sec.General_Adjoint_method}

\subsection{Finding Lagrangians of (Non-conservative) Initial Value Problems}

Using Lagrangians (variational principles) to find equations of motion that describe non-conservative elastic deformation is not routinely done in the geophysical community. On the contrary, non-conservative forces are inserted, when possible, in an ad hoc way in the equations of motion. There exists, however, a Lagrangian structure for non-conservative systems and there are methods for extracting important information, such as conservation laws and approximate solutions for such cases \citep[e.g.][]{oden1983variational}.

Finding a Lagrangian that describes certain initial value problem is called \textit{the inverse problem of the calculus of variations} and its implications have been well studied over the years \citep{anderson1972role,prasad1972adjoint,anderson1973application,djukic1975noether,vujanovic1989variational,davis1928inverse,vujanovic1989variational,davis1929inverse,tonti1982general,oden1983variational}. \cite{gurtin1963variational,gurtin1963variational1,Gurtin1964,gurtin1964variational} introduced an approach for constructing variational principles for the linear theory of viscoelasticity, that involves reducing the given initial value problem to an equivalent boundary-value problem using the idea of convolutions, what is known as \textit{a convolution bilinear form}. The resulting Euler-Lagrange equations are integro-differential equations equivalent to the original partial differential equations, and contain the initial conditions implicitly. Variational self-adjointness with respect to the convolution bilinear form, are necessary and sufficient conditions for the existence of a Lagrangian \citep{santilli1977necessary}. 

\cite{tonti1973variational} presented a general and simple physical and mathematical understanding of variational initial value methods using the convolution bilinear form. According to \cite{tonti1973variational}, early methods can be classified in three groups:
\begin{enumerate}
	\item The method of formally self-adjoint operators.
	\item The method of adding the adjoint equation.
	\item Gurtin's method of convolutions \citep{gurtin1963variational,gurtin1963variational1,gurtin1964variational,Gurtin1964}.
\end{enumerate}

We here will not attempt to describe how to find Lagrangians of initial value problems, but how the ideas introduced by \cite{gurtin1963variational,gurtin1963variational1,tonti1973variational} are the basis of the adjoint method used for Earth sciences applications. To do so, we only focus on the methods of self-adjoint operators and the adjoint equation which, for our purposes, are basically the same. We step aside of Gurtin's method of convolutions because it does not lead to the same forward and adjoint equations, which is one of the main advantages for geophysical applications as we will see.

\subsection{Initial Value Problems and Their Adjoints}
\label{sec.Inital_value_problems_adjoints}

To present the method introduced by \cite{tonti1973variational} and to show its direct connection with the adjoint method in seismology, we present the derivation of the adjoint equation method to gradually complex problems. 

\paragraph{Initial Value Problem 1:}

Consider the following \textit{initial value} problem
\begin{align}
	\Linop u = \partial_t^2 u (t) -f(t) = 0, \quad \text{with} \quad u(0) = 0, \quad \partial_t u(0) = 0 , \quad 0\leq t \leq T,
	\label{eq.IVP1}
\end{align}
where the symbol $\Linop$ refers to a general linear operator over $u$. If we compute a weighted average of eq. \eqref{eq.IVP1} using a continuous function $v$, after integrating by parts twice we can write 
\begin{align}
\begin{aligned}
	 \left\langle v,\Linop u \right\rangle_t  = &  \int_0^{T} v (t) \left[\partial_t^2 u(t) - f (t)\right] \dif t  = \int_0^{T} \left[\partial_t^2 v (t)  - \frac{v(t)}{u(t)} f(t) \right] u(t) \dif t \\
	& + \eval{v(t) \partial_t u(t)}_0^{T} - \eval{\partial_t v(t)   u(t)}_0^{T}  =  \left\langle \adLinop v,u \right\rangle_t =  0 ,
	\label{eq.IVP_adjoint_eq_calc}
\end{aligned}
\end{align}
which leads to the following \textit{boundary value} problem
\begin{align}
	\adLinop v = \partial_t^2 v (t) - g (t)=0, \quad \text{with} \quad v(T) = 0, \quad \partial_t v(T) = 0, \quad g (t)= v(t)/u(t)f(t) .
\label{eq.IVP2}
\end{align}
where the symbol $\adLinop$ refers to the adjoint operator of $\Linop$ (see Sec. \ref{sec.notational_agreement}). Note that we have converted the initial value problem eq. \eqref{eq.IVP1} $(u(0) = \partial_t u(0) = 0)$ to a boundary value problem eq. \eqref{eq.IVP2} with final conditions $(v(T) = \partial_t v(T) = 0)$. In formal mathematical terms, we have found the \textit{adjoint equation} to the initial value problem eq. \eqref{eq.IVP1} using the following \textit{bilinear form}
\begin{align} 
  \left\langle u,v \right\rangle_t = \int_{0}^{T} u (t) v (t) \dif t ,
  \label{eq.bilinear_form}
\end{align}
which can be formally written as follows
\begin{equation}
\Linop u = \begin{cases}
\partial_t^2 u (t) = f(t), \\
 u(0) = \partial_t u(0) = 0 ,
\end{cases} \qquad \text{whose adjoint equation is} \qquad
\adLinop v  = \begin{cases}
\partial_t^2 v (t) = g(t), \\
v(T) = \partial_t v(T) = 0 .
\end{cases}
\label{eq.value_problem1}
\end{equation}

Note that we have omitted the definition of the source $g(t)$, since at this point it is unimportant for our purposes. The operator eq. \eqref{eq.IVP1} is defined as self-adjoint because its formal part, i.e. $\partial_t^2$ is equal to the formal part of its adjoint, but the domain of the operator is neither equal nor contained in that of its adjoint, and therefore it is not symmetric \citep{dunford1988linear}. The problem that arises is the physical justification of the boundary constraints of the adjoint equation, since we have to satisfy the ending conditions at time $t=T$. This clearly has no physical sense and it can rarely be fulfilled. 

\cite{gurtin1963variational,gurtin1963variational1,Gurtin1964} was the pioneer that proposed to solve this problem by applying a convolution bilinear form to the diffusion equation in combination with the Laplace transform, being able to find variational formulation for non-conservative problems which adjoint equation was completely different from the initial one. In response to this complex and intricate mathematical procedure, \cite{tonti1973variational} showed that the use of the Laplace transform proposed by \cite{Gurtin1964} was not necessary and proposed a formal methodology to find the variational formulation of the diffusion equation. 

The contribution made by \cite{tonti1973variational} consisted in finding the adjoint equations with respect to the following \textit{convolution bilinear form}
\begin{align}
\left\langle u,v \right\rangle^c_t = \int_{0}^{T} v (T-t) u (t) \dif t .
\label{eq.convolution_bilinear_form}
\end{align}
If we apply eq. \eqref{eq.convolution_bilinear_form} to the initial value problem given in eq. \eqref{eq.IVP1} and compute the adjoint equation, as done in eq. \eqref{eq.IVP_adjoint_eq_calc}, we can write
\begin{align}
\begin{aligned}
\left\langle v,\Linop u \right\rangle_t^c = &  \int_0^{T} v (T-t) \left[\partial_t^2 u(t) - f (t)\right] \dif t  =  \int_0^{T} \left[ \partial_t^2 v (t) - \frac{v(T-t)}{u (T-t)} f(t) \right] u (T-t) \dif t \\
& +  \eval{v(T-t) \partial_t u (t) }_0^{T} - \eval{\partial_t v(T-t)   u (t)}_0^{T}  = \left\langle \adLinop u, v \right\rangle_t^c =  0 .
\label{eq.IVP_conv_adjoint_eq_calc}
\end{aligned}
\end{align}
Note that the boundary term can be written as follows
\begin{align}
\begin{aligned}
 \eval{v(T-t) \partial_t u (t) }_0^{T} = v(0)\partial_t u(T) - v(T)\partial_t u(0), \quad \eval{\partial_t v(T-t)   u (t)}_0^{T} = \partial_t v(0)   u (T) - \partial_t v(T)   u (0) ,
\end{aligned}
\end{align}
which leads to the following \textit{initial value} problem
\begin{align}
	\adLinop v = \partial_t^2 v (t) - g (t)=0, \quad \text{with} \quad v(0) = 0, \quad \partial_t v(0) = 0, \quad g (t)= v(T-t)/u (T-t)f(t) .
	\label{eq.IVP3}
\end{align}
Note that by considering the convolution bilinear form eq. \eqref{eq.convolution_bilinear_form}, we have converted the initial value problem eq. \eqref{eq.IVP1} $(u(0)  = \partial_t u(0) = 0)$ into another initial value problem eq. \eqref{eq.IVP3} $(v(0) = \partial_t v(0) = 0)$ with the same mathematical expression with the only difference in the source term. Formally this can be written as follows 
\begin{equation}
\Linop^{\dagger}v = \begin{cases}
\partial_t^2 v (t) = \overleftarrow{g} (t), \\
 v(0) = \partial_t v(0) = 0,
\end{cases}
\label{eq.formal_IVP3}
\end{equation}
where, borrowing notation from \cite{Tarantola1988}, $\overleftarrow{g} (t)= v(T-t)/u (T-t) f(t)$ and we have used the commutative property of the convolution $(u*v=v*u)$. Note that we used $\overleftarrow{g}$ to emphasize that the source in the adjoint equation (adjoint source) runs backward in time. 

\paragraph{Initial Value Problem 2:}

Consider now the scalar wave equation with homogeneous initial conditions, given by the following expression
\begin{equation}
\Linop u = \begin{cases}
\partial_t^2 u = c^2 \partial_x^2 u + f , \\
u(x,0) = \partial_t u(x,0) = 0 .
\end{cases}
\label{eq.scalar_1dD_wave}
\end{equation}
where $c$ is the wave speed. If we apply the following bilinear form 
\begin{align}
	  \left\langle u,v \right\rangle_{x,t} = \int_{\Omega} \int_{0}^{T} u (x,t) v (x,t) \dif t \dif x ,
	  \label{eq.Spatial_Bilinear_Form}
\end{align}
to the initial value problem eq. \eqref{eq.scalar_1dD_wave},  i.e., $\left\langle v, \Linop u \right\rangle_{x,t}$, and perform the necessary algebraic calculations just as done in eq.  \eqref{eq.IVP_adjoint_eq_calc}, we obtain the following adjoint equation
\begin{equation}
\adLinop v = \begin{cases}
\partial_t^2 v = c^2 \partial_x^2 v + \overrightarrow{g} , \\
v(x,T) = \partial_t v(x,T) = 0 ,
\end{cases}
\label{eq.Bilinear_1D_wave}
\end{equation}
where we have omitted the definition of $g$ and instead used $\overrightarrow{g}$ to emphasize that the adjoint source of eq. \eqref{eq.Bilinear_1D_wave} runs forward in time. Note that the adjoint equation obtained has the same mathematical structure as the original scalar wave equation. However, we have converted the initial value problem $(u(x,0) =\partial_t u(x,0)=0)$ into a boundary value problem with ending conditions $(v(x,T)=\partial_t v(x,T) = 0)$, without being able to provide physical justification. Following \cite{tonti1973variational}, consider the following convolution bilinear form
\begin{align}
\left\langle u,v \right\rangle^c_{x,t} = \int_{\Omega} \int_{0}^{T} u (x,t) v (x,T-t) \dif t \dif x .
\label{eq.Spatial_Convoltion_Bilinear_Form}
\end{align}
Applied to the scalar wave equation eq. \eqref{eq.scalar_1dD_wave} we find 
\begin{equation}
\adLinopc v = \begin{cases}
\partial_t^2 v = c^2 \partial_x^2 v + \overleftarrow{g} , \\
v(x,0) = \partial_t v(x,0) = 0 .
\end{cases}
\end{equation}
Note that the adjoint equation is the same initial wave equation with the only difference in the source term, running backward in time $(\overleftarrow{g})$. We are able to obtain this result by the commutative property of the convolution, i.e.,
\begin{align}
\int_{\Omega}\int_{\Sigma} u(\x,t) v(\x,T-t) \dif t \dif \x= \int_{\Omega} \int_{\Sigma} v(\x,t) u(\x,T-t) \dif t \dif \x.
\end{align}   

\paragraph{Initial Value Problem 3:}

\cite{tonti1973variational} showed that an important case to study is the following initial value problem
\begin{align}
	\Linop u =\partial_t u -f = 0,\quad \text{with} \quad u(0)=0.
\end{align}
The adjoint equations using the bilinear $\left\langle v, \Linop u \right\rangle_{t}$ (eq. \eqref{eq.bilinear_form}) and convolution bilinear $\left\langle v,\Linop u \right\rangle^c_{t}$ (eq. \eqref{eq.convolution_bilinear_form}) forms are given by\footnote{note that we explicitly write calculations here to allow readability, since these results have profound consequences as we will see in the next sections.}
\begin{align}
	\begin{aligned}
	\left\langle v, \Linop u \right\rangle_t & = \int_0^T \left[\partial_t u (t) - f (t)\right]  v (t) \dif t \\
	& = \int_0^T  \left[- \partial_t v (t) -  v(t)/u(t) f(t)\right] u(t) \dif t+ u(T)v(T) - u(0)v(0) , \\
	\left\langle v, \Linop u \right\rangle^c_t & = \int_0^T \left[\partial_t u (t) - f(t)\right]  v (T-t)\dif t = \int_0^T \left[\partial_t u (T-t) - f(T-t)\right]  v (t)\dif t \\
    & = \int_0^{T}  \left[\partial_t v (t)  - v (t)/u (T-t)f(T-t)\right] u (T-t) \dif t + u(T)v(0) - u(0)v(T) ,
    \label{eq.Tonti_EX3}
	\end{aligned}
\end{align}
which leads to
\begin{equation}
\adLinop v = \begin{cases}
- \partial_t v = \overrightarrow{g}, \\
v(T) = 0 .
\end{cases}
\qquad  \adLinopc v =
\begin{cases}
\partial_t v = \overleftarrow{g}, \\
v(0) = 0 ,
\end{cases}
\end{equation}
where we have made use of the property of the convolution $\partial_t(u*v)=\partial_t u * v = u * \partial_t v$. To understand eqs. \eqref{eq.Tonti_EX3}, simply note that $u(T-t) \dif t=-u(t')\dif t'$ with $T-t=t'$. 

Just like in the case of the scalar wave equation, the use of the convolution bilinear form allows us to find the original initial value problem with the difference in the source term only. This proves that the convolution bilinear form makes symmetric the operator $\partial_t$ with homogeneous initial conditions. This has profound consequences when considering dissipation effects in wave propagation. 

\paragraph{Initial Value Problem 4:}

Consider the initial value problem of the scalar wave equation with dissipation given by the following expression
\begin{equation}
\Linop u = \begin{cases}
\partial_t^2 u = c^2 \partial_x^2 u + \alpha \partial_t u + f , \\
u(x,0) = \partial_t u(x,0) = 0 ,
\end{cases}
\end{equation}
where $c$ is the wave speed and $\alpha$ is a dissipation constant. The adjoint equations using the bilinear $\left\langle v,\Linop u \right\rangle_{x,t}$ and convolution bilinear $\left\langle v,\Linop u \right\rangle^c_{x,t}$ forms are respectively given by
\begin{equation}
\adLinop v = \begin{cases}
\partial_t^2 v = c^2 \partial_x^2 v - \alpha \partial_t v + \overrightarrow{g} , \\
v(x,T) = \partial_t v(x,T) = 0 ,
\end{cases}
\qquad  \adLinopc v =
\begin{cases}
\partial_t^2 v = c^2 \partial_x^2 v + \alpha \partial_t u + \overleftarrow{g} , \\
v(x,0) = \partial_t v(x,0) = 0 .
\end{cases}
\end{equation}

The advantages of using the convolution bilinear form eq. \eqref{eq.Spatial_Convoltion_Bilinear_Form} to find variational formulations of initial value problems become evident at this point, since we are able to find adjoint equations which resemble the exact original equation with the same initial conditions even under the presence of dissipation effects, unlike the bilinear form eq. \eqref{eq.bilinear_form} which converts the initial value problem into a boundary value problem with boundary conditions with no physical meaning and changing the sign of the dissipation terms in the adjoint equations. Note that these results hold for any odd-order term in an initial value problem.      

\paragraph{Initial Value Problem 5:}

Consider the following \textit{initial value} problem
\begin{align}
	\Linop u = u \partial_t u = 0, \quad \text{with} \quad u(0) = 0 , \quad 0\leq t \leq T,
	\label{eq.IVP5}
\end{align}
where the symbol $\Linop$ refers to a general operator over $u$. The adjoint equations using the bilinear $\left\langle v^2, \Linop u \right\rangle_{t}$ and convolution bilinear $\left\langle v^2,\Linop u \right\rangle^c_{t}$ forms are given by the following
\begin{align}
	\begin{aligned}
	\left\langle v^2, \Linop u \right\rangle_t & = \int_0^T \left[u (t) \partial_t u (t)\right]  v^2 (t) \dif t \\
	 & =  - \int_0^T \left[v (t) \partial_t v (t)\right]  u^2 (t) \dif t + \frac{u^2(T)v^2(T)}{2} - \frac{u^2(0)v^2(0)}{2} \\
	& = - \left\langle u^2, \Linop v \right\rangle_t+ \frac{u^2(T)v^2(T)}{2} - \frac{u^2(0)v^2(0)}{2} = 0, \\
	\left\langle v^2, \Linop u \right\rangle^c_t & = \int_0^T \left[u (t) \partial_t u (t)\right]  v^2 (T-t) \dif t \\
	& =  \int_0^T \left[v (T-t) \partial_t v (T-t)\right]  u^2 (t) \dif t + \frac{u^2(T)v^2(T)}{2} - \frac{u^2(0)v^2(0)}{2} \\
	& = \left\langle u^2, \Linop v \right\rangle_t^c + \frac{u^2(T)v^2(0)}{2} - \frac{u^2(0)v^2(T)}{2} = 0,
	\end{aligned}
\end{align}
which leads to
\begin{equation}
\adLinop v = \begin{cases}
- v \partial_t v = 0, \\
v(T) = 0 .
\end{cases}
\qquad  \adLinopc v =
\begin{cases}
v \partial_t v = 0, \\
v(0) = 0 .
\end{cases}
\end{equation}

\paragraph{Initial Value Problem 6:}

Consider the following initial value problem
\begin{align}
	\Linop u =\partial_x u(x) -f(x) = 0,\quad \text{with} \quad u(0)=u(L)=0 \quad \text{and} \quad x\in[0,L].
\end{align}

While this case can be considered similar to the previous case, note that, unlike the temporal initial condition at rest $u(t=0)=0$, the spatial initial condition is a boundary condition. This means that both adjoint equations using the bilinear $\left\langle v, \Linop u \right\rangle_{x}$ (eq. \eqref{eq.bilinear_form}) and convolution bilinear $\left\langle v,\Linop u \right\rangle^c_{x}$ (eq. \eqref{eq.convolution_bilinear_form}) forms are physically justified. After a little algebra, we can write the adjoint equations as follows
\begin{equation}
\adLinop v = \begin{cases}
- \partial_x v = g_1, \\
v(L) = 0 .
\end{cases}
\qquad  \adLinopc v =
\begin{cases}
\partial_x v = g_2, \\
v(0) = 0 .
\end{cases}
\end{equation}

Note that the use of the convolution bilinear form allows us to find the original initial value problem with the difference in the source term only.

\paragraph{Physical interpretation:}

The physical meaning of the adjoint equation obtained with the convolution bilinear form becomes evident at this point. \cite{tonti1973variational} found that in order to answer the question of the physical meaning of the functional $v$, while not having previous information of it, it seems natural to define
\begin{align}
v(x,t) \stackrel{\text{def}}{=} u(x,t) ,
\label{eq.Adjonint_variable_V0}
\end{align}  
therefore the functional $v$, when evaluated by the convolution bilinear form eq. \eqref{eq.convolution_bilinear_form}, whose physical meaning did not appear to be found up to now, it is simply the original variable $u$ evaluated at the instant $(T-t)$. Note, however, that this \textit{is an arbitrary choice} because we do not impose any other knowledge on the variable $v$. The time reversed property $(T-t)$ comes from the definition of the convolution bilinear operator. We could rightfully impose 
\begin{align}
	v(x,t)\stackrel{\text{def}}{=} \mathcal{F} u(x,t) ,
\end{align}
where $\mathcal{F}$ is \textit{any} certain linear functional (that satisfies the equation in study), and the properties previously explained will still hold. Adjoints of some common linear operators are summarized in Table \ref{tb.adj_op}.
	\begin{table}
	\begin{center}
	\caption{Adjoints of some linear operators}
	{\renewcommand{\arraystretch}{1.5} 
	\begin{tabular}{ | c | c | c |}
		\hline
		$\Linop $ & $\adLinop $ & $\adLinopc $  \\ \hline
		$\partial_x $ and $\partial_t $ & $-\partial_x $ and $-\partial_t $ & $\partial_x $ and $\partial_t $  \\ \hline
		$\partial_x  \partial_y $ & $\partial_y  \partial_x $ & $\partial_y  \partial_x $ \\ \hline
		$\partial^2_x $ and $\partial^2_t $ & $\partial^2_x $ and $\partial^2_t $ & $\partial^2_x $ and $\partial^2_t $  \\ \hline
	\end{tabular} }
\label{tb.adj_op}
	\end{center}
	\end{table}

\section{The Adjoint (Equation) Method in Seismology}

For geophysical applications, with the help of the adjoint method, we look to minimize the waveform misfit or error function $\E$ \citep{menke2012geophysical} given by the differences between data $(\d)$ and synthetics $(\u)$ defined as a quadratic loss function as follows 
\begin{align}
\begin{aligned}
\E & = \frac{1}{2} \sum_{r} \int_{0}^{T} \left(\d - \u \right)^2 \delta (\x-\x^r) \dif t, \\
\text{E}_i & =  \frac{1}{2} \sum_{r} \int_{0}^{T} \left(d_i - u_i \right)^2 \delta (x_i-x_i^r) \dif t,
\end{aligned}
\label{eq.chi_squared}
\end{align} 
where $d_i$ represents $i$th spatial component of the data (displacement, velocity or acceleration seismograms) at certain receiver locations $(x^r_i)$ and $u_i$ the synthetics at the same position and the interval $[ 0,T]$ denotes the time series of interest. 

We assume that the data can (accurately enough) be described by the (linear) elastic wave equation with homogeneous initial conditions given by
\begin{equation}
\begin{cases}
	\Linop \textbf{u} = \den \partial_t^2 \textbf{u} - \dvg \stress - \f = 0 , \qquad \text{with} \quad \left. \stress \cdot \hat{\text{n}} \right|_{\Gamma} = 0 , \\
	\u(\x,0) = \partial_t \u(\x,0) = 0 ,
	\end{cases}
	\label{eq.Elastic_wave}
\end{equation}
where $\den$ is the material density, $\dvg$ is the divergence operator $(\partial_i \sigma_{ij})$, $\hat{\text{n}}$ is unit outward normal on the surface $(\Gamma)$ and $\stress$ the second-order stress tensor defined by \citep{Slawinski2010}
\begin{align}
	\sigma_{ij} = \mathbb{C}_{ijkl} \varepsilon_{kl}, \qquad \varepsilon_{ij} = \frac{1}{2} \left(\partial_i u_j + \partial_j u_i\right) ,
    \label{eq.stress_and_strain}
\end{align}
where $\mathbb{C}_{ijkl}$ is the fourth-order tensor of elastic constants, $\varepsilon_{ij}$ the second-order strain tensor. In symbol notation we can write
\begin{align}
   \stress  = \mathbb{C} : \boldsymbol{\varepsilon}, \qquad \boldsymbol{\varepsilon} = \frac{1}{2} \left( \nabla \u + (\nabla \u)^T\right) .
\end{align}
Due to the symmetry of the strain tensor $\varepsilon_{ij}=\varepsilon_{ji}$, the stress tensor can be written as follows
\begin{align}
	\sigma_{ij} = \mathbb{C}_{ijkl} \nabla_k u_l ,\qquad \text{or in symbol notation} \quad  \stress  = \mathbb{C} : \nabla \u .
\end{align} 
The term $f$ in eq. \eqref{eq.Elastic_wave} represents a point source given by \citep{Tromp1998,aki2002quantitative}
\begin{align}
	\f = - \text{M} \cdot \nabla \delta(\textbf{x}-\textbf{x}^s) S(t),
	\label{eq.Moment_tensor_source}
\end{align}
where $\text{M}$ is the moment tensor, $\textbf{x}^s$ the position of the source, and $S(t)$ certain source time function. 

\paragraph{Lagrange Minimization}

We here employ the method of Lagrange multipliers to minimize the error function $\E$ (eq. \eqref{eq.chi_squared}) subject to the constraint that the data can be described by the wave equation eq.  \eqref{eq.Elastic_wave}. This minimization problem can be written as follows
\begin{align}
	\chi = \int_{0}^{T} \int_{\Omega}  \left[ \frac{1}{2} (\textbf{d} -\textbf{u})^2 \sum_{r} \delta (\textbf{x}-\textbf{x}^r) - \Lm \left(\den \partial_t^2 \textbf{u} - \dvg \stress - \textbf{f}\right)\right] \dif^3 \x \dif t ,
	\label{eq.Adjoint_General_Functional}
\end{align}
where $\Omega$ is the volume of certain Earth model, $\Lm$ is a vector called the Lagrange multiplier. Note that we call the Lagrange multiplier $\Lm$ instead of the common notation $\lambda$ to distinguish it from the Lam\'e parameter. 

Note that the expression relating the wave equation in eq. \eqref{eq.Adjoint_General_Functional} resembles a variational problem with respect to a certain chosen bilinear form, i.e, $\left\langle \Lm,\Linop \u \right\rangle_{\x,t}$ and/or $\left\langle \Lm,\Linop \u \right\rangle^c_{\x,t}$. In general operator form, and using the convolution bilinear form, we thus can write eq. \eqref{eq.Adjoint_General_Functional} as follows
\begin{align}
\chi = \frac{1}{2}\left\langle \e,\e \right\rangle_{\x,t} + \left\langle \Lm ,\Linop(\m) \u - \f \right\rangle^c_{\x,t} , \qquad \text{with} \quad \e = (\textbf{d} -\textbf{u}) \sum_r\delta(\x-\x^r) ,
\label{eq.operator}
\end{align}
where $\m$ refers to the model parameters and $\Linop$ to the linear wave equation operator eq. \eqref{eq.Elastic_wave}. 

\paragraph{Karush–Kuhn–Tucker Conditions}

To minimize the difference between data and synthetics $(\e)$ with the elastic wave equation eq. \eqref{eq.Elastic_wave} as a constraint, the Karush–Kuhn–Tucker (KKT) conditions must be satisfied \citep{boyd2004convex}. KKT conditions in this case can be written as follows
\begin{align}
\left\{
\begin{aligned}
\partial_\m \chi(\m,\Lm,\u,\d) & = 0 , \\
\partial_{\Lm} \chi(\m,\Lm,\u,\d) & = 0 , \\
\partial_\d \chi(\m,\Lm,\u,\d) & = 0 , \\
\partial_\u \chi(\m,\Lm,\u,\d) & = 0 ,
\end{aligned} \right.
\label{eq.KKT_cond}
\end{align}
which lead to the following expressions \citep{menke2012geophysical}
\begin{align}
	\begin{aligned}
	\partial_\m \chi = \left\langle \Lm,\delta_{\m} \Linop \u \right\rangle^c_{\x,t} = 0, \qquad \partial_\Lm \chi = \Linop \u - \f= 0, \qquad \partial_\u \chi = - \e + \Linop^* \Lm = 0 , \qquad \partial_\d \chi = 0.
	\end{aligned}
\end{align}

If we assume (or guarantee) that the elastic wave equation eq. \eqref{eq.Elastic_wave} is satisfied, i.e., the data are well enough described by it, and that the adjoint wavefield $\Lm$ (which runs backward in time due to the convolution bilinear form) satisfies $\Linop^* \Lm = \e$, which implies that $\partial_\Lm \chi = \partial_\u \chi = 0$, then one can note that the expression for $\partial_\m \chi$ represents the Fr\'echet derivative of $\chi$. Additionally, note that imposing $\partial_\Lm \chi = \partial_\u \chi = 0$ implies that $ \left\langle \Lm ,\Linop(\m) \u - \f \right\rangle^c_{\x,t}=0$, which allow us to write
\begin{align}
	\chi = \frac{1}{2}\left\langle \e,\e \right\rangle_{\x,t} = \E .
\end{align} 
Thus the expression
\begin{align}
	\partial_\m \chi=\partial_\m \E=\left\langle \Lm,\delta_{\m} \Linop \u \right\rangle^c_{\x,t} ,
\end{align}
can be understood as the Fr\'echet derivative of the error or misfit function $\E$ (see eq. \eqref{eq.chi_squared}) \citep{menke2012geophysical}. To illustrate the presented theory, in the next section we first simplify the problem to the 1D elastic case.

\subsection{The 1D Elastic Adjoint Equations}

We consider the 1D shear wave equation given by the following expression 
\begin{align}
\Linop u = \rho \partial^2_t u - \partial_x \left(\mu \partial_x u\right) - f \delta(x-x^s) =0 , \qquad \text{with} \qquad u(x,0)=\partial_t u(x,0) = 0,
\label{eq.1D_wave_equation}
\end{align}
where $\mu$ is the shear modulus, $\rho$ the density, $x_s$ the source location and $f$ certain time dependent function. We can write the constrained Lagrange minimization problem eq. \eqref{eq.Adjoint_General_Functional} as follows
\begin{align}
\chi = \int_{0}^{T} \int_{\Omega}  \left\{ \frac{1}{2} (d  - u)^2 \sum_{r} \delta (x-x^r) - \Lm \left(\rho \partial^2_t u - \partial_x \left(\mu \partial_x u\right) - f  \right)\right\}\dif x \dif t ,
\label{eq.1D_Constrained_Action}
\end{align}
where the vector Lagrange multiplier $\Lm(x,t)$ (which minimizes the functional $\chi$) remains to be determined. Note that the expression containing the wave equation in eq. \eqref{eq.1D_Constrained_Action} resembles a variational problem applied to the scalar wave equation eq. \eqref{eq.scalar_1dD_wave}. 

Taking the first variation $\delta$ of the functional eq. \eqref{eq.1D_Constrained_Action} with respect to model parameters $\m=(\rho,\mu)$, the force term $f$ and displacement $u$,  and using KKT conditions eq. \eqref{eq.KKT_cond} and assuming that the data are accurately enough described by the wave equation, i.e., $\partial_\Lm \chi = \Linop \u - \f= 0$, we can write the following
\begin{align}
\begin{aligned}
\delta_\m \chi = &  -  \int_{0}^{T}  \int_{\Omega}  \Lm \left[\delta \rho \partial_t ^2 u - \partial_x \left( \delta \mu \partial_x u\right) - \delta f \right] \dif x  \dif t  = 0 ,\\
\delta_u \chi = &  \int_{0}^{T} \int_{\Omega}  \sum_{r} \delta(x-x^r) (d  - u)\delta u  \dif x  \dif t   - \int_{0}^{T} \int_{\Omega} \Lm \left[\rho \partial_t ^2 \delta u - \partial_x\left( \mu \partial_x \delta u\right)\right] \dif x \dif t = 0 .
\end{aligned}
\label{eq.1D_General_Constrained_Action}
\end{align}
It is important to note that eq. \eqref{eq.1D_General_Constrained_Action} ignores the second-order terms, meaning that, it is equivalent to the Born approximation. Note also that assuming $\Lm=\Lm(x,t)$, the expressions involving the Lagrange multiplier $\Lm$ for $\delta_u \chi$ in eqs. \eqref{eq.1D_General_Constrained_Action} are similar to the bilinear form eq. \eqref{eq.Spatial_Bilinear_Form} applied to the 1D scalar wave equation eq. \eqref{eq.scalar_1dD_wave}. The adjoint equations for eq. \eqref{eq.1D_General_Constrained_Action} will convert the initial value problem given by eq. \eqref{eq.1D_wave_equation} into a boundary value problem, just like in eq. \eqref{eq.Bilinear_1D_wave}. It is thus convenient to choose $\Lm=\Lm(x,T-t)$ to obtain an adjoint equation with correct initial conditions. Therefore, integrating by parts (twice) the terms involving spatial and temporal derivatives of the displacement $u$ and the variation $\delta u$, we obtain
\begin{align}
\begin{aligned}
\delta_u \chi = & \int_{0}^{T} \int_{\Omega} \sum_{r} \delta(x-x^r) (d  - u)\delta u \dif x \dif t \\
& - \int_{0}^T \int_{\Omega} \left[\rho \partial^2_t \Lm - \partial_x \left(\mu \partial_x\Lm\right) \right] \delta u \dif x \dif t = 0 , \qquad \text{with} \qquad \Lm(x,0)=\partial_t \Lm(x,0) = 0 ,
\end{aligned}
\label{eq.1D_variation_action_integration_parts} 
\end{align}
where again $\Lm=\Lm(x,T-t)$. In the absence of perturbations in the model parameters $(\delta \rho = \delta \mu = \delta f = 0)$, i.e., $\delta_\m \chi =0$, the variation in the action $\delta \chi$ is reduced to
\begin{align}
\begin{aligned}
\delta \chi = \delta_u \chi = &  \int_{0}^{T} \int_{\Omega} \left\{\sum_{r} \delta(x-x^r) (d  - u) - \left[\rho \partial^2_t \Lm - \partial_x \left( \mu \partial_x\Lm\right) \right]\right\}\delta u \dif x \dif t =  0, \\
& \text{with} \quad \Lm(x,0)=\partial_t \Lm(x,0) = 0.
\end{aligned}
\label{eq.1D_reduced_variation_action_integration_parts} 
\end{align}
The variation in the action $\chi$ is stationary $(\delta_\m \chi + \delta_{\Lm} \chi + \delta_u \chi =0)$ if
\begin{flalign}
\rho \partial^2_t \Lm - \partial_x \left(\mu \partial_x\Lm\right)  = \sum_{r}  \delta (x^r - x)(d  - u) , \qquad \text{with} \qquad \Lm(x,0)=\partial_t \Lm(x,0) = 0 .
\label{eq.1D_adjoint_equation_with_lambda}
\end{flalign}
Note that the expression involving the Lagrange multiplier $(\Lm)$ is equal to the initial wave equation eq. \eqref{eq.1D_wave_equation} with a different source term. Without having previous information of the adjoint wavefield, it seems reasonable to assume
\begin{flalign}
\Lm (x,T-t) \stackrel{\text{def}}{=} u (x,T-t) = u^* (x,t),
\label{eq.Reversed_displacement_adjoint}
\end{flalign}
thus, the adjoint wavefield $u^*$ is equal to the time-reversed wavefield $u (x,T-t)$. Note that we name the adjoint wavefield $u^*$ to keep the original notation used by \citet[p. 298]{Morse1953}. Note also that we could have defined the adjoint wavefield in any other way
\begin{align}
\Lm (x,T-t)\stackrel{\text{def}}{=} \mathcal{F} u(x,T-t) = u^* (x,t)
\label{eq.Adjoint_displacement_adjoint}
\end{align}
where $\mathcal{F}$ is \textit{any} certain linear functional, and the properties previously explained will still hold. For example, we could have rightfully chosen $\Lm(x,T-t)\stackrel{\text{def}}{=} \partial_t u(x,T-t)$. 

The new defined adjoint wavefield must satisfy the equations of motion given for $\Lm$ (eq. \eqref{eq.1D_adjoint_equation_with_lambda}), that is,
\begin{align}
\rho \partial^2_t u^* - \partial_x \left(\mu \partial_x u^*\right)    = \sum_{r}  \delta (x^r - x) (d  - u) . 
\label{eq.1D_elastic_adjoint_wave_equation}
\end{align}
Note that the underlying assumption is the linear dependence of the equations on the model, meaning that second-order terms are ignored. Note that the adjoint source is determined by the differences between data and synthetics $(d  - u)$ at certain receiver locations $\sum_{r}  \delta (x^r - x)$.

\subsubsection{1D Fr\'echet Derivatives}

In the following, the dependence of $u^*$ on the model is ignored, i.e., we assume that the adjoint wavefield $u^*$ perfectly satisfies the adjoint equation of motion eq. \eqref{eq.1D_adjoint_equation_with_lambda} and there are perturbations in the model parameters $(\delta \rho \neq \delta \mu \neq \delta f \neq 0)$. We can write the variation in the action eq. \eqref{eq.1D_General_Constrained_Action} as follows 
\begin{align}
\begin{aligned}
\delta_\m \chi &= - \int_{0}^{T} \int_{\Omega} u^*  \left[\delta \rho \partial_t ^2 u - \partial_x \left( \delta \mu \partial_x u\right) - \delta f\right] \dif x \dif t , \\
& = - \int_{0}^{T} \int_{\Omega}  \left[u^* \delta \rho  \partial_t ^2 u +  \delta \mu \partial_x u^* \partial_x u - u^*\delta f\right] \dif x \dif t, \\
& = \int_{\Omega}  \left(\delta \ln \rho \, K_{\rho} + \delta \ln \mu \text{K}_{\mu} + \delta \ln f \, K_{f}\right)  \dif x = 0, 
\end{aligned}
\label{eq.1D_elastic_variation_action_kernels}
\end{align}
where we have defined the following sensitivity kernels
\begin{align}
K_{\rho}  = - \int_{0}^{T} \rho u^* \partial^2_t u \dif \tau = \int_{0}^{T} \rho \partial_t u^* \partial_t u \dif \tau,\qquad  \text{K}_{\mu} = - \int_{0}^{T} \mu \partial_x  u^* \partial_x u \dif \tau, \qquad K_{f}  = \int_{0}^{T} u^* f \dif \tau .
\label{eq.1D_Adjoint_Kernels}
\end{align}

The kernels given in eqs. \eqref{eq.1D_Adjoint_Kernels} give us the change in the misfit function due to changes in the model parameters $(\rho,\mu)$ and source $(f)$, in terms of the original $(u)$ and adjoint $(u^*)$ wavefields.

\subsection{General 3D (Linear) Elastic Adjoint Equations}

Taking the first variation $\delta$ of the functional $\chi$ given in eq. \eqref{eq.Adjoint_General_Functional} with respect to model parameters $\m=(\den,\mathbb{C})$, the force term $\f$ and displacement $\u$, using KKT conditions eq. \eqref{eq.KKT_cond} and assuming that the data are accurately enough described by the wave equation, i.e., $\partial_\Lm \chi = \Linop \u - \f= 0$, leads to \citep{Liu01122006}
\begin{align}
\begin{aligned}
\delta_\u \chi = & \int_{0}^{T} \int_{\Omega} (\textbf{d}  - \textbf{u}) \sum_{r} \delta(\x-\x^r) \delta \u \dif^3 \x\dif t - \int_{0}^{T} \int_{\Omega} \Lm\left(\rho \, \partial_t^2 \delta \textbf{u} - \dvg (\mathbb{C} : \nabla \delta  \textbf{u})\right) \dif^3 \x \dif t = 0  \\
\delta_\m \chi = &  - \int_{0}^{T} \int_{\Omega} \Lm \left(\delta \boldsymbol{\rho} \partial_t^2 \textbf{u} - \dvg \left(\delta \mathbb{C} : \nabla \textbf{u}\right) - \delta \f\right) \dif^3 \x \dif t = 0.
\end{aligned}
\label{eq.General_Constrained_Action}
\end{align}

Assuming $\Lm=\Lm(\x,T-t)$, the expressions involving the Lagrange multiplier $\Lm$ and the wave equation operator in eq. \eqref{eq.General_Constrained_Action} resemble the convolution bilinear form eq. \eqref{eq.Spatial_Bilinear_Form}, which adjoint equation will be an initial value problem just like the original elastic equation of motion (eq. \eqref{eq.Elastic_wave}). The variation $\delta_\u \chi $ is obtained by integrating by parts (twice) the terms involving spatial and temporal derivatives of the displacement $\textbf{u}$ and the variation $\delta \textbf{u}$, 
\begin{align}
\begin{aligned}
\delta_\u \chi = &\int_{0}^{T} \int_{\Omega} (\d  - \u) \sum_{r} \delta(\textbf{x}-\textbf{x}^r) \delta \u \dif^3 \mathbf{x} \dif t  - \int_{0}^T \int_{\Omega} \left(\den \partial^2_t \Lm - \dvg ( \mathbb{C} : \nabla \Lm)\right) \delta \u \dif^3 \x \dif t ,
\end{aligned}
\label{eq.variation_action_integration_parts} 
\end{align}
subject to the following conditions
\begin{align}
	\left.\left(\mathbb{C} : \nabla \Lm\right) \cdot \hat{\text{n}}\right|_{\Gamma} = 0, \qquad \Lm(\x,0)=\partial_t\Lm(\x,0) = 0 .
	\label{eq.Boundary_3D_conditions}
\end{align}
In the absence of perturbations in the model parameters $(\delta \den = \delta \mathbb{C} = \delta \f = 0)$, i.e., $\delta_\m \chi =0$, the variation in the action eq. \eqref{eq.variation_action_integration_parts} reduces to
\begin{align}
\delta \chi =\delta_\u \chi = \int_{0}^T \int_{\Omega} \cbr[3]{(\d-\u) \sum_{r}  \delta (\x^r - \x)- \left[\den \partial^2_t \Lm - \dvg ( \mathbb{C} : \nabla \Lm)\right] } \delta \u \dif^3 \x \dif t = 0.
\label{eq.reduced_variation_action_integration_parts} 
\end{align}
The variation in the action $\delta \chi$ is stationary $(\delta_\m \chi = \delta_{\Lm} \chi = \delta_\u \chi=0)$ with respect to $\delta \u$, if
\begin{align}
\den \partial^2_t \Lm - \dvg ( \mathbb{C} : \nabla \Lm)  = (\d-\u) \sum_{r}  \delta (\x^r - \x),
\label{eq.adjoint_equation_with_lambda}
\end{align}
subject to conditions given by eq. \eqref{eq.Boundary_3D_conditions}. Note that the equation involving the Lagrange multiplier $\Lm$ is equal to the elastic wave equation (eq. \eqref{eq.Elastic_wave}) with a different source term.  Without having previous information of the adjoint wavefield, it seems reasonable to assume
\begin{flalign}
\Lm (\x,T-t) \stackrel{\text{def}}{=} \u (\x,T-t) = \u^* (\x,t),
\label{eq.3D_Reversed_displacement_adjoint}
\end{flalign}
thus, the adjoint wavefield $\u^*$ is equal to the time-reversed wavefield $u (\x,T-t)$. Note again that we could have defined the adjoint wavefield in any other way
\begin{align}
\Lm (\x,T-t)\stackrel{\text{def}}{=} \mathcal{F} \u(\x,T-t) = \u^* (\x,t) ,
\end{align}
where $\mathcal{F}$ is \textit{any} certain linear functional, and the properties previously explained will still hold. The new defined adjoint wavefield must satisfy the equations of motion given for $\Lm$, that is,
\begin{align}
\den \partial^2_t \u^* - \dvg \stress^*  = (\d-\u) \sum_{r}  \delta (\x^r - \x), 
\label{eq.adjoint_wave_equation}
\end{align}
where we have defined the adjoint stress tensor 
\begin{align}
\stress^* = \mathbb{C} : \nabla  \u^* ,
\end{align}
and subject to free surface and homogeneous initial conditions respectively
\begin{align}
\left.\stress^*\cdot  \hat{\text{n}}\right|_{\Gamma} = 0, \qquad \text{and} \quad \u^*(\x,0) = \partial_t \u^*(\x,0) = 0.
\end{align}

\subsubsection{3D Fr\'echet Derivatives}

If we assume that the adjoint wavefield $(\u^*)$ satisfies the adjoint equation of motion eq. \eqref{eq.adjoint_wave_equation} and there are perturbations in the model parameters $(\delta \den \neq \delta \mathbb{C} \neq \delta \f \neq 0)$, we can write the variation in the action as follows \citep{Liu01122006}
\begin{align}
\delta \chi = \delta_\m \chi  = \int_{\Omega}  \left(\delta \ln \den \, K_{\rho} + \delta \mathbb{C}:: \text{K}_{\mathbb{C}}\right) \dif^3 \x + \int_{0}^{T} \int_{\Omega}  \u^* \delta \f \dif^3 \x \dif t = 0, 
\label{eq.variation_action_kernels}
\end{align}
where we have defined the kernels
\begin{align}
K_{\rho} (\x) = - \int_{0}^{T} \den \u^* \partial^2_t \u \dif t = \int_{0}^{T} \den \partial_t \u^* \partial_t \u \dif t,\qquad  \text{K}_{\mathbb{C}} (\x) = - \int_{0}^{T} \nabla \u^*  : \nabla \u \dif t.
\label{eq.Adjoint_Kernels}
\end{align}
The perturbation to the point source given by eq. \eqref{eq.Moment_tensor_source} appearing in the source kernel in eq. \eqref{eq.variation_action_kernels}, can be written as follows \citep{Liu01122006}
\begin{align}
	\begin{aligned}
\delta \f = &  - \delta \text{M} \cdot \nabla \delta (\x-\x^s) S(t) - \text{M} \cdot \nabla \delta (\x-\x^s-\delta \x^s ) S(t) \\
& - \text{M}\cdot \nabla \delta (\x-\x^s)\delta S(t) + \text{M}\cdot \nabla \delta (\x-\x^s)S(t),
	\end{aligned}
\end{align}
where $\delta \text{M}$ denotes the perturbed moment tensor, $\delta \x^s$ the perturbed point source location and $\delta S(t)$ the perturbed source-time function. In an isotropic Earth model we can write the fourth-order tensor of elastic constants as follows \citep{Slawinski2010}
\begin{align}
\mathbb{C}_{ijkl} = (\boldsymbol{\kappa} - 2/3 \, \boldsymbol{\mu}) \, \delta_{jk} \delta_{lm} + \boldsymbol{\mu} (\delta_{jl} \delta_{km} + \delta_{jm} \delta_{kl}),
\end{align}
where $\mu$ and $\kappa$ denote the shear and bulk moduli, respectively. We thus can write
\begin{align}
\delta \mathbb{C} :: \text{K}_{\mathbb{C}} (\x) = \delta \ln \boldsymbol{\mu} \, K_{\mu} (\x) + \delta \ln \boldsymbol{\kappa} \, K_{\kappa} (\x),
\end{align}
where the isotropic kernels $K_{\mu}(\x)$ and $K_{\kappa}(\x)$ represent \textit{Fr\'echet derivatives} with respect to the relative bulk and shear moduli perturbations $\delta \ln \boldsymbol{\kappa}$ and $\delta \ln \boldsymbol{\mu}$, respectively. These isotropic kernels are given by \citep{tromp2005seismic,Liu01122006}
\begin{align}
\begin{aligned}
K_{\mu} (\x) & = - \int_0^T 2 \mu (\x) \text{D} ^{\dagger} (\x,T-t) \text{D}(\x,t) \dif t , \\ 
K_{\kappa} (\x) & = - \int_0^T \kappa(\x) \intcc[1]{\nabla \cdot u^*(\x,T-t)} \intcc[1]{\nabla \cdot u(\x,t)} \dif t,
\end{aligned}
\label{eq.Kernels_expressions}
\end{align}
where
\begin{align}
\text{D} = \frac{1}{2} \left[\nabla \u + (\nabla \u)^T\right] - \frac{1}{3} \tr \left(\nabla  \u\right), \qquad \text{D}^{\dagger} = \frac{1}{2} \left[\nabla \u^* + (\nabla \u^*)^T\right] - \frac{1}{3} \left(\nabla \cdot \u^*\right), 
\end{align}
denote the traceless strain deviator and its adjoint, respectively. Additionally, we can express sensitivity kernels in an isotropic earth model in terms of relative variations in mass density $\delta \ln \den$, shear-wave speed $\delta \ln \boldsymbol{\beta}$, and compressional-wave speed $\delta \ln \boldsymbol{\alpha}$ based upon the wave speeds expressions
\begin{align}
\boldsymbol{\alpha}^2 = \frac{\boldsymbol{\kappa} + \frac{4}{3}\boldsymbol{\mu}}{\den} \qquad \text{and} \qquad \boldsymbol{\beta}^2 = \frac{\boldsymbol{\mu}}{\den},
\end{align}
which leads to \citep{Liu01122006}
\begin{align}
\delta \ln \den K_{\rho} (\x) + \delta \ln \boldsymbol{\mu} K_{\mu} (\x) + \delta \ln \boldsymbol{\kappa} K_{\kappa} (\x) &= \delta \ln \den K_{\rho'}(\x)  + \delta \ln \boldsymbol{\beta} K_{\beta} (\x) + \delta \ln \boldsymbol{\alpha} K_{\alpha} (\x),
\end{align}
where
\begin{align}
K_{\rho'} (\x) & = K_{\rho} (\x) + K_{\kappa} (\x) + K_{\mu}(\x) , \qquad K_{\beta} (\x)= 2 \intoo[3]{K_{\mu}(\x) - \frac{4}{3}\frac{\boldsymbol{\mu}}{\boldsymbol{\kappa}} K_{\kappa}(\x)}, \qquad K_{\alpha} (\x)= 2 \intoo[3]{\frac{\boldsymbol{\kappa} + \frac{4}{3}\boldsymbol{\mu}}{\boldsymbol{\kappa}}} K_{\kappa} (\x).
\end{align}

\subsection{Viscoelastic Attenuation}

Seismic waves that propagate inside the Earth sense a natural attenuation effect through the entire seismic frequency range (0.01--100 Hz) \citep{Liu1976,kjartansson1979constant,Romanowicz1999,romanowicz2015}. This attenuation is manifested in the seismic records by physical dispersion and dissipation \citep{Zhu2013,Durand2013,Widmer1991,Bezada2017,Ford2012}.

Seismic attenuation measured by seismic records is a combination of the intrinsic (or inelastic) attenuation and the extrinsic (or elastic) attenuation. Intrinsic attenuation is the energy loss caused by materials' microstructure, like defects and degrees of freedom, as well as others factors \citep{Romanowicz1999,Jackson2007}. Extrinsic attenuation, also called apparent attenuation, is the apparent energy loss due to the redistribution or scattering of the wavefield. The combination of both intrinsic and extrinsic attenuation can be taken into account by choosing certain rheology of wave propagation.

Different sensitivity kernels can be found depending on the chosen rheology; thus, we limit our attention to find general adjoint equations of the viscoelastic problem. The 3D elastic equation of motion in presence of viscoelastic effects can be written as follows \citep{borcherdt2009viscoelastic}
\begin{align}
\Linop \u = \den \partial_t^2 \u - \dvg \left(\int_{-\infty}^T \mathcal{R}(t-\tau) : \partial_t \boldsymbol{\varepsilon} (\tau) \dif \tau\right) - \f = 0 , \qquad \text{with} \quad \u(\x,0) = \partial_t \u(\x,0) = 0,
\label{eq.viscoelastic_wave}
\end{align}
where the stress tensor in general viscoelastic media is given by \citep{borcherdt2009viscoelastic}
\begin{align}
	\stress = \int_{-\infty}^T \mathcal{R}(t-\tau) \partial_t \boldsymbol{\varepsilon} (\tau) \dif \tau, \qquad \text{with} \qquad \boldsymbol{\varepsilon}=\frac{1}{2} \left(\nabla \u + \left(\nabla \u\right)^T\right) , 
\end{align}
where $\mathcal{R}$ is a time-dependent tensor-valued kernel called the stress relaxation function \citep{mainardi2010fractional}. Using the bilinear form $\left\langle \u^*,\Linop \u \right\rangle_{\x,t}$ (eq. \eqref{eq.Spatial_Bilinear_Form}), we find the adjoint equations to eq. \eqref{eq.viscoelastic_wave} \citep{Tarantola1988,fichtner2006adjoint}
\begin{align}
\begin{aligned}
\adLinop v & = \begin{cases}
\den \partial_t^2 \u^* = \dvg \left(\displaystyle \int_{-\infty}^T \mathcal{R}(\tau-t) : \partial_t \boldsymbol{\varepsilon}^* (\tau) \dif \tau\right) + \overrightarrow{\textbf{g}} , \\
\u^*(\x,T) = \partial_t \u^*(\x,T) = 0 ,
\label{eq.viscoelastic_adjoint_1}
\end{cases}
\end{aligned}
\end{align}
where we have defined the adjoint strain as
\begin{align}
	\boldsymbol{\varepsilon}^* = \frac{1}{2} \left(\nabla \u^* + \left(\nabla \u^*\right)^T\right) .
\end{align}

The adjoint eq. \eqref{eq.viscoelastic_adjoint_1} corresponds to a wave equation with negative attenuation (note the change in sign in the stress relaxation function). If we use the convolution bilinear form $\left\langle \u^*,\Linop \u \right\rangle_{\x,t}^c$ (eq. \eqref{eq.Spatial_Convoltion_Bilinear_Form}), it is straightforward to find
\begin{align}
\begin{aligned}
\adLinopc \u^* & =
\begin{cases}
\den \partial_t^2 \u^* = \dvg \left(\displaystyle \int_{-\infty}^T \mathcal{R}(t-\tau) : \partial_t \boldsymbol{\varepsilon}^{*} (\tau) \dif \tau\right) + \overleftarrow{\textbf{g}} , \\
\u^*(\x,0) = \partial_t \u^*(\x,0) = 0 .
\label{eq.viscoelastic_adjoint_2}
\end{cases}
\end{aligned}
\end{align}

Note that the adjoint eq. \eqref{eq.viscoelastic_adjoint_2} has the same mathematical structure as the original viscoelastic wave equation eq. \eqref{eq.viscoelastic_wave}. In other words, using the convolution bilinear form eq. \eqref{eq.Spatial_Convoltion_Bilinear_Form}, the obtained adjoint equation does not correspond to a wave equation with negative attenuation but to the original one. 

To make this statement more readable, consider the 1D shear wave with viscoelastic attenuation (Stokes wave equation) given by \citep{ricker1977transient}
\begin{align}
\begin{aligned}
\Linop u= 
\begin{cases}
\rho \partial_t^2 u = \partial_x \left(\mu \partial_x u + \eta \partial^2_{x,t} u \right)+ f,  \\
u(x,0) = \partial_t u(x,0) = 0 .
\end{cases}
\label{eq.Stokes_eq}
\end{aligned}
\end{align}
where $\eta$ is a viscosity coefficient. In homogeneous media, the adjoint equations using the bilinear forms $\left\langle v,\Linop u \right\rangle_{x,t}$ (eq. \eqref{eq.Spatial_Bilinear_Form}) and $\left\langle v,\Linop u \right\rangle_{x,t}^c$ (eq. \eqref{eq.Spatial_Convoltion_Bilinear_Form}) are given by
\begin{align}
\begin{aligned}
\adLinop v& =
\begin{cases}
\rho \partial_t^2 v = \mu \partial^2_x v - \eta \partial^3_{x,x,t} v + \overrightarrow{g} , \\
v(x,T) = \partial_t v(x,T) = 0 ,
\end{cases}
\qquad \adLinopc v =
\begin{cases}
\rho \partial_t^2 v = \mu \partial^2_x v + \eta \partial^3_{x,x,t} v  + \overleftarrow{g} , \\
v(x,0) = \partial_t v(x,0) = 0 .
\end{cases}
\end{aligned}
\end{align}

In agreement with \cite{Tarantola1988,fichtner2006adjoint,yang2016review} (and several others), the adjoint wave equation with negative attenuation corresponds to the bilinear form eq. \eqref{eq.Spatial_Bilinear_Form}. However, using the convolution bilinear form eq. \eqref{eq.Spatial_Convoltion_Bilinear_Form}, the adjoint equation corresponds to the original one (positive attenuation). 

\section{A Mathematical Inconsistency in the Viscoelastic Adjoint Equations}

A mathematical inconsistency appears at this point since \cite{Tarantola1988,fichtner2006adjoint,yang2016review} (and several others), correctly claim to find the viscoelastic adjoint equation using the bilinear form eq. \eqref{eq.Spatial_Bilinear_Form}, however, in order to satisfy initial conditions, they arbitrarily choose the adjoint wavefield as the time reversed displacement, without taking into account that this assumption changes the adjoint equations in case of viscoelastic/dissipative effects. 

Our claims can already be seen in previous works using more complex mathematical machinery than the one presented here. For instance, in \cite{chen2011full} the author assumes the adjoint wavefield to be time reversed from the beginning of the mathematical development, and therefore there is no change in sign in the adjoint equations. In a different work from the same author \cite{chen2015full}, it is found that when not having the same initial assumption, there is the need to introduce an anti-causal rate-of-relaxation function just as done in \cite{Tarantola1988,fichtner2006adjoint}. This mathematical inconsistency is far from trivial and it has profound consequences.

\paragraph{How do We Model Intrinsic Attenuation?} 

Up today, the mathematical description of the physical mechanism causing intrinsic attenuation of seismic waves is not yet well understood. Dispersion and dissipation effects are commonly introduced in plane wave solutions of the elastic wave equation by using approximations \citep{aki2002quantitative}, with very little connection to the physics that causes these effects. There is no consensus among seismologists on how the effects of intrinsic attenuation should be quantified and measured \citep{Zhu2013}. The mathematical description of seismic intrinsic attenuation is routinely taken into account by a macroscopic, phenomenological point of view, and no attempt to inquire into the microscopic solid-state mechanisms responsible for seismic attenuation is usually done \citep{Tromp1998,FichtnerQ,moczo2014finite}.

A measure of internal friction (or anelasticity) is given by the following expression \citep{aki2002quantitative}
\begin{align}
	\frac{1}{Q} = - \frac{1}{2\pi} \frac{\delta A}{A},
\end{align}
where $A$ is the wave amplitude and $-\delta A$ is the amplitude lost due to imperfections in the elasticity of the material. Assuming a plane wave of the form
\begin{align}
	A(x,t)= A_0 e^{\imag(\omega t + k x)} ,
	\label{eq.Plane_wave_Solution}
\end{align}
where $A_0$ is the amplitude, $\omega$ the angular frequency, $k$ the wavenumber, $t$ the travel-time, $x$ the distance and $c=\omega/k$ is the phase velocity. We can include amplitude lost (or reduction) in the plane wave solution eq. \eqref{eq.Plane_wave_Solution}, it is commonly assumed that if spatial attenuation exists then the wavenumber $k$ is a complex quantity $(k=k_R+\imag k_I)$ \citep{Anderson2007}. The imaginary part of $k$ $(k_I)$ is called the attenuation coefficient.  

Taking into account dissipation after a single wave period (i.e. $k_R \, x=2\pi$), the attenuation coefficient is $ \exp ( - k_I x) = \exp (- 2 \pi k_I / k_R)$ and the difference in amplitude is $\delta A = 1 - \exp (- 2 \pi k_I / k_R)$. The exact expression for the quality factor $Q$ may thus be written as follows
\begin{align}
	Q^{-1}=-\frac{1}{2\pi}(e^{-2\pi k_I/k_R}-1) .
\end{align}
For a complex wave number $k$, the plane wave solution eq. \eqref{eq.Plane_wave_Solution} can thus  be written as follows
\begin{align}
	A(x,t) & =  A_0  e^{ -\omega x/(2 Q c)} e^{\imag\omega ( t + x/c)} .
	\label{eq.Attenuated_plane_wave}
\end{align}
Note that assuming a complex wavenumber $k$ has led to an exponentially decreasing plane wave with an attenuation factor given by $e^{-\omega x / 2c Q}$, with $x$ being measured along the propagation direction. Note also that $Q$ is inversely related to the strength of the attenuation.

\paragraph{A Change in Sign?}

Results found by \cite{Tarantola1988,fichtner2006adjoint,yang2016review}  include a change in the sign of the viscoelastic attenuation. This means that eq. \eqref{eq.Attenuated_plane_wave} is written as follows
\begin{align}
	A(x,t) & =  A_0  e^{ \omega x/(2 Q c)} e^{\imag\omega ( t + x/c)} ,
	\label{eq.Positive_attenuated_plane_wave}
\end{align} 
which implies that instead a decrease in energy (amplitude), the seismic wave will increase its energy, and this increase of energy will be exponential. From the practical point of view, it represents a challenge to numerically implement an increase of energy due to the instabilities emerging from the positive exponential. This increase of energy has been justified by conservation of energy.

\paragraph{Conservation of Energy?}

\cite{fichtner2006adjoint} pointed out the similarity between the original and the adjoint equations in the absence of viscoelastic attenuation. The authors have related this similarity to the conservation of energy. However, for the case of viscoelastic attenuation they claim: \textit{it is therefore not surprising that the spatial derivative term is different in the adjoint equations because it incorporates the loss of elastic energy in the form of a time-dependent rheology.}

As a consequence, in the case of viscoelastic attenuation, we can interpret the change in sign of the adjoint equations as a result of the conservation of energy, i.e., the forward equations lose energy and the adjoint equation regains the energy lost, thus leading to a full conservation of energy. We have shown, however, that this interpretation is not correct: Forward and adjoint equations are the same including any kind of energy lost or not.

As we have mentioned in Section  \ref{sec.General_Adjoint_method}, the adjoint equation is a mathematical object that allows us to obtain Lagrangians for non-conservative systems (treating them as if they were conservative) that describe certain initial value problem, allowing to find conservation laws and approximate solutions \citep{anderson1972role,prasad1972adjoint,anderson1973application,djukic1975noether,vujanovic1989variational,davis1928inverse,vujanovic1989variational,davis1929inverse,tonti1982general,oden1983variational}.

\paragraph{Numerical Example} A solution of Stokes equation eq. \eqref{eq.Stokes_eq} given by the following expression \citep{ricker1977transient}
\begin{align}
   u(x,t) = A_0 e^{-\eta x} \cos\left(t-\frac{x}{c}\right).
   \label{eq.Stokes_sol}
\end{align}

Consider wave propagation in viscous air at 25$^{\circ}$C, $c=331.3 \, \text{m/s},\eta=18.46\, \mu\text{Pa}\cdot\text{s}$ \citep{kestin1959absolute} and $A_0=0.01$ m.  According to \cite{Tarantola1988,fichtner2006adjoint,yang2016review}, the adjoint field $u^*$ is the time reversed signal including a positive attenuation, i.e.,  $u^*(x,t)=e^{\eta x}u(x,T-t)$, while according to the theory presented in this work is $u^*(x,t)=e^{-\eta x}u(x,T-t)$. 

\begin{figure}
    \begin{center}
        \includegraphics[width=1\textwidth]{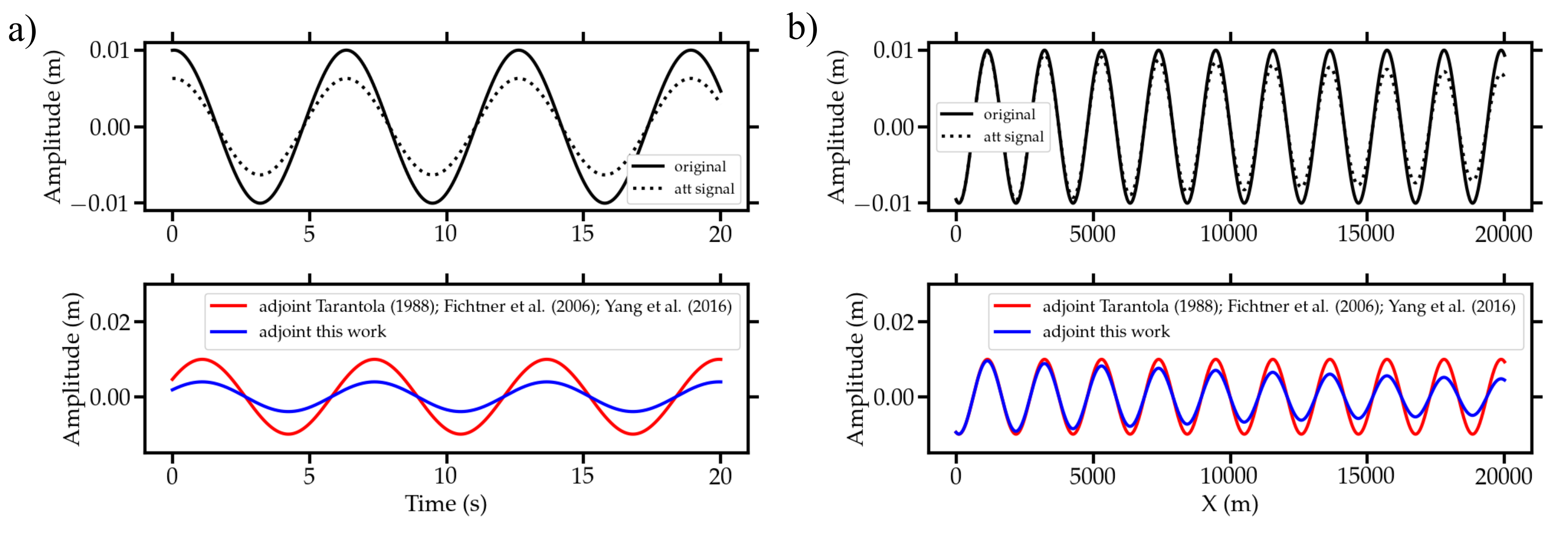}
        \caption{Comparison between theories previously proposed in the literature and the new one proposed here.}
        \label{Fig.seismograms_comp}
    \end{center}
\end{figure}

Results at a location of $x=25$ km are presented in Fig. \ref{Fig.seismograms_comp} a), where we can observe the dissipation suffered by the plane wave (top panel) and the differences in adjoint fields (bottom panel). According to \cite{Tarantola1988,fichtner2006adjoint,yang2016review}, the energy of the adjoint field at this location should have increased and, for the parameters chosen, it will be almost the same as the original wavefield (0.01 m). However, according to the theory presented here, it should have considerably decreased. Similar results are observed after 60 s of wave propagation (Fig. \ref{Fig.seismograms_comp}--b). We can observe the decrease in energy of the wavefield (top panel) and the difference in amplitudes of the adjoint wavefields (bottom panel). According to \cite{Tarantola1988,fichtner2006adjoint,yang2016review}, the spatial attenuation will increase the energy of the (original) attenuated wave resulting in lack of lost of energy at all. On the contrary,  according to the theory presented here, the energy of the wave will decreasing according to the spatial location. As we can see the theory presented by \cite{Tarantola1988,fichtner2006adjoint,yang2016review} and here represent two fundamentally different points of view. 

\paragraph{Attenuation Only?}

Our claims can also be found in \cite{liu2008finite} where the authors claim that in global wave propagation and assuming an anelastic earth model, \textit{the adjoint wavefield is determined by exactly the same equations as the regular wavefield, with the exception of the source term and the sense of rotation}. In another work by the same authors \cite{tromp2005seismic} state that: \textit{the adjoint momentum equation for a rotating, self-gravitating earth model involves an earth model that rotates in the opposite direction}.

We attempt to explain this interpretation because odd-order spatial derivatives appear in the Coriolis term of the equations of motion when considering the rotation of the Earth and we assume that the authors interpret that these terms change to opposite sign in the adjoint equations.

\cite{liu2008finite} correctly claim that forward and adjoint equations are the same even in the case of viscoelastic effects, however, they fail to see that this should also happen for any other term as well. As we have seen (see Table \ref{tb.adj_op}), there should not be change in sign in the adjoint equation, even for odd-order spatial derivative terms. We agree that considering the adjoint equations, obtained using the bilinear form, includes a change in the sign in the spatial derivatives with odd powers, and despite there is no mathematical error here, we find that there is no physical justification for not keeping the same mathematical structure of the original equations, since using a convolution bilinear form as previously showed, we are able to obtain always the same mathematical expression (we refer the reader to Section \ref{sec.Inital_value_problems_adjoints} problems 5 and 6). As a consequence, we believe that there is no need and/or physical justification to make the Earth change its sense of rotation in the adjoint equations. We further justify this argument with solid physical basis when we compare the adjoint equation method against time reversal imaging in Section \ref{sec.TimeReversal}. 

Having clarified mathematical inconsistencies and physical interpretations, we now turn our attention to the numerical implementation of the method.

\section{Numerical Implementation}

We next discuss the numerical implementation of the adjoint method in the time domain. To simplify the illustration we limit this to the 1D case, since the generalization to the 3D case is straightforward. 

\subsection{1D Equations of Motion}

Depending on the discretization of the wave equation chosen, the adjoint method has two fundamental ways of being implemented in the time domain. Let us consider two formulations of the 1D shear wave equation: i) the displacement $(u)$ formulation given by \citep{moczo2014finite}
\begin{align}
\rho \partial^2_t u =  \mu \partial^2_x u + \delta(x-x^s)f , \qquad \text{with} \quad u(x,0) = \partial_t u(x,0) = 0,
\end{align}
and ii) the velocity $(v)$-stress $(\sigma)$ formulation given by \citep{moczo2014finite}
\begin{align}
\begin{aligned}
\rho \partial_t v &= \partial_x \sigma+ \delta(x-x^s) f , \\
\partial_t \sigma &= \mu \partial_x v ,  \qquad \text{with} \quad v(x,0) = \sigma(x,0) = 0 .
\end{aligned}
\end{align}

The main difference between the displacement and the velocity-stress formulations are that time and spatial derivatives are expressed as second and first-order respectively. The calculation of Fr\'echet derivatives (eqs. \eqref{eq.1D_Adjoint_Kernels}) requires the interaction between the forward wavefield and adjoint wavefield which runs backward in time. To compute this interaction, the forward wavefield is first computed to be later reconstructed backward in time, which is possible thanks to the symmetry of the wave equation. 

\subsection{Forward Modeling}

To illustrate the procedure, we assume second- and first-order FD discretizations for the time derivatives of the wave equation in displacement $(u)$ and velocity $(v)$-stress $(\sigma)$ formulation as follows
\begin{align}
u_x^{t+\Delta t} = 2u_x^t - u_x^{t-\Delta t} + \frac{\Delta t^2}{\rho} \left(\mu D^2_x u + \delta(x-x^s)f \right) + \error \left(\Delta t^2\right), 
\label{eq.Numerical_Second_order}
\end{align}
and
\begin{align}
\begin{aligned}
v_x^{t+\Delta t} &= v_x^t + \frac{\Delta t}{\rho }\left(D_x \sigma+ \delta(x-x^s) f\right)  + \error \left(\Delta t\right) , \\
\sigma_x^{t+\Delta t} &= \sigma_x^t + \mu \Delta t D_x v  + \error \left(\Delta t\right),
\label{eq.Numerical_First_order}
\end{aligned}
\end{align}
where $D_x,D_x^2$ are general spatial derivative operators, which will be chosen depending on the numerical method used. We do not discuss the spatial derivatives since they do not affect the general implementation of the adjoint method.

\subsection{Backward Reconstruction}

The backward reconstruction of the forward wavefield has two fundamental differences in eqs. \eqref{eq.Numerical_Second_order}--\eqref{eq.Numerical_First_order}, while the backward reconstruction of the wavefield using eq. \eqref{eq.Numerical_First_order} can be found by simply a sign change $(\Delta t\to- \Delta t)$, that is,
\begin{align}
\begin{aligned}
v_x^{t+\Delta t} &= v_x^t - \frac{\Delta t}{\rho } D_x \sigma , \\
\sigma_x^{t+\Delta t} &= \sigma_x^t  - \mu \Delta t D_x v , \qquad \text{with} \qquad v(x,0) = v_N(x), \quad \sigma(x,0) = \sigma_N(x) ,
\label{eq.Adjoint_Numerical_First_order}
\end{aligned}
\end{align} 
where $v_N,\sigma_N$ are the last snapshots stored from the forward simulation eq. \eqref{eq.Numerical_First_order}. The sign change $(\Delta t\to-\Delta t)$, on the contrary, will leave eq. \eqref{eq.Numerical_Second_order} exactly the same. The backward reconstruction can be done by a flip in the initial conditions as follows 
\begin{align}
\begin{aligned}
u_x^{t+\Delta t} = 2u_x^t - u_x^{t-\Delta t} + \frac{\Delta t^2}{\rho} \mu D^2_x u , \qquad \text{with} \qquad u(x,0) = u_N(x), \quad u(x,\Delta t) = u_{N-1}(x) ,
\label{eq.Adjoint_Numerical_Second_order}
\end{aligned}
\end{align} 
where $u_N$ is the last snapshot $(N)$ and $u_{N-1}$ the penultimate $(N-\Delta t)$ snapshot of the forward wavefield. 

\subsection{Storing Boundary Wavefields}

When considering absorbing boundary conditions, e.g., \citep{clayton1977absorbing,stacey1988improved}, and/or any type of boundary condition different from rigid, it is necessary to store the boundary information at each time level in order to be able to backward reconstruct the forward wavefield appropriately. This comes with an increase in storage and memory requirements. Different methodologies can be used, e.g., by not storing all time levels and trying to reconstruct the forward wavefield as accurately as possible. In common occasions, absorbing boundary conditions like perfectly matched layers (PMLs) \citep{komatitsch2003perfectly,duru2012well,wang2003finite} are not only implemented at the boundaries of the computational domain but also at interior grid points. In this case, it becomes necessary to store the wavefield at interior points too, increasing even more the computational requirements. We can of course, accepting a decrease in accuracy, not store all grid points and all time levels. This will depend on how accurate the computation of the sensitivity kernels we expect to have and the computational power we have available.

\subsection{Fr\'echet Kernel Calculations}

As previously stated, the calculation of Fr\'echet derivatives (eqs. \eqref{eq.1D_Adjoint_Kernels}) requires the interaction between the forward and adjoint wavefields. After having run the forward simulation and the last snapshots stored, the required misfit (see Sec. \ref{se.Mistif_fuctions}) is computed and the adjoint source injected at the corresponding receivers locations. Two simulations, one forward reconstructed backward in time and a another adjoint, run in parallel and allow the computation of sensitivity kernels.

\begin{figure}
    \begin{center}
        \includegraphics[width=1\textwidth]{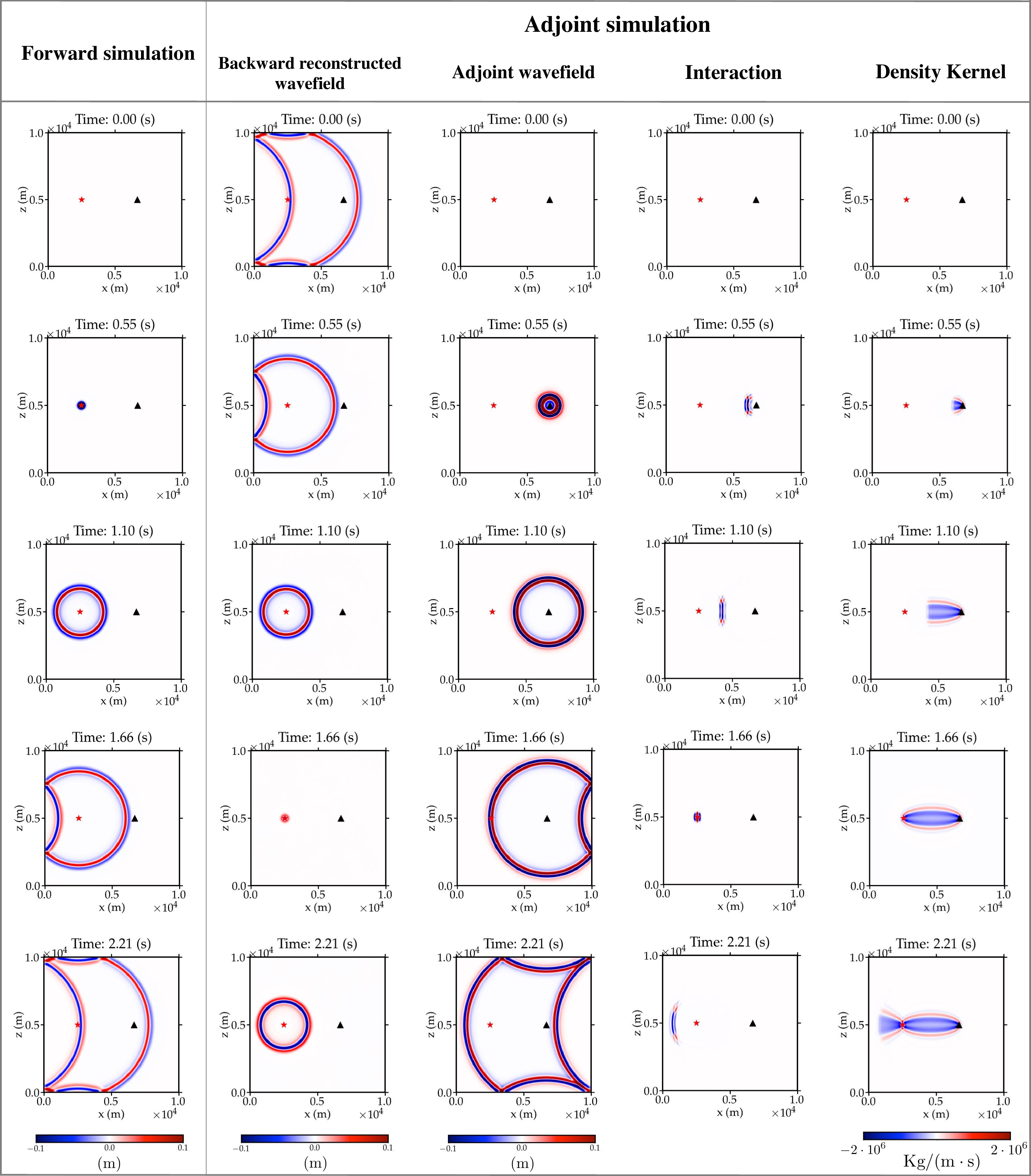}
        \caption{Sensitivity kernel calculation process.}
        \label{Fig.kernels}
    \end{center}
\end{figure}

To illustrate this, consider a 2D SH wave propagation in a domain of 10$\times$10 km$^2$, with uniform velocity of 3.2 km/s. We inject a Ricker source time function with dominant frequency of 3 Hz, amplitude of 1 m and a time delay of $t_s=0.56$ s (red star location, see Fig. \ref{Fig.kernels}). We run the simulation for a total time $T$ of $2.21$ s (see Fig. \ref{Fig.kernels}) and the last two snapshots are stored and used for the backward reconstruction using the displacement formulation of the wave equation (eq. \eqref{eq.Adjoint_Numerical_Second_order}).

We next compute waveform sensitivity kernels. To do so, we simply inject the time reversed recorded velocity field at the receiver location (black triangle location, see Fig. \ref{Fig.kernels}) and run the backward reconstruction of the forward wavefield (see Fig. \ref{Fig.kernels}--Backward reconstructed wavefield). It is important to note, however, that there is in reality no simulation running backward in time: While the backward reconstructed wavefield is simply a forward simulation with certain initial conditions (see eq. \eqref{eq.Adjoint_Numerical_Second_order}), the adjoint wavefield is just a forward simulation with the required (velocity reversed) source time function injected at the receiver location. It is also important to note that the adjoint and backward reconstructed simulation do not run the same total time $T$ that the initial forward simulation did. Instead, we need to run the adjoint simulation to a total time equal to the total forward simulation time $T$ minus the time delay of the initial source time function $(T-t_s)$; otherwise, we will obtain a larger (undesired) sensitivity at the source side (see Fig. \ref{Fig.kernels}--Density $(\rho)$ Kernel).         

\section{The Adjoint (Equation) Method and its Equivalence to Time Reversal}
\label{sec.TimeReversal}

\begin{figure}
	\begin{center}
		\includegraphics[width=0.8\textwidth]{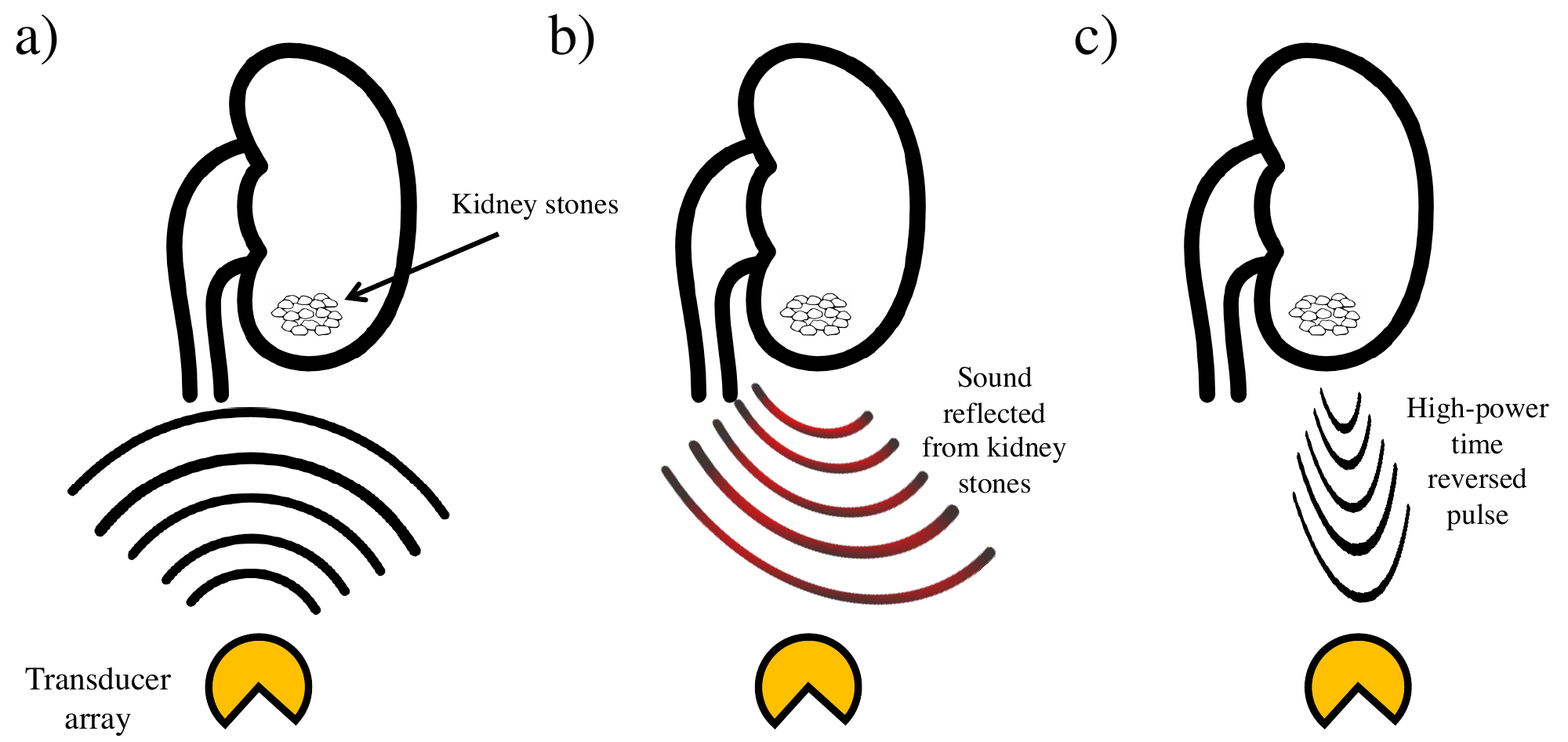}
		\caption{Time reversal techniques used for medical procedures (after \cite{fink1999time}). Same physics of wave propagation in forward and time reversed signals.}
		\label{Fig.time_reversal_medical}
	\end{center}
\end{figure}

Time reversal acoustics (TRA) is a well-established technique that is used to focus acoustic (or elastic) energy to a selected location in space \cite{fink1992time,cassereau1992time}. The technique can be used for multiple applications \citep[e.g.][]{fink2003time,wang2022review,borcea2003theory}. In the particular case of biomedical imaging \citep{fink2003time}, it is mostly used for the destruction of stones in kidneys and gallbladders \citep{fink1999time}. The process is illustrated in Fig. \ref{Fig.time_reversal_medical}, where in a first step an ultrasound source is sent toward the human body and the reflections produced by the kidney stones are recorded by the transducer array (see Figs. \ref{Fig.time_reversal_medical}--a and b). In a next step, an amplified time reversed recorded signal is sent back. This signal is going to focus on the original scatterer (this case the kidney stones) breaking it up. Iterating this process improves the energy focusing and also allows real-time tracking as the stones move \citep{fink1999time}.

We can observe the similarities between the time reversal and the adjoint simulation, where the wavefield is sent back in time and focuses at the source location (see Fig. \ref{Fig.kernels}--Backward reconstructed wavefield). The high power time reversed pulse is the equivalent to the time reversed forward wavefield. Note that in the whole process of time reversal experiments, waves travel in a medium with equivalent physical properties. In other words, attenuation (or any other property) is never changed; the medium attenuate energy in all circumstances. This is the reason why it is a requirement for the success in the application of time reversal that the sound wave must propagate without losing too much energy to heat \citep{fink1999time}. We claim in this work that the same physical analogy can be understood for the adjoint equation method: In forward and backward simulation, the physical properties of the medium remain the same. Attenuation should never be changed in sign, i.e., the medium never injects new energy to the system (positive attenuation).

Returning to the statement made by \cite{liu2008finite} where the authors show that the adjoint wavefield is determined by exactly the same equations as the regular wavefield, with the exception of the source term and the change in the sense of rotation of the Earth. Making an analogy to time reversal, we have observed that from a strict physical point of view this should not happen, since all physical properties of the medium should remain exactly the same. Thus the Earth should never change its sense of rotation.
	
We now turn the attention to the missing ingredient of the adjoint method for geophysical applications: the choice of the misfit function and construction of the adjoint source.

\section{The Misfit Function and Construction of the Adjoint Source}
\label{se.Mistif_fuctions}

Up to now we have considered a general definition of the error or misfit function $\E$ as the difference between data and synthetics at certain spatial locations (eq. \eqref{eq.chi_squared}). We can, however, define a large number of misfit functions depending on the information that we would like to (or can) extract from each seismogram.

\begin{figure}
	\begin{center}
		\includegraphics[width=0.6\textwidth]{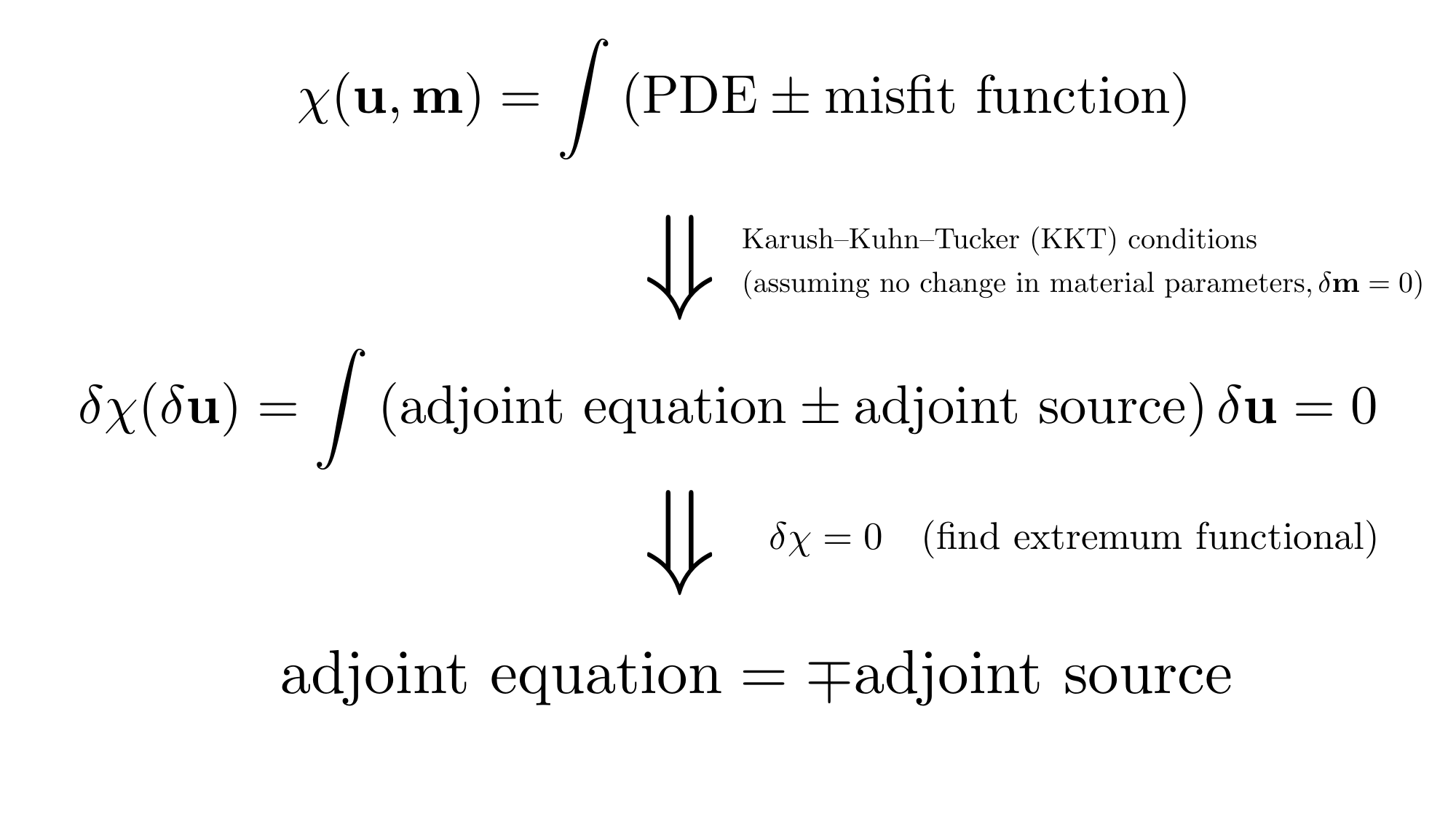}
		\caption{Outline of the Lagrangian formalism used to find the adjoint equation and source. Note the need to isolate $\delta \u$ in both adjoint equation and source.}
		\label{Fig.Lagrangian_Formalism}
	\end{center}
\end{figure}

Using the Lagrangian formalism, the general procedure to obtain the correct adjoint source from the chosen misfit function is always the same: 
\begin{enumerate}[a)]
	\item Define the misfit function $\E$.
	\item Compute the variation of the misfit function $\delta \E$, leaving explicit the term $\delta \u$. This allows us to isolate the adjoint equation and its corresponding source of motion (see Fig. \ref{Fig.Lagrangian_Formalism}).
\end{enumerate}

To illustrate this procedure, we next illustrate the construction of the adjoint source for several misfit measures using 2.5D SH wave propagation in the whole mantle.

\subsection{Global 2.5D SH Wave Propagation}

While most of the time full 3D simulations of wave propagation in the whole Earth's mantle become not feasible due to the large computational power required, Cartesian 2D simulations do not capture the physics of 3D geometrical spreading. A popular alternative is called 2.5D simulations, where the 2D elastic equations of motion (SH and/or P-SV) are solved in cylindrical or spherical coordinates \citep[e.g.][]{igel1999wave,igel1995sh,igel1996p,thomas2000acoustic,takenaka2003quasi,toyokuni2005quasi,jahnke2008global,nissen2007two,nissen20082,nissen2007spherical,nissen2014axisem}. This allows to naturally take into account 3D geometrical spreading effects while only requiring computational resources of a 2D simulation. This kind of simulations is commonly referred to as axisymmetric simulations, since one can obtain the same waveforms by running 2.5D simulations in certain 2D domain and running a full 3D simulation while considering the same 2D domain rotated along certain chosen axis (this of course applies only to certain kind of sources). 

Applications of 2.5D simulations are mostly used to understand the physics of the observations \citep[e.g.][]{schlaphorst2016investigation,ma2020small,ma2019localized,saki2019causes,schumacher2018out} and to a lesser extent to perform inversions at very short periods, allowing to exploit the full frequency information of the data, without the incredibly demanding computational resources of full 3D methods and without the simplicity of ray theory, helping, among other things, to avoid errors coming from crustal corrections \citep[e.g.][]{baker2014full,okamoto2009waveform,smithyman2012waveform,smithyman2013waveform}. 

To illustrate the different sensitivity kernels obtained when considering different misfit functions, we use the second-order in time and fourth-order in space finite-difference method presented by \cite{jahnke2008global} to solve the (forward and adjoint) velocity $(v_{\phi})$, stress $(\sigma_{r\varphi}, \sigma_{\theta\varphi})$ 2D SH equations of motion in spherical coordinates (2.5D simulation) given by the following expressions \citep{igel1999wave,slaughter2012linearized}
\begin{align}
\begin{aligned}
\rho \partial_t v_\varphi & =  \partial_r \sigma_{r\varphi} + \frac{1}{r} \partial_\theta \sigma_{\theta\varphi} + \frac{1}{r}(3 \sigma_{r\varphi} + 2\sigma_{\theta\varphi}\; \cot \theta) + f_\varphi (t)\delta(\x-\x_s)\\
\partial_t \sigma_{r\varphi} & = \mu \left(\partial_r v_{\varphi} - \frac{v_{\varphi}}{r}\right) , \qquad  \partial_t \sigma_{\theta\varphi} = \frac{\mu}{r} \left(\partial_{\theta} v_{\varphi} - v_{\varphi}\cot \theta\right) ,
\label{eq.SH_25D_equations_of_motion}
\end{aligned}
\end{align}
where $r$ is the distance from the Earth's center, $\varphi$ is the azimuth angle and $\theta$ the polar angle, $f_\varphi (t)$ is certain source time function and $\delta(\x-\x_s)$ Dirac delta distribution related to the spatial location of the source $(\x \in(r,\theta))$.

The density kernel $K_{\rho}$ and shear wave kernel $K_{\beta}$ in SH media are given by the following expressions \citep{tromp2005seismic,fichtner2009sensitivity}
\begin{align}
\begin{aligned}
K_{\rho} & = - \int_0^T \rho(\x) \partial_t u^{*}(\x,T-t) \cdot \partial_t u (\x,t) \dif t , \\
K_{\beta} & = 2 K_{\mu} = - 4 \int_0^T \mu(\x) D^{*}(\x,T-t) : D (\x,t) \dif t ,
\label{eq.sensitity_kernels}
\end{aligned}
\end{align}
where $D$ and $D^{*}$ denote the traceless strain deviator and its waveform adjoint, respectively.

\subsection{Waveforms}

Probably, the most evident misfit measure $(\E)$ between data $(\u^{\text{obs}})$ and synthetics $(\u)$ is the waveform difference given by eq. \eqref{eq.chi_squared}. Following the procedure suggested in Sec. \ref{se.Mistif_fuctions}, after having defined the misfit measure, the next step is to find its first variation $(\delta \E)$, in this case given by the following expression
\begin{align}
\delta \E = \sum_{r} \int_{0}^{T} \int_{\Omega} \underbrace{(\u^{\text{obs}}  - \u) \sum_{r} \delta(\x-\x^r)}_{\text{adjoint source}} \delta \u \dif^3 \x \dif t,
\label{eq.waveform_mistfit_variation}
\end{align}  
which allows us to identify the waveform adjoint source $f^{*}$ given by (see eq. \eqref{eq.adjoint_wave_equation} and Fig. \ref{Fig.Lagrangian_Formalism})
\begin{align}
f^{*} = \frac{1}{M_r} (\u^{\text{obs}} - \u) \sum_{r} \delta(\x-\x^r) w_r ,
\label{eq.Waveform_AdjS}
\end{align}
i.e., the adjoint source $f^{*}$ is the term inside the integral of eq. \eqref{eq.waveform_mistfit_variation} excluding $\delta \u$. Note that we have introduced $w_r$ as a time window that allows us the possibility of only using part of the waveform, and $M_r$ defined as the energy of the data
\begin{align}
	M_r=\int_0^T w_r(t)(\u^{\text{obs}})^2(\x,t)\dif t ,
	\label{eq.Mr_def}
\end{align}
which equalizes amplitudes of different time windows in the inversion. When lacking observations, we can assume that $\u^{\text{obs}}=0$ and by normalizing by the energy of the synthetics, we can write the waveform adjoint source eq. \eqref{eq.Waveform_AdjS} as follows
\begin{align}
	f^{*}= \frac{\int_{t_1}^{t_2} \u  \dif t}{\int_{t_1}^{t_2} \u^2 \dif t} \sum_{r} \delta(\x-\x^r)  ,
	\label{eq.Waveform_noobs_AdjS}
\end{align} 
where we have selected the waveform between certain time window $[t_1,t_2]$. Note that, if we omit a scaling factor, the adjoint source is simply a selected time window of a portion of interest of the seismogram.

\subsection{Travel-Time: Cross-correlation}

Cross-correlation travel-time measurements provide robust estimates in full-waveform inversion, provided that synthetic and observed waveforms are similar enough.  Assume that we look to minimize the cross-correlation $(\circledast)$ travel-time misfit measure, between synthetics $(\u)$ and observations $(\u^{\text{obs}})$, given by the following expression
\begin{align}
\begin{aligned}
\E & = \frac{1}{2} \int_{\Omega} \int_0^T  \left[ \u^{\text{obs}}\circledast \u \right]^2 \sum_r \delta(\x-\x^r) \dif t \dif^3 \x , \\ 
& = \frac{1}{2} \int_{\Omega} \int_0^T  \left[  \int_{t_1}^{t_2}\u(t+\tau) \u^{\text{obs}}(t) \right]^2 \sum_r \delta(\x-\x^r) \dif t  \dif t \dif^3 \x .
\end{aligned}
\label{eq.crosscorrelation_tt_misfit}
\end{align} 
Taking the variation of eq. \eqref{eq.crosscorrelation_tt_misfit}, we can write the following
\begin{align}
\delta \E = \int_{\Omega} \int_0^T \int_{t_1}^{t_2} \left[ \u(t+\tau) \u^{\text{obs}}(t) \right] \delta \u(t+\tau) \sum_r \delta(\x-\x^r) \dif t\dif t \dif^3 \x .
\end{align}
Let us define $t'=t+\tau$, i.e., $t=\tau-t'$, to get
\begin{align}
\begin{aligned}
\delta \E &= \int_{\Omega} \int_0^T \int_{t_1}^{t_2} \left[ \u(t') \u^{\text{obs}}(\tau-t') \right] \delta \u(t') \sum_r \delta(\x-\x^r) \dif t' \dif t \dif^3 \x , \\
&= \int_{\Omega} \int_0^T \u^{\text{obs}} \star \u \sum_r  \delta(\x-\x^r) \delta \u \dif t \dif^3 \x 
\end{aligned}
\end{align}
where the symbol $(\star)$ denotes the convolution operator. We can identify the adjoint source as
\begin{align}
f^* = \u^{\text{obs}} \star \u  \sum_r   \delta(\x-\x^r) .
\label{eq.cross-correlation_adj_source}
\end{align}
Equation \eqref{eq.cross-correlation_adj_source} is not very useful when we do not have data to compare and at the same time we look to gain some insight into the travel-time finite-frequency dependence of the waves of interest. To overcome this limitation, we assume the following misfit function \citep{marquering1999three,dahlen2000frechet,bos2011proof}
\begin{align}
	\E(\tau) = \u^{\text{obs}} \circledast \u \sum_{r}  \delta (\x-\x^r) = \int_{t_1}^{t_2} \u(t-\tau) \u^{\text{obs}}(t) \dif t \sum_{r}  \delta (\x-\x^r) .
	\label{eq.crosscorrelation}
\end{align}
Assuming that \citep{marquering1999three,dahlen2000frechet}
\begin{align}
	\u^{\text{obs}}(t) = \u(t) + \delta \u(t) ,
\end{align}
the cross-correlation eq. \eqref{eq.crosscorrelation} may be written as 
\begin{align}
	\E(\tau) = \gamma(\tau) + \delta \gamma(\tau),
\end{align}
with
\begin{align}
	\gamma(\tau) = \int_{t_1}^{t_2} \u(t-\tau) \u(t) \dif t , \qquad \delta \gamma(\tau) = \int_{t_1}^{t_2} \u(t-\tau) \delta \u(t) \dif t .
\end{align}
The travel-time shift $(\delta T= T^{\text{obs}}-T)$ between the observed and the synthetic data are determined by finding the maximum of $\E(\tau)$ using the Taylor series expansion \citep{bos2011proof}
\begin{align}
	\E(\tau) = \gamma(0) + \partial_{\tau} \gamma(0) \delta \tau + \frac{1}{2}\partial_{\tau\tau} \gamma(0) (\delta \tau)^2 + \delta \gamma(0) + \partial_{\tau} \delta \gamma(0) \delta \tau + \frac{1}{2}\partial_{\tau\tau} \delta \gamma(0) (\delta \tau)^2 + ...
\end{align}
A critical point can be found using
\begin{align}
\partial_{\tau} \E(\tau) = \partial_{\tau} \gamma(0) + \partial_{\tau\tau } \gamma(0) \delta \tau + \partial_{\tau} \delta \gamma(0) + \partial_{\tau \tau} \delta \gamma(0) \delta \tau + ... = 0,
\end{align}
rearranging we have
\begin{align}
	\delta \tau \approx - \frac{\partial_{\tau} \gamma(0) + \partial_{\tau} \delta \gamma(0)}{\partial_{\tau \tau} \delta \gamma(0) + \partial_{\tau \tau} \gamma(0) }.
\end{align}
Setting $\delta_{\tau} \gamma (0)=0$ and $\delta_{\tau \tau} \delta \gamma (0)=0$ (see \cite{bos2011proof} for further details), we thus write
\begin{align}
	\delta \tau \approx - \frac{\partial_{\tau} \delta \gamma(0)}{\partial_{\tau\tau } \gamma(0)} .
\end{align}
Upon making the identification $\delta \tau = \delta T$ we can write 
\begin{align}
	\delta T = - \frac{\int_{t_1}^{t_2} \partial_t \u  \dif t}{\int_{t_1}^{t_2}   \u   \partial^2_t \u  \dif t} \delta \u ,
\end{align}
where negative travel-time shift, $\delta T < 0$, corresponds to an advance in the arrival of the observed waveform $\u^{\text{obs}}(t)$ with respect to the synthetic waveform $\u(t)$, whereas a positive travel-time shift $\delta T > 0$, corresponds to a delay. We can identify the travel-time adjoint source as
\begin{align}
	f^{*} = - \frac{\int_{t_1}^{t_2} \partial_t \u \dif t}{\int_{t_1}^{t_2} \u  \partial^2_t \u  \dif t} \sum_{r} \delta(\x-\x^r)  = \frac{\int_{t_1}^{t_2} \partial_t \u \dif t}{\int_{t_1}^{t_2} \partial_t \u  \partial_t \u  \dif t} \sum_{r} \delta(\x-\x^r),
    \label{eq.TravelTime_AdjS}
\end{align}
which is independent of the observations and it can be used to gain insight into the nature of finite-frequency sensitivity kernels of the waves of interest.

\subsection{Amplitude}

Following \cite{dahlen2002frechet}, we can define the synthetic and observed body wave amplitudes to be the root mean square (rms) averages of the corresponding time-domain waveforms $\u_{\text{syn}}(t)$ and $\u_{\text{obs}}(t)$ over the arrival interval $t_1\leq t\leq t_2$ as follows
\begin{align}
A_{\text{syn}} =  \sqrt{\frac{1}{t_2-t_1} \int_{t_1}^{t_2} \u_{\text{syn}}^2(t) \dif t },\qquad A_{\text{obs}} = \sqrt{\frac{1}{t_2-t_1} \int_{t_1}^{t_2} \u_{\text{obs}}^2(t) \dif t} .
\label{eq.amplitude_definitions}
\end{align}

The amplitude misfit function can thus be conveniently defined as follows 
\begin{align}
\E = \frac{1}{2} \int_{0}^{T}  \int_{\Omega} \left[\frac{A_{\text{obs}}-A_{\text{syn}}}{A_{\text{syn}}}\right]^2 \sum_r \delta(\x-\x^r) \dif^3 \x \dif t . 
\label{eq.amplitude_misfit}
\end{align}
The variation of the misfit function eq. \eqref{eq.amplitude_misfit} is given by
\begin{flalign}
\delta \E & = \int_{0}^{T} \int_{\Omega} \left[\frac{A_{\text{obs}}}{A_{\text{syn}}}-1\right] \delta \ln A_{\text{syn}} \sum_r \delta(\x-\x^r) \dif^3 \x \dif t .
\label{eq.amplitude_misfit_func_grad}
\end{flalign}
Using the definition of $A_{\text{syn}}$ given by eq. \eqref{eq.amplitude_misfit} and dropping the mnemonic subscript upon the synthetic displacement we can write the following
\begin{align}
	\delta \ln A_{\text{syn}} = \frac{\delta A_{\text{syn}}}{A_{\text{syn}}}  = \frac{\int_{t_1}^{t_2} \u  \dif t}{\int_{t_1}^{t_2} \u^2 \dif t}  \delta \u ,
\end{align}
substituting allows us to isolate the term $\delta \u$, which is needed to identify the adjoint source. We thus write
\begin{align}
\delta \E & = \int_{0}^{T} \int_{\Omega} \left[ \left(\frac{A_{\text{obs}}}{A_{\text{syn}}}-1\right) \frac{\int_{t_1}^{t_2} \u \dif t}{\int_{t_1}^{t_2} \u^2  \dif t}  \sum_r \delta(\x-\x^r) \right]\delta \u  \dif^3 \x \dif t,
\end{align}
where we can identify the amplitude adjoint source given by the following expression
\begin{align}
f^{*}= \left(\frac{A_{\text{obs}}}{A_{\text{syn}}}-1\right) \frac{\int_{t_1}^{t_2} \u  \dif t}{\int_{t_1}^{t_2} \u^2 \dif t} \sum_{r} \delta(\x-\x^r)  .
\label{eq.Amplitude_AdjS}
\end{align}

We can now start to draw connections between waveform, travel-time and amplitude adjoint sources.

\subsection{Mathematical Equivalence Between Waveform, (Cross-correlation) Travel-Time and Amplitude Fr\'echet Kernels}

If we compare the adjoint sources for waveforms eq. \eqref{eq.Waveform_noobs_AdjS}, travel-times eq. \eqref{eq.TravelTime_AdjS} and amplitudes eq. \eqref{eq.Amplitude_AdjS}, we can observe the similarities in the mathematical structures. If we assume that we have no observations $(\u_{\text{obs}}=0)$, the adjoint sources for waveforms and amplitudes are (omitting scaling factors) mathematically the same:
\begin{align}
    f^{*}_{\text{wavef.}} (\u)= f^{*}_{\text{ampli.}} (\u)= \frac{1}{M_r} \int_{t_1}^{t_2} \u  \dif t  \sum_{r} \delta(\x-\x^r) ,
\end{align}
where $M_r$ is certain normalization factor given by the squared norm of the integration variable. In the same way (omitting scaling factors), the travel-time adjoint source eq. \eqref{eq.TravelTime_AdjS} can be written as follows
\begin{align}
    f^{*}_{\text{travel t.}} (\u) = f^{*}_{\text{wavef.}} (\partial_t \u) = f^{*}_{\text{ampli.}} (\partial_t \u)= \frac{1}{M_r} \int_{t_1}^{t_2} \partial_t \u  \dif t  \sum_{r} \delta(\x-\x^r) ,
\end{align}
where we have applied the adequate normalization factor $M_r$ given by the squared norm of the integration variable.

For computing sensitivity kernels we need to have access to the forward $u(x,t)$ and adjoint $u^*(x,t)$ wavefields. We have shown that the adjoint wavefield can be defined as a linear-functional applied to the forward wavefield (eq.  \eqref{eq.Adjoint_displacement_adjoint}). As previously suggested \citep[e.g.][]{tarantola1984inversion,Tarantola1988,tromp2005seismic,fichtner2006adjoint}, we can define
\begin{align}
    u^*(\x,t) =  u(\x,T-t) 
    \label{eq.adjoint_displ}
\end{align}
and compute waveform, travel-time and amplitude kernels. However, using eq. \eqref{eq.Adjoint_displacement_adjoint}, we can also rightfully choose
\begin{align}
u^*(\x,t) =  \Delta^{+}_{t}  u(\x,T-t) , \qquad \text{and/or} \qquad u^*(\x,t) =  \Delta^{+}_{\x}  u(\x,T-t),
\label{eq.adjoint_vel}
\end{align}
where the forward operators $\Delta^{+}$ applied in time and/or space to a continuous function $f(\x,t)$ are defined as follows
\begin{align}
\begin{aligned}
\Delta^{+}_{t} f(\x,t) & = f(\x,t + \Delta t) - f(\x,t) , \\
\Delta^{+}_{\x} f(\x,t) & = f(\x+\Delta \x,t ) - f(\x,t) .
\end{aligned}
\label{eq.operators}
\end{align}

Note that applying the forward operator in time and/or space to the adjoint variable (eq. \eqref{eq.adjoint_vel}) does not change its units, therefore, travel-time (density) kernels computed using eq.  \eqref{eq.adjoint_displ} are equivalent (up to a certain scaling value) to waveform and/or amplitude kernels. This means that, the Fr\'echet sensitivity kernels for waveforms, amplitudes and travel-times are fundamentally equivalent and depend of the choice of the adjoint wavefield. This is physically sound since, by modifying waveforms, we implicitly modify amplitudes and travel-times. We expect no new different information to be obtained when using by separate any of these adjoint sources.

\paragraph{Illustrative Example} To develop intuition and illustrate the previous statement, we compute density $(\rho)$ amplitude sensitivity kernels (eq. \eqref{eq.sensitity_kernels}) in 2.5D SH media for the S, ScS and sS and sScS waves. We assume a Gaussian wavelet as a source time function, i.e., $f_t=\exp(-2\pi^2 f_0^2 (t-t_0)^2)$, where $t_0=30$s is the time delay, and with a dominant period of $1/f_0=30$s. We place the source at a depth of 600 km and at 45$^{\circ}$ from the symmetry axis in order to avoid high sensitivity close to the source which can considerably affect the shape of the sensitivity kernels, due to the boundary conditions required by the axisymmetric model \citep{jahnke2008global}. 

\begin{figure}
    \begin{center}
        \includegraphics[width=0.8\textwidth]{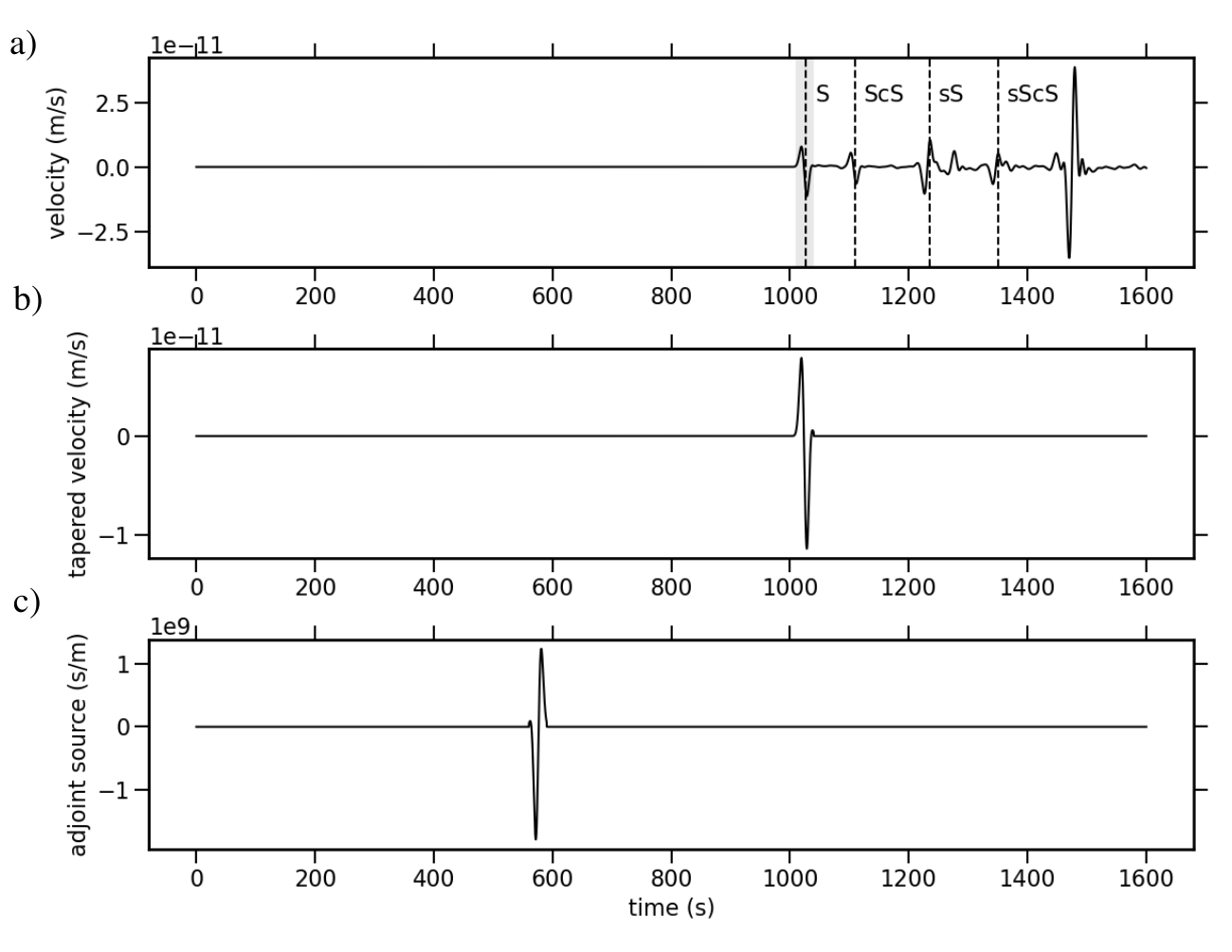}
        \caption{Steps taken in the construction of the adjoint source time function needed for the calculation of the S wave sensitivity kernel. a) Transverse velocity seismogram recorded at a distance of 60$^{\circ}$ from an earthquake at a depth of 600 km. b) Isolated S wave used for computing the Fr\'echet sensitivity kernel. c) Time reversed source time function used in the adjoint simulation constructed using the isolated S wave scaled by the inverse squared norm of itself.}
        \label{Fig.seismograms}
    \end{center}
\end{figure}

We record the velocity field ($v_{\varphi}$, see eqs. \eqref{eq.SH_25D_equations_of_motion}) at a distance of 60$^{\circ}$, where we can observe the clear arrival of the S, ScS and sS and sScS waves, see Fig. \ref{Fig.seismograms}--a, where theoretical travel-times are represented with dotted vertical lines which correspond to PREM \citep{DZIEWONSKI1981297} and are computed using the TauP toolkit \citep{crotwell1999taup} implemented in Obspy \citep{Krischer2015}. \\

\begin{figure}
	\begin{center}
		\includegraphics[width=1\textwidth]{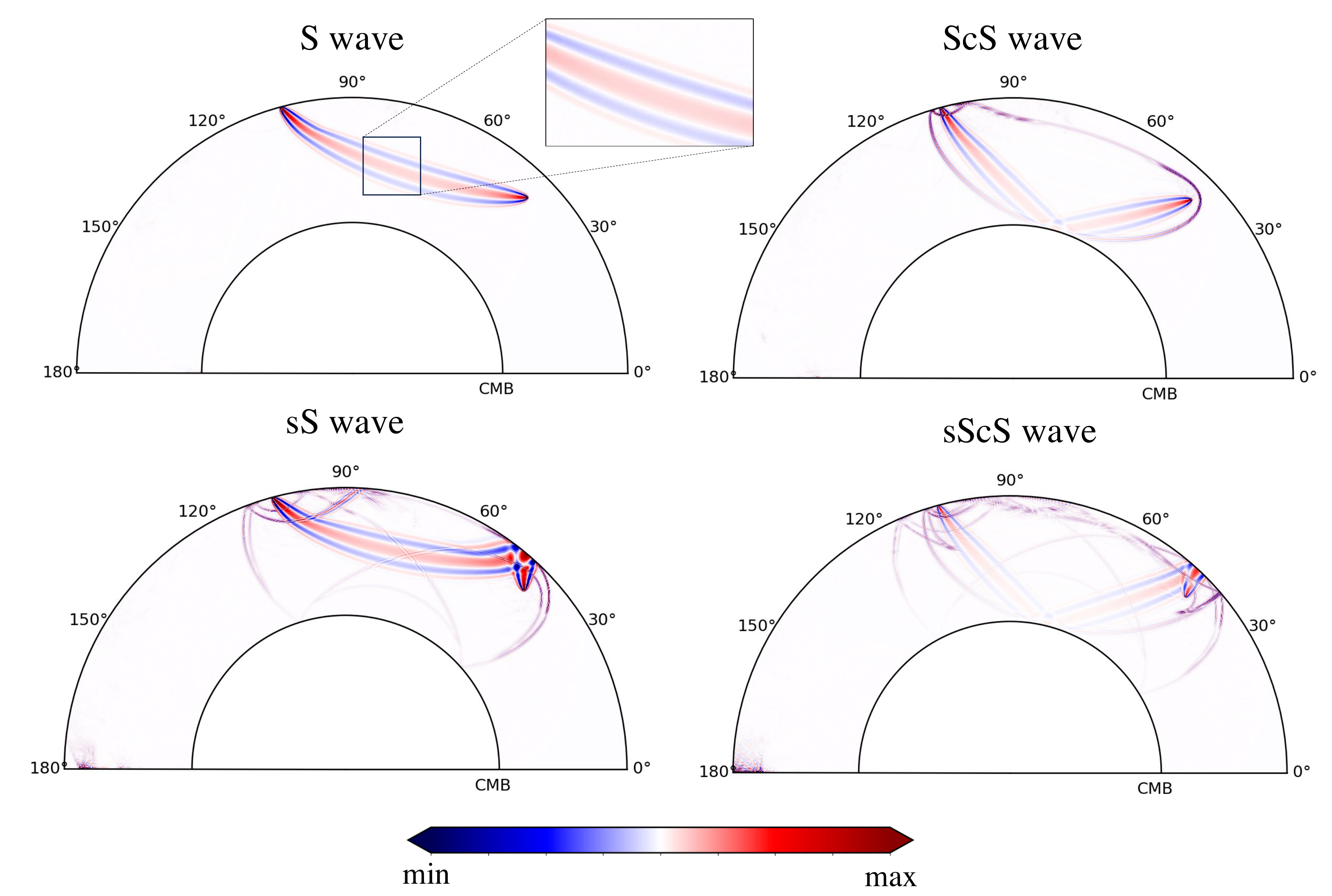}
		\caption{Normalized amplitude density $\rho$ kernels for the S, ScS, sS and sScS waves placing the source at 45$^{\circ}$.}
		\label{Fig.SH_kernels}
	\end{center}
\end{figure}

We first compute amplitude sensitivity (Fr\'echet) kernels assuming the adjoint wavefield to be the time reversed displacement recorded at the receiver, i.e., $u^*(x,t) = u(x,T-t)$. For the calculation of each sensitivity kernel we isolate the wave of interest (see Fig. \ref{Fig.seismograms}--b for an example of the S wave), we then normalize it using eq. \eqref{eq.Mr_def} and time reverse it (see Fig. \ref{Fig.seismograms}--c) in order to use it for the adjoint calculation (see Fig. \ref{Fig.kernels}). The obtained results are presented in Fig. \ref{Fig.SH_kernels}, where we can observe many characteristic features of these types of sensitivity kernels. A most evident characteristic property is the very large sensitivity at the source and the receiver locations. This is related to the single point approximation made to the Dirac delta distribution in order to implement the point source that excites the forward and adjoint wavefields. We can also observe that the largest sensitivity is always observed along the ray path of the wave (center along the sensitivity). We can also observe highly oscillatory sensitivity is clearly visible especially for the ScS, sS and sScS waves (see Fig. \ref{Fig.SH_kernels}). This is because these waves are not as isolated as the S wave due to the presence of crustal reverberation and/or precursors and/or other waves that arrive at the same time of the main waves.   

\begin{figure}
	\begin{center}
		\includegraphics[width=1\textwidth]{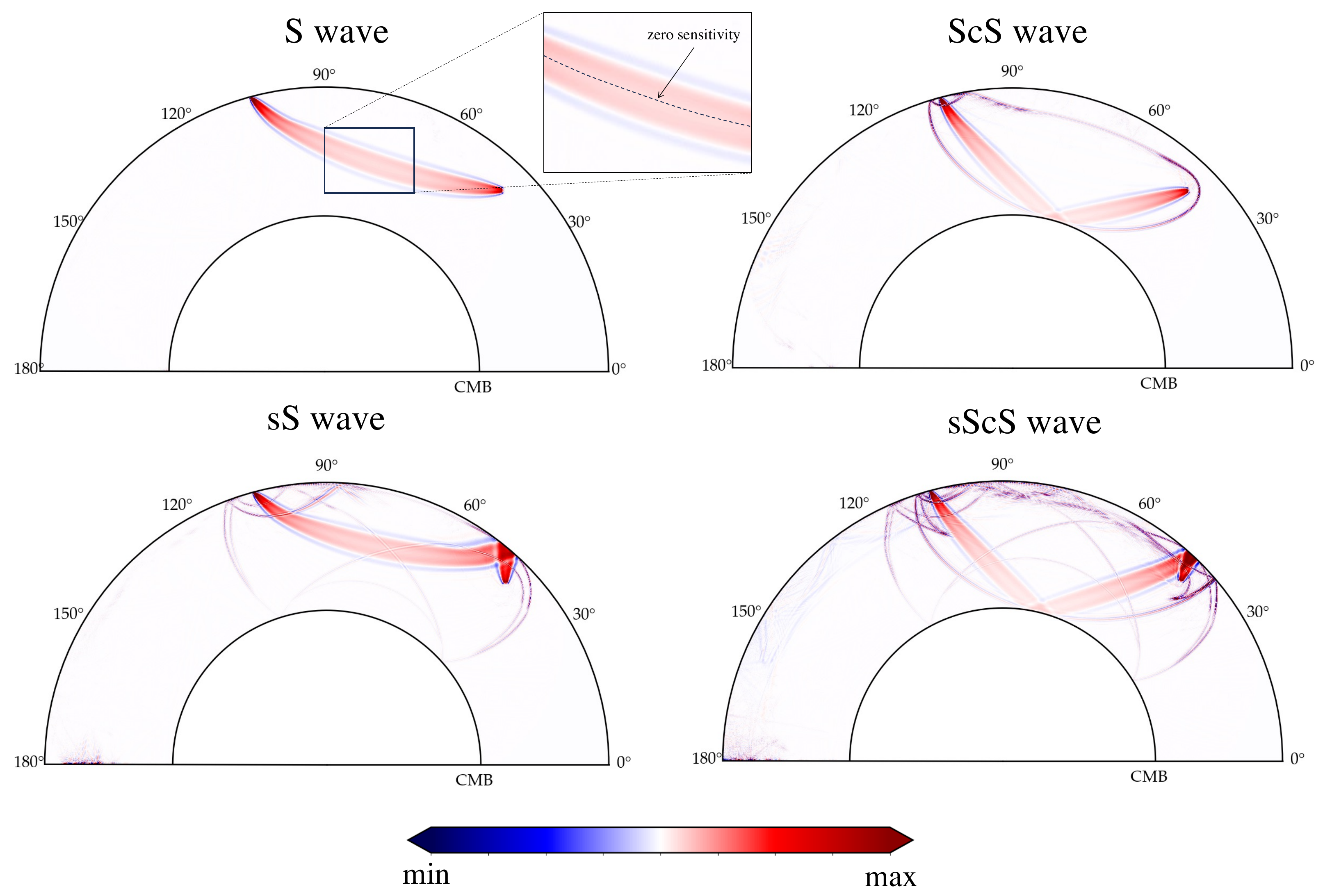}
		\caption{Normalized travel-time density $\rho$ kernels for the S, ScS, sS and sScS waves placing the source at 45$^{\circ}$.}
		\label{Fig.SH_kernels2}
	\end{center}
\end{figure}

We now compute travel-time sensitivity kernels assuming the adjoint wavefield to be the time derivative reversed displacement recorded at the receiver, i.e., $u^*(x,t) =  \partial_t u(x,T-t)$. Note that up to a certain scaling factor, this is equivalent to assume that $u^*(x,t) = \Delta^{+}_{t} u(x,T-t)$. Fig. \ref{Fig.SH_kernels2} shows the obtained results.

We can observe that, for this case, the zero sensitivity is located along the ray path for the four waves. This, well known feature, contradicts our physical intuition since one can hardly expect that the Fr\'echet kernel has no sensitivity where the ray is located. \cite{marquering1999three} described the shape of the sensitivity kernel as resembling a hollow banana for the 2D case, and as a doughnut in the 3D case. Since then, the expression banana-doughnut (sensitivity) kernel has been used and (very efficiently) spread among the seismological community to refer to these types of Fr\'echet kernels. The physical interpretation of the zero sensitivity inside the kernel has been a topic of controversy. A well known dispute was published almost two decades ago \citep{de2005sensitivity,van2005banana,de2005reply,montelli2006comment,van2006reply}. 

We can observe that computing waveform and travel-time sensitivity kernels is mathematically equivalent (up to certain scaling factor) and depend on the selected adjoint source. However, the obtained Fr\'echet kernels are fundamentally different in nature. However, to relate these differences to only changes in travel-time and/or amplitude, seems to depend on how we define the adjoint wavefield  $u^*(x,t)$, the particular type of seismic wave, the frequency content and the near receiver geology \citep{fichtner2010book}. In Sec. \ref{sec.banana_doughnut_paradox} we show how we can reconcile all these different points of view into a single misfit function, thus solving the famous banana-doughnut paradox  \citep{de2005sensitivity,van2005banana,de2005reply,montelli2006comment,van2006reply}.

\subsection{Instantaneous Phase and Envelope}

Separating phases from amplitudes is required to overcome the excessive nonlinearity introduced by the waveform functional eq. \eqref{eq.waveform_mistfit_variation} \citep{fichtner2008theoretical}. Using the analytic signal concept, \cite{bozdaug2011misfit} proposed a simple and elegant way to use instantaneous phase and envelope differences in the misfit function in the time domain. The analytic signal $f_a(t)$ is constructed from a real signal $f(t)$ and its Hilbert transform $\mathcal{H} (f(t))$ as follows 
\begin{align}
f_a(t) = f(t) + \imag \mathcal{H} (f(t)) ,
\end{align}
where $\imag$ is the imaginary unit. The analytic signal can be written in a polar form
\begin{align}
f_a(t) = \Env(t) e^{\imag \phi(t)},
\end{align}
where the instantaneous amplitude or envelope $E(t)$ and the instantaneous phase or phase angle $\phi(t)$ are respectively given by
\begin{align}
\Env(t) = \sqrt{\RealPart(f_a)^2 + \ImagPart(f_a)^2}, \qquad \phi(t) = \arctan \frac{\ImagPart(f_a)}{\RealPart(f_a)} .
\label{eq.Envelope_Phase_Def}
\end{align}

\paragraph{Instantaneous Phase:} 

The squared instantaneous phase misfit is defined as
\begin{align}
\E = \frac{1}{2} \int_{\Omega} \int_{0}^{T} \left[\phi^{\text{obs}}-\phi^{\text{syn}}\right]^2 \sum_r \delta(\x-\x^r) \dif t  \dif^3 \x .
\label{eq.Phase_misfit}
\end{align} 
After some algebra and dropping the mnemonic notation for the synthetic instantaneous phase, envelope and displacement, the variation in the misfit function eq. \eqref{eq.Phase_misfit} can be written as follows (see \cite{bozdaug2011misfit} for further details)
\begin{align}
\delta \E = \int_{\Omega} \int_{0}^{T} \left[\left(\phi^{\text{obs}}-\phi\right) \frac{\mathcal{H}(\u)}{\Env^2} + \mathcal{H} \left\{\left(\phi^{\text{obs}}-\phi\right) \frac{\u}{\Env^2}\right\}\right] \delta \u\sum_r \delta(\x-\x^r) \dif t \dif^3 \x ,
\end{align}
where we can identify the phase misfit adjoint source as
\begin{align}
f^{*} = \left[\left(\phi^{\text{obs}}-\phi\right) \frac{\mathcal{H}(\u)}{\Env^2} + \mathcal{H} \left\{\left(\phi^{\text{obs}}-\phi\right) \frac{\u}{\Env^2}\right\}\right] w_r \sum_{r} \delta(\x-\x^r) ,
\end{align}
where we have introduced $w_r$ as a time window that allows us the possibility of using only part of the waveform.

\paragraph{Envelope:} 

The envelope misfit is defined as the squared logarithmic ratio of the envelopes of the observed $\Env^{\text{obs}}$ and synthetic $\Env^{\text{syn}}$ waveforms as follows
\begin{align}
\E = \frac{1}{2} \int_{\Omega} \int_{0}^{T} \left[\ln \frac{\Env^{\text{obs}}}{\Env^{\text{syn}}}\right]^2 \sum_r \delta(\x-\x^r) \dif t \dif^3 \x .
\label{eq.Envelope_mmisfit}
\end{align}
After some algebra and dropping the mnemonic notation for the synthetic envelope and displacement, the variation in the misfit function eq. \eqref{eq.Envelope_mmisfit} can be written as (see \cite{bozdaug2011misfit} for further details)
\begin{align}
\delta \E = - \int_{\Omega} \int_{0}^{T} \left[\ln \left(\frac{\Env^{\text{obs}}}{\Env}\right) \frac{\u}{\Env^2} - \mathcal{H} \left\{ \ln \left(\frac{\Env^{\text{obs}}}{\Env}\right) \frac{\mathcal{H}(\u)}{\Env^2}\right\}\right] \delta \u \sum_r \delta(\x-\x^r) \dif t \dif^3 \x,
\end{align}
where we can identify the phase misfit adjoint source as
\begin{align}
f^{*} = \left[\ln \left(\frac{\Env^{\text{obs}}}{\Env}\right) \frac{\u}{\Env^2} - \mathcal{H} \left\{ \ln \left(\frac{\Env^{\text{obs}}}{\Env}\right) \frac{\mathcal{H}(\u)}{\Env^2}\right\}\right] w_r \sum_{r} \delta(\x-\x^r) ,
\end{align}
where we have introduced $w_r$ as a time window that allows us the possibility of using only part of the waveform.

\subsection{Travel-time Fr\'echet Kernels Using Envelopes: The Banana-Doughnut Paradox Solved}
\label{sec.banana_doughnut_paradox}

\paragraph{The Problem with Cross-correlation Travel-Time Measurements:} Travel-time measurements obtained using a cross-correlation error function (eq. \eqref{eq.crosscorrelation_tt_misfit}) are only useful when observed and numerical waveforms are similar enough. This is not an easy precondition to fulfill since it requires an accurate earth model that is able to produce realistic synthetic waveforms. It thus becomes a circular problem in adjoint tomography: to have an accurate (initial) earth model that allows us to use the cross-correlation error function (eq. \eqref{eq.crosscorrelation_tt_misfit}) to be able to invert for a \textit{more accurate} earth model. Several evident questions arise: (i) how accurate the initial model has to be? (ii) how accurate the final (inverted) model can/will be? (iii) does it worth (at all) the effort? These questions have been and are still under debate and do not have (yet) a unique answer.

\begin{figure}
    \begin{center}
        \includegraphics[width=1\textwidth]{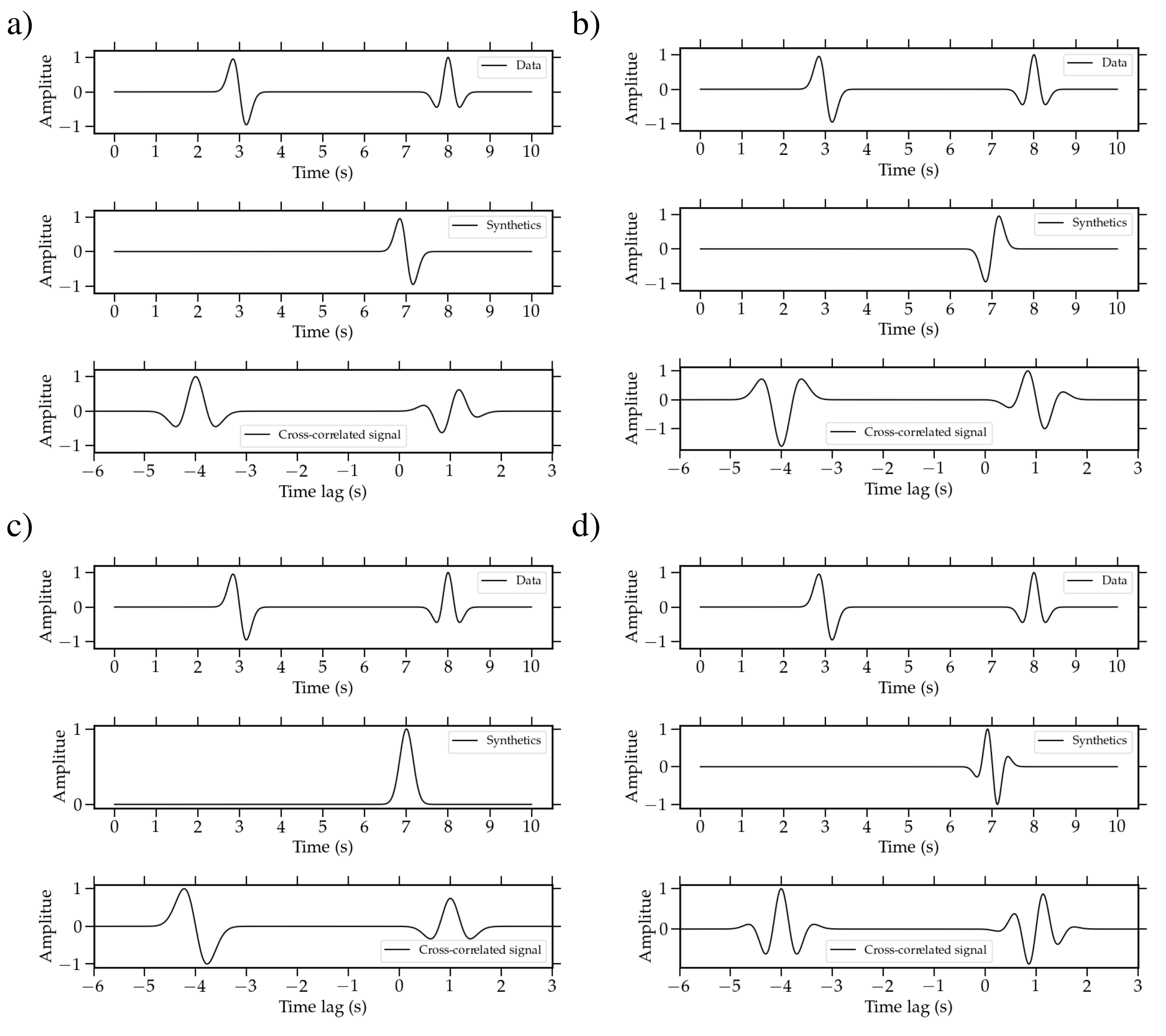}
        \caption{Cross-correlation travel-time measurements obtained when using different synthetic data.}
        \label{Fig.Cross_correlation_examples}
    \end{center}
\end{figure}

To illustrate the requirement of similarity between data and synthetics, let us consider that the observations are a time series with two main signals: The first one the derivative of a Gaussian function located at 3s and the second one a Ricker function located at 8s with both having a dominant period of 1s and a unit maximum amplitude (see Fig. \ref{Fig.Cross_correlation_examples}). As a first example, let us consider that synthetics are given by the derivative of a Gaussian function located at 7s with dominant period of 1s and a unit maximum amplitude. Fig. \ref{Fig.Cross_correlation_examples}--a shows the data, synthetics and travel-time cross-correlation measurements predicted. We can observe that the maximum of the cross-correlation predicts two arrivals at -4s and at 1.25s. The first one corresponds to the first arrival in the data if we have measured the travel-time of the wave at the center of the waveform. The second arrival corresponds to the second wave when we measure the travel-time at the maximum of the amplitude of the data. This demonstrates the ambiguity that exists when we measure travel-times of data using cross-correlations: should we pick the travel-time at the beginning? at the maximum amplitude? or at the center? of the wave. Depending on the choice, the maximum of the cross-correlation will give different results when waveform of data and synthetics at different. As a second example, let us consider that data and synthetics have the same waveform but with different polarities (see Fig. \ref{Fig.Cross_correlation_examples}--b). The travel-time now predicted by the maximum of the cross-correlation signal shows two main picks: the first one at -4s which correctly predicts the travel-time of the data and the second one of 0.75s that underestimates the correct travel-time of 1s of the second wave. Figure \ref{Fig.Cross_correlation_examples}--c,d shows two additional examples when waveforms between data and synthetics are completely different. In both cases we are unable to predict a consistent travel-time difference between data and synthetics.

Therefore, we have demonstrated that it is essential having comparable waveforms between data and synthetics to compute reliable travel-time differences using cross-correlations. In the next, we overcome this limitation.

\paragraph{An Envelope Based Travel-Time Misfit Function} To overcome the requirement/limitation of an accurate earth model that allows us to compute synthetics that have waveforms that are similar (enough) to data, which constrains the use of cross-correlation travel-time measurements in realistic applications, we define the travel-time misfit measure as the time difference at which the maximum of the envelopes of the observed and synthetic waveforms is found. This can be mathematically written as follows
\begin{align}
\E = \left[\max_t \left(\Env^{\text{obs}} \right) -\max_t \left(\Env^{\text{syn}} \right) \right] \sum_r \delta(\x-\x^r) ,
\label{eq.traveltime_misfit}
\end{align} 
where the symbol $\displaystyle \max_t ()$ refers to the time that the maximum of $()$ occurs and $\Env$ refers to the envelope defined in eq. \eqref{eq.Envelope_Phase_Def}. We note that the misfit function eq. \eqref{eq.traveltime_misfit} is, unlike the cross-correlation misfit function eq. \eqref{eq.crosscorrelation_tt_misfit}, a scalar function. We thus map the two data points difference given in eq. \eqref{eq.traveltime_misfit} to a time dependent function as follows
\begin{align}
\E = \frac{1}{2}\int_{\Omega} \int_{0}^{T} \left[\max_t \left(\Env^{\text{obs}} \right) \delta(t-t_{\text{max}}^{\text{obs}}) - \max_t \left(\Env^{\text{syn}} \right) \delta(t-t_{\text{max}}^{\text{syn}}) \right]^2 \sum_r \delta(\x-\x^r) \dif t  \dif^3 \x ,
\label{eq.traveltime_misfit2}
\end{align} 
where 
\begin{align}
t_{\text{max}}^{\text{obs}}=\max_t \left(\Env^{\text{obs}} \right), \qquad t_{\text{max}}^{\text{syn}}=\max_t \left(\Env^{\text{syn}} \right) .
\end{align}

Note that the mapping of the two travel time scalars to the function given in eq. \eqref{eq.traveltime_misfit2} is one choice of a functional mapping space among many others. This methodology is equivalent to the well known technique of Reproducing Kernel in Hilbert Spaces (RKHS) used in machine learning \citep[e.g.][]{paulsen2016introduction,berlinet2011reproducing}. The first variation $\delta \E$ is given by
\begin{align}
\delta \E = -\int_{\Omega} \int_{0}^{T} \left[\max_t \left(\Env^{\text{obs}} \right) \delta(t-t_{\text{max}}^{\text{obs}}) -\max_t \left(\Env^{\text{syn}} \right)  \delta(t-t_{\text{max}}^{\text{syn}})  \right] \max_t \delta \left(\Env^{\text{syn}} \right) \delta(t-t_{\text{max}}^{\text{syn}}) \sum_r \delta(\x-\x^r) \dif t  \dif^3 \x .
\end{align} 
Dropping the mnemonic subscript, the envelope of the synthetic seismogram is $\Env^{\text{syn}} = \sqrt{\u^2+ \mathcal{H} (\u)^2}$, thus $\delta \left(\Env^{\text{syn}} \right)$ becomes (see \cite{bozdaug2011misfit} Appendix)
\begin{align}
\begin{aligned}
\int_{0}^{T} \delta \left(\Env^{\text{syn}} \right) \dif t & = \int_{0}^{T} \frac{\u \delta \u + \mathcal{H} (\u)  \delta \mathcal{H} (\u)}{\sqrt{\u^2+ \mathcal{H} (\u)^2}} \dif t\\
& =\int_{0}^{T} \left[\frac{\u }{\sqrt{\u^2+ \mathcal{H} (\u)^2}} - \mathcal{H}\left(\frac{ \mathcal{H} (\u)}{\sqrt{\u^2+ \mathcal{H} (\u)^2}} \right)\right] \delta \u \dif t .
\end{aligned}
\end{align}
Substituting we write
\begin{align}
\begin{aligned}
\delta \E = &  -\int_{\Omega} \int_{0}^{T} \left[\max_t \left(\Env^{\text{obs}} \right) \delta(t-t_{\text{max}}^{\text{obs}}) -\max_t \left(\Env^{\text{syn}} \right)\delta(t-t_{\text{max}}^{\text{syn}})  \right] \\
&  \max_t \left[\frac{\u}{\sqrt{\u^2+ \mathcal{H} (\u)^2}} - \mathcal{H}\left(\frac{ \mathcal{H} (\u)}{\sqrt{u^2+ \mathcal{H} (\u)^2}} \right)\right] \delta(t-t_{\text{max}}^{\text{syn}}) \delta \u \sum_r \delta(\x-\x^r) \dif t  \dif^3 \x ,
\end{aligned}
\end{align} 
where we can identify the (envelope) travel-time adjoint source as
\begin{align}
f^* = \max_t \left(\Env^{\text{obs}} \right) \delta(t-t_{\text{max}}^{\text{obs}}) -\max_t \left(\Env^{\text{syn}} \right)\delta(t-t_{\text{max}}^{\text{syn}}) .
\label{eq.travel_time_adjoint_src_env}
\end{align}

\paragraph{Numerical Implementation of the Dirac Distribution} a numerical approximation of the Dirac delta distribution is required to implement the (envelope) travel-time adjoint source given by eq. \eqref{eq.travel_time_adjoint_src_env}. This can be done in numerous ways \citep[e.g.][]{di2021remarks,yang2009smoothing,engquist2005discretization,tornberg2004numerical,hosseini2016regularizations,min2008robust,smereka2006numerical}. In general, it is assumed that the discrete Dirac delta distribution $\delta_{\Delta x}$ is represented by a tensor product of a single-variable kernel $\phi_{\Delta x}$ 
\begin{align}
\delta_{\Delta x} (\x) = \frac{1}{\Delta x} \prod_{i=1}^{n} \phi_{\Delta x} \left( \frac{x_i}{\Delta x}\right), \qquad \x=(x_1,x_2,...,x_n)^T,
\end{align}
where $\Delta x$ is the grid size. Common choices of $\phi_{\Delta x}$ are given by the following expression \citep{di2021remarks}
\begin{align}
\phi_{\Delta x} = \frac{1}{2^s} \sum_{l=0}^{s}\binom{s}{l} \psi(r-l),
\end{align}
where $\psi$ is a function of compact support (bump function). Examples of $\delta_{\Delta x} $ are
\begin{align}
\delta_{\Delta x}^L (x) = \begin{cases}
\frac{1}{(n/2)^2} \left( \frac{n}{2} - \abs{x} \right) \qquad \text{if} \quad \abs{\frac{x}{\Delta x}} \leq \frac{n}{2} \\
 0 \qquad \text{otherwise} \end{cases}, \qquad \delta_{\Delta x}^{\cos} (x) = \begin{cases}
\frac{1}{n} \left( 1 - \cos \left( \frac{2\pi x}{n} + \pi \right) \right) \qquad \text{if} \quad \abs{\frac{x}{\Delta x}} \leq \frac{n}{2} \\
 0 \qquad \text{otherwise} .
\end{cases}
\label{eq.Dirac_delta_approx}
\end{align}

\begin{figure}
    \begin{center}
        \includegraphics[width=1\textwidth]{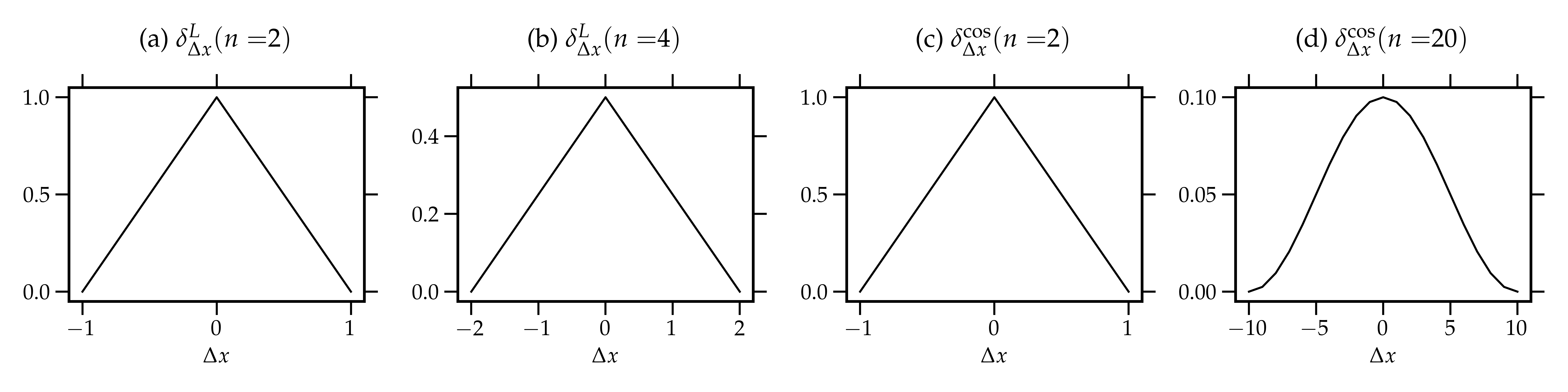}
        \caption{Different approximations of the Dirac delta distribution (see eqs. \eqref{eq.Dirac_delta_approx}).}
        \label{Fig.Dirac_delta}
    \end{center}
\end{figure}

Fig. \ref{Fig.Dirac_delta} shows two different scenarios with $n=2$ and $n=6$ for the $\delta_{\Delta x}^L $ and $\delta_{\Delta x}^{\cos}$ approximations. We can observe that both approximations vary their amplitudes depending on the choice of the number of grid spacing $\Delta x$ taken. This is because the total area of the Dirac delta distribution must be equal to one, independently of the choice of $n$. We can also observe that the cosine approximation $\delta_{\Delta x}^{\cos}$ represents a smoother version of the Dirac delta distribution compared to the linear version $\delta_{\Delta x}^L$ only when $n$ is large enough. For the case $n=2$ both approximations are equivalent (see Fig. \ref{Fig.Dirac_delta} (a) and (c)). The choice between one over the other (or any other one) in numerical simulations of wave propagation has, however, not been studied in detail up today.  

\paragraph{Illustrative Example:} We first assume that we do not have observations; therefore, the (envelope) travel-time adjoint source (eq. \eqref{eq.travel_time_adjoint_src_env}) is reduced to
\begin{align}
f^* = -\max_t \left(\Env^{\text{syn}} \right)\delta(t-t_{\text{max}}^{\text{syn}}) .
\label{eq.noobs_travel_time_adjoint_src_env}
\end{align}

\begin{figure}
    \begin{center}
        \includegraphics[width=1\textwidth]{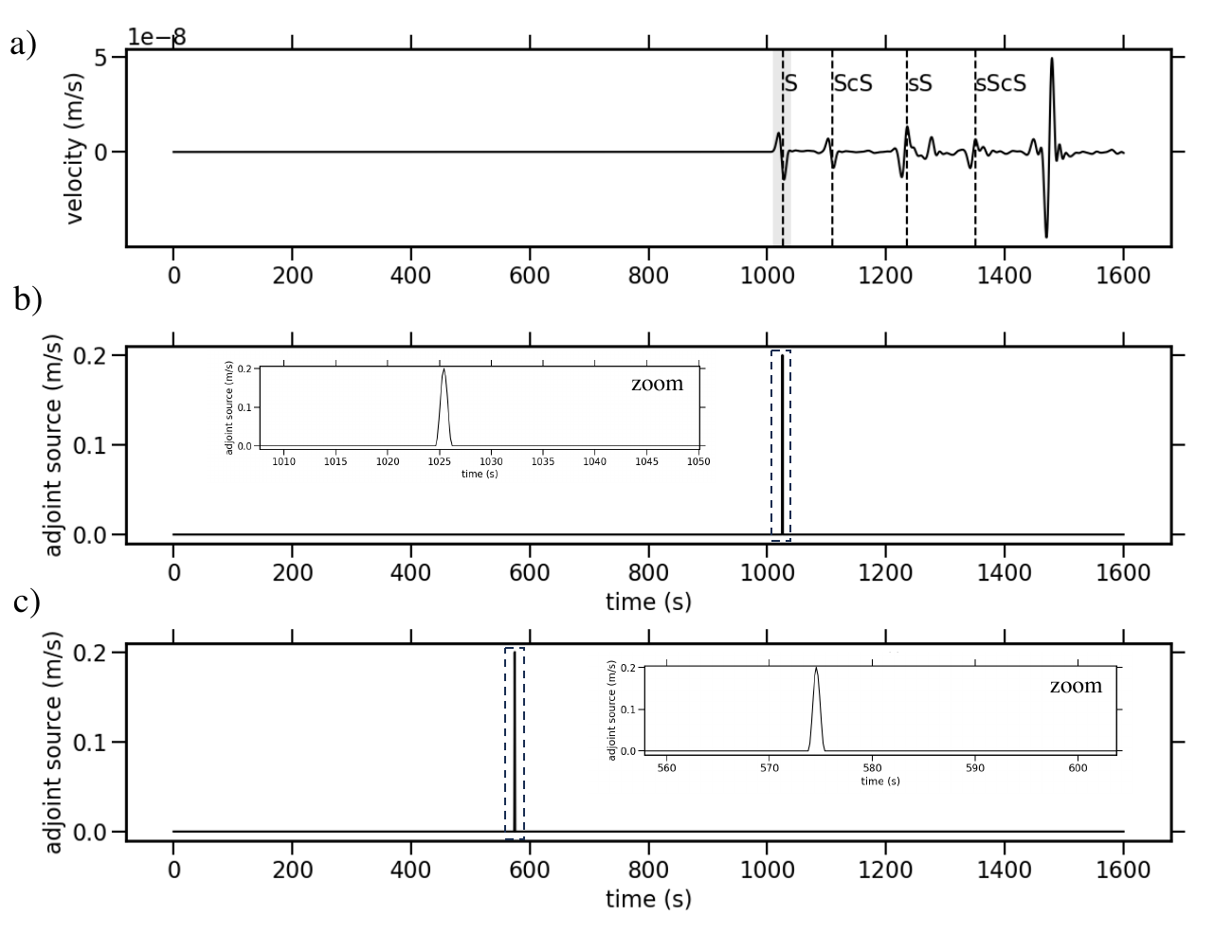}
        \caption{Steps taken in the construction of the Dirac adjoint source time function needed for the calculation of the S wave envelope travel-time sensitivity kernel. a) Transverse velocity seismogram recorded at a distance of 60$^{\circ}$ by an earthquake at a depth of 600 km. b) Dirac delta cosine approximation using $n=20$  (see eq. \eqref{eq.Dirac_delta_approx}) located at the maximum of the S wave envelope and used for computing the Fr\'echet sensitivity kernel. c) Time reversed source time function used in the adjoint simulation.}
        \label{Fig.Dirac_seismograms}
    \end{center}
\end{figure}

As before, to compute Fr\'echet kernels we assume global SH wave propagation in the whole mantle. Figure \ref{Fig.Dirac_seismograms} shows the construction of the adjoint Dirac delta source time function using the S wave travel-time. We approximate the Dirac delta distribution with the cosine approximation using $n=20$ (see eq. \eqref{eq.Dirac_delta_approx}). This guaranties that a Gaussian shape like of the Dirac delta distribution is obtained (see Fig. \ref{Fig.Dirac_delta}) and translates into a $20\Delta t = 3.2$s, which is much smaller compared to the dominant period of the simulation $\sim 30$s. We have additionally also tested using a smaller number $n$ and the sensitivity kernels obtained are the same.

\begin{figure}
    \begin{center}
        \includegraphics[width=1\textwidth]{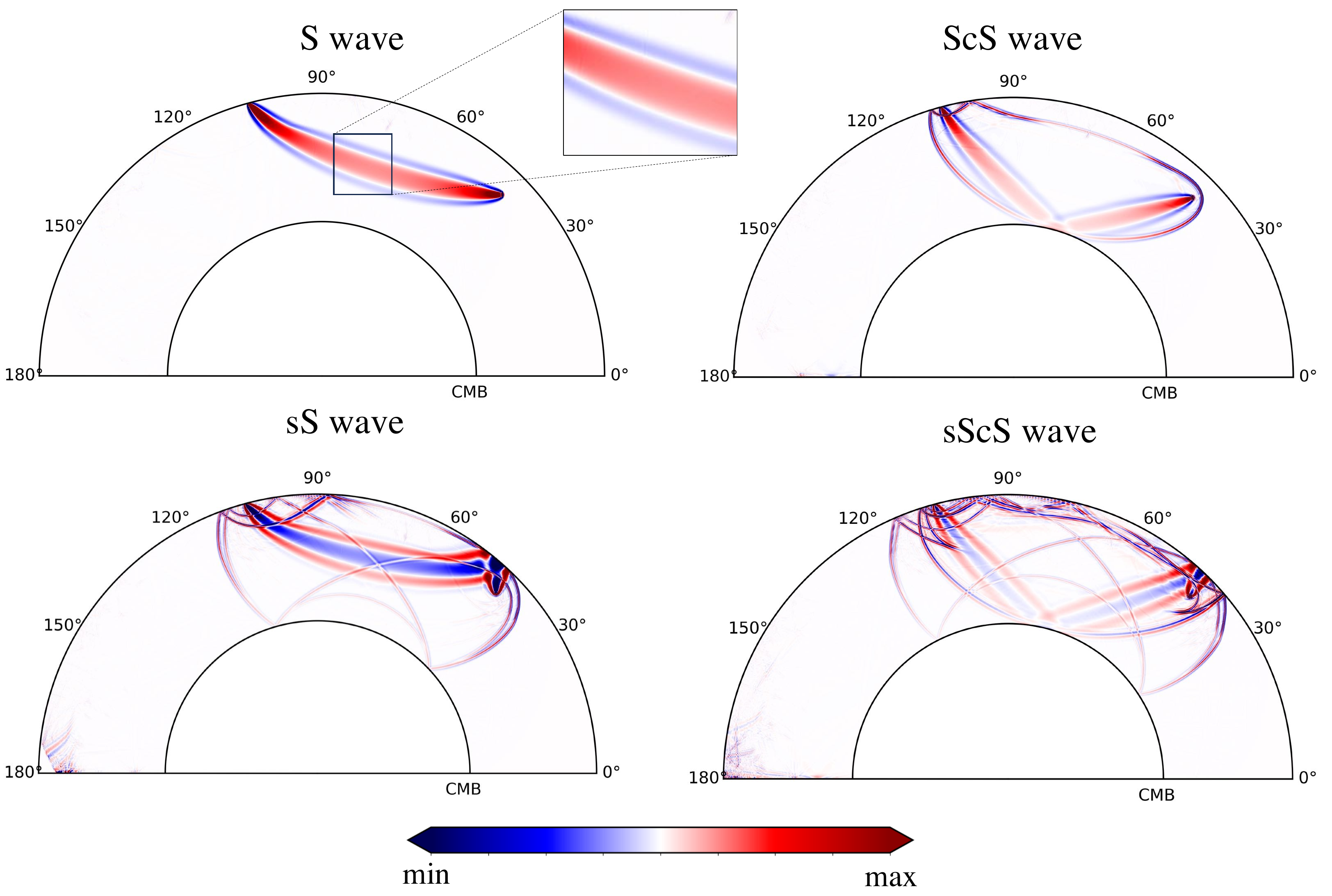}
        \caption{Normalized envelope travel-time density $\rho$ kernels for the S, ScS, sS and sScS waves placing the source at 45$^{\circ}$.}
        \label{Fig.SH_Dirac_kernels}
    \end{center}
\end{figure}

Figure \ref{Fig.SH_Dirac_kernels} shows the obtained density $\rho$ Fr\'echet kernels. We can observe that, in contrast to travel-time sensitivity kernels predicted using cross-correlation, an extreme (maximum or minimum) is concentrated along the ray path. In contrast to the sensitivity obtained with travel-time cross-correlation measurements, these results are physically sound since they explain that the largest sensitivity of the travel-time measurement is precisely concentrated along the ray-path. Whether it is a maximum and/or a minimum depends on the choice of the sign and the polarization of the wave in study with respect to the polarization chosen for the Dirac delta. For example, S and ScS waves have the same polarization of the Dirac delta distribution (see Fig. \ref{Fig.Dirac_seismograms}), unlike sS and sScS which show opposite polarization thus the change in the sign of the obtained sensitivity.
When synthetics travel-times are close to observations $(\text{syn}\to \text{obs})$ we can write the following
\begin{align}
\lim_{\text{syn}\to \text{obs}} f^* = \delta(t-t_{\text{max}}^{\text{obs}}) - \delta(t-t_{\text{max}}^{\text{syn}}) \approx \partial_t \delta_{\Delta t}^{\cos} \Delta t = \begin{cases}
- \frac{2 \pi}{n^2} \sin \left( \frac{2 \pi t}{n} \right) \Delta t \qquad \text{if} \quad \abs{\frac{t}{\Delta t}} \leq \frac{n}{2} \\
 0 \qquad \text{otherwise} . \end{cases}
\label{eq.close_obs_travel_time_adjoint_src_env}
\end{align}

\begin{figure}
    \begin{center}
        \includegraphics[width=1\textwidth]{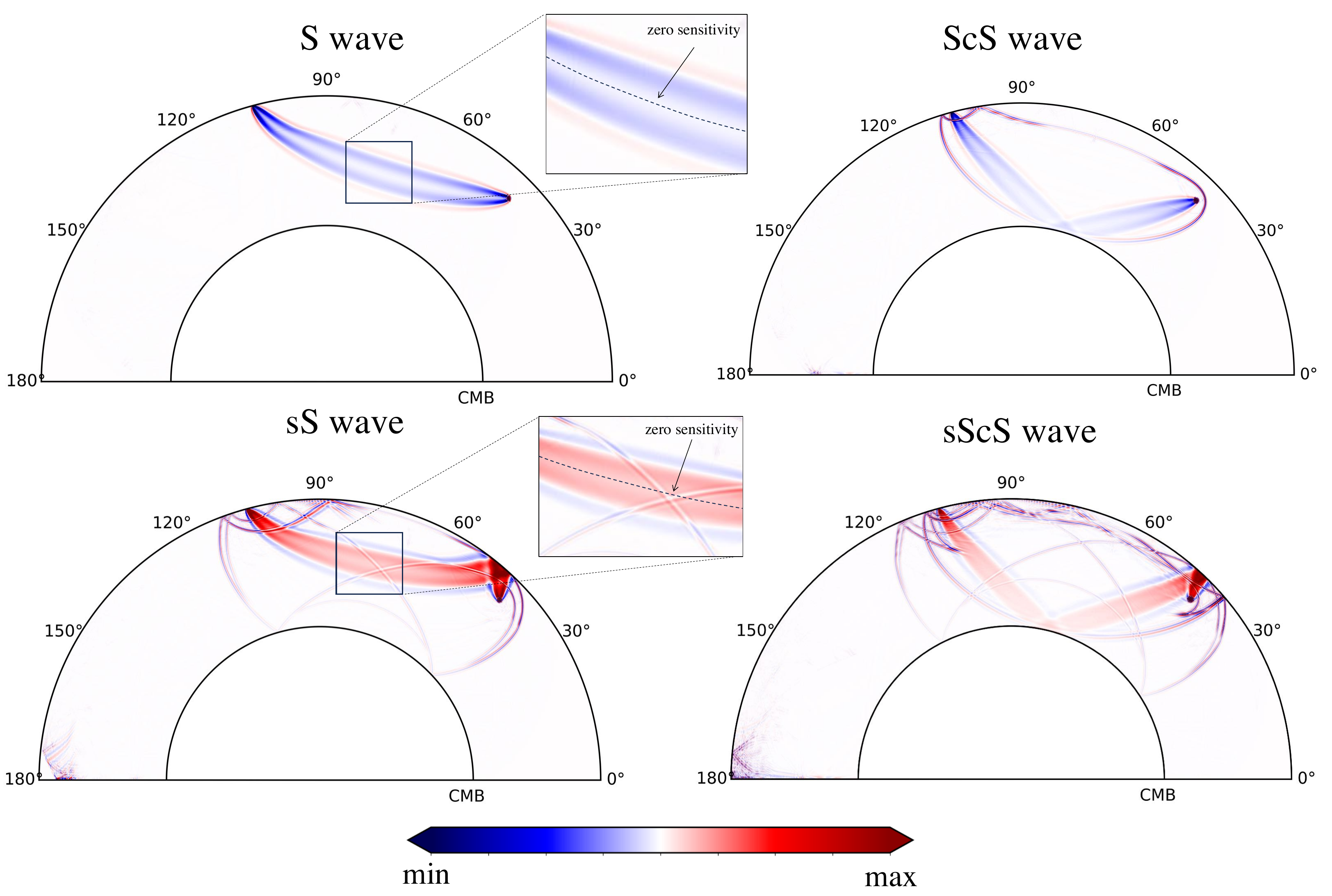}
        \caption{Normalized travel-time envelope density $\rho$ kernels for the S, ScS, sS and sScS waves placing the source at 45$^{\circ}$ considering that data and synthetics travel-times are similar.}
        \label{Fig.SH_Dirac_diff_kernels}
    \end{center}
\end{figure}

This means that the adjoint envelope travel-time will be comparable to the derivative of a function of compact support (bump function). Figure \ref{Fig.SH_Dirac_diff_kernels} shows the Fr\'echet kernels obtained, where we can observe zero sensitivity along the ray path. This is physically sound since it explains that, when data and synthetics are similar, the ray-path has no influence anymore in the travel-time and the sensitivity will be located around it. This explains and clarifies the long standing debate of the physical interpretation of the zero sensitivity along the ray path for cross-correlation travel-time Fr\'echet kernels \citep{dahlen2000frechet,de2005sensitivity,van2005banana,de2005reply,montelli2006comment,van2006reply}. The reason is simply because in order to compute travel-time differences using cross-correlation between data and synthetic waveforms, one has to assume that data and synthetics are similar enough. Otherwise, the cross-correlation cannot be performed. The assumption of the similarity between data and synthetics is thus a limitation of cross-correlation travel-time measurements. The advantage of the envelope travel-time misfit function introduced in this work is that, unlike cross-correlation, it is independent of the similarity between data and synthetic waveforms. This, of course, comes with the requirement that an accurate source location is needed as well as taking into account accurate variations in attenuation \citep[e.g.][]{karaouglu2018inferring}.

\begin{figure}
    \begin{center}
        \includegraphics[width=1\textwidth]{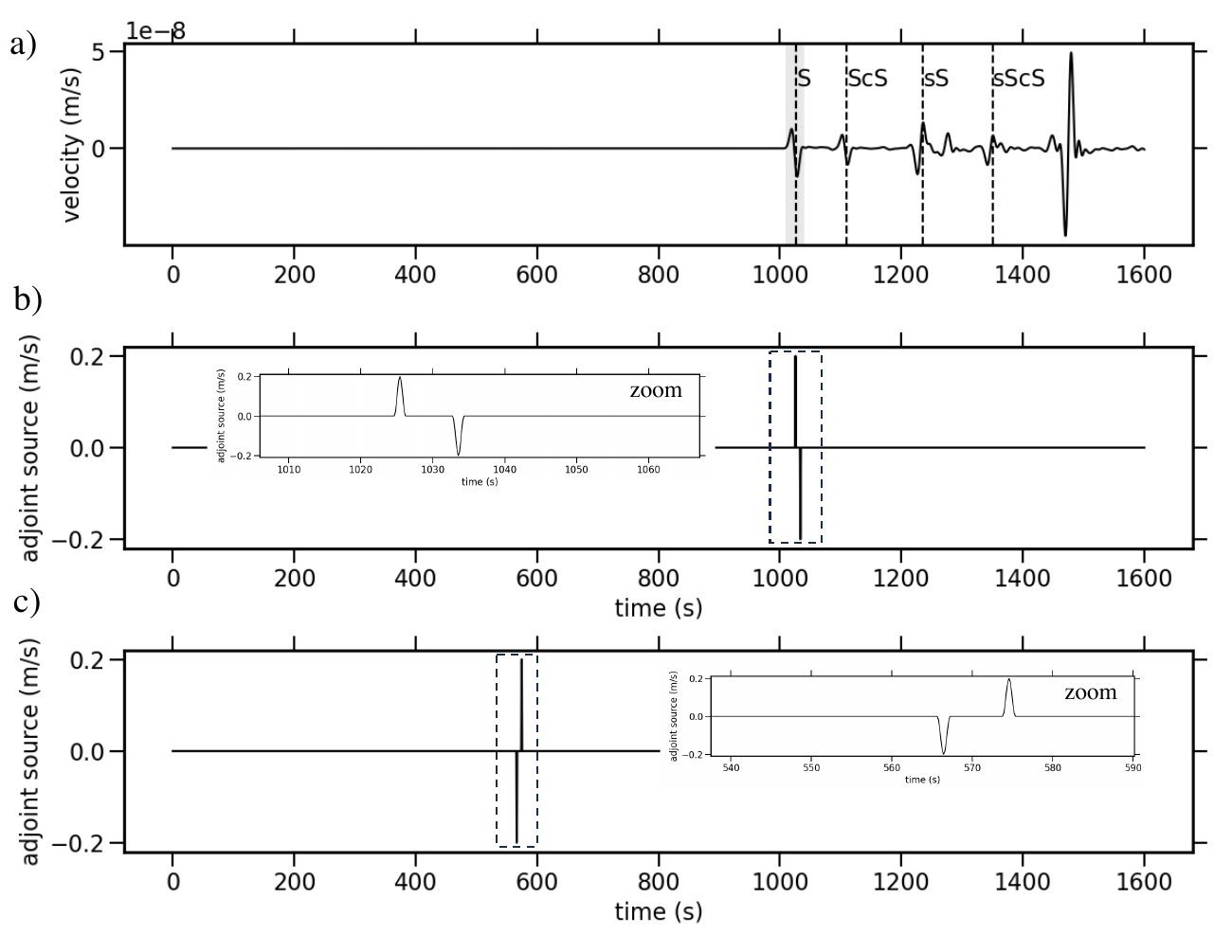}
        \caption{Steps taken in the construction of the Dirac adjoint source time function needed for the calculation of the S wave envelope travel-time sensitivity kernel when considering the presence of data. a) Transverse velocity seismogram recorded at a distance of 60$^{\circ}$ by an earthquake at a depth of 600 km. b) Dirac delta cosine approximation using $n=20$ (see eq. \eqref{eq.Dirac_delta_approx}) located at the maximum of the S wave envelope and used for computing the Fr\'echet sensitivity kernel, including an observation Dirac delta distribution located at -9 s of the synthetic S wave. c) Time reversed source time function used in the adjoint simulation.}
        \label{Fig.Dirac_seismograms_with_data}
    \end{center}
\end{figure}

When considering the presence of observations that are not close to synthetics, we need to take into account the differential travel-time between both signals. This results into placing two Dirac distributions at the required temporal locations. Figure \ref{Fig.Dirac_seismograms_with_data} shows the construction of the adjoint source considering that the time difference between data and synthetics is of 9 s. We can observe two Dirac distributions with opposite signs.

\begin{figure}
	\begin{center}
		\includegraphics[width=0.8\textwidth]{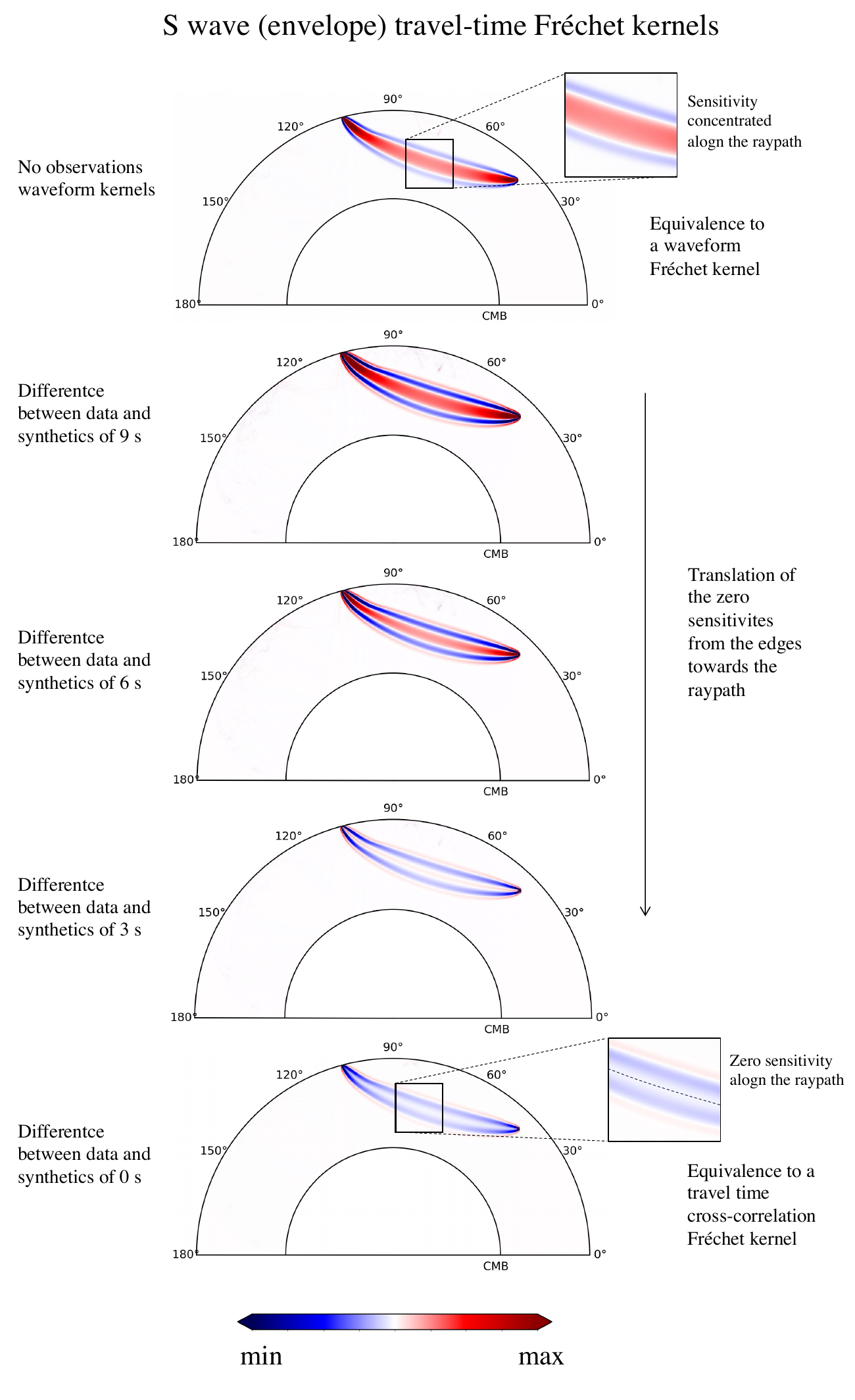}
		\caption{Normalized travel-time envelope density $\rho$ kernels for the S when considering the presence of data at different time locations.}
		\label{Fig.data_envelope_S_kernels}
	\end{center}
\end{figure}

Evidently, the resulting travel-time envelope Fr\'echet kernels will depend on how far or close observations and synthetics are. Figure \ref{Fig.data_envelope_S_kernels} shows the resulting kernels when considering different scenarios. We can observe that when no data are considered, an extremum of the Fr\'echet kernel is observed along the ray path. This, as previously shown, is equivalent to a waveform Fr\'echet kernel. When considering that observations are far from the synthetics (9s) we can observe that the observed Fr\'echet kernel still resembles waveform one. Interestingly, when the travel-time of data are close enough to synthetics, the sensitivity along the ray path disappears up to the point that the Fr\'echet kernel resembles a cross-correlation Fr\'echet kernel. Note that the extremum observed along the ray-path on the Fr\'echet kernels depends whether the sign chosen for the Dirac distribution corresponds to the same polarization (positive or negative) of the wave in study. 
We have thus clarified why cross-correlation travel-time measurements show a zero sensitivity along the ray-path and the equivalence between waveform and travel-time sensitivity measurements.

\subsection{Cross-convolution}

The cross-convolution measure \citep{menke2003cross} is designed to be applied to seismic phases that have complicated shapes, e.g., closely spaced arrivals. It retains information about the Earth structure but it is insensitive to source properties. It is defined as follows
\begin{align}
\begin{aligned}
\mathtt{w} & = u_j^{\text{obs}}(t) \star u_i(\mathbf{m},t) - u_i^{\text{obs}}(t) \star u_j(\mathbf{m},t) , \\
& = \int_{t_1}^{t_2} u_j^{\text{obs}}(\tau) u_i(t-\tau) \dif \tau - \int_{t_1}^{t_2} u_i^{\text{obs}}(\tau) u_j(t-\tau) \dif \tau ,
\label{eq.cross-convolution_measure}
\end{aligned}
\end{align} 
where the symbol $(\star)$ denotes the convolution operator and $u_i,u_j$ are two components of the synthetic seismograms and $u^{\text{obs}}_i,u^{\text{obs}}_j$ are two components of the observed seismograms. 

The method has been used to invert observations of shear wave splitting \citep{menke2003cross}, crustal reverberations for Earth structure \citep{bodin2014inversion} and sensitivity kernels have been studied in \cite{menke2017sensitivity}. The squared cross-convolution misfit function is defined as
\begin{align}
	\E = \frac{1}{2} \int_{\Omega} \int_{0}^{T} \left[\mathtt{w} \right]^2 \sum_r \delta(\x-\x^r) \dif t \dif^3 \x .
	\label{eq.crossconvolution_misfit}
\end{align} 
Using eq. \eqref{eq.normal_vector_displ}, the variation of the cross-convolution misfit function eq. \eqref{eq.crossconvolution_misfit} can be written as follows
\begin{align}
	\delta \E = \int_{\Omega} \int_{0}^{T} \int_{t_1}^{t_2} \left[\int_{t_1}^{t_2} u_j^{\text{obs}}(\tau) \u (t-\tau) - u_i^{\text{obs}}(\tau) \u (t-\tau) \right] \delta \u(t-\tau) \sum_r \delta(\x-\x^r) \dif \tau  \dif t \dif^3 \x ,
\end{align}
where we have assumed that $\delta u_i \to 0$ and $\delta u_j\to 0$ and that $\delta u_i = \hat{n}_i \delta\u$ and $ \delta u_j=\hat{n}_j \delta \u$. Let us define $t'=t-\tau$, i.e., $\tau=t'-t$, to get
\begin{align}
\begin{aligned}
\delta \E & = - \int_{\Omega} \int_{0}^{T} \int_{t_1}^{t_2} \left[ u_j^{\text{obs}}(t'-t) \u (t') - u_i^{\text{obs}}(t'-t)  \u(t')\right] \delta \u(t') \sum_r \delta(\x-\x^r) \dif t'  \dif t \dif^3 \x \\
& = - \int_{\Omega} \int_{0}^{T} \left[ \u \circledast u_j^{\text{obs}} -  \u \circledast u_i^{\text{obs}} \right] \delta \u  \sum_r \delta(\x-\x^r) \dif t\dif^3 \x ,
\end{aligned}
\end{align}
where we can identify the cross-convolution adjoint source as
\begin{align}
f^{*} = - w_r \left[ \u \circledast u_j^{\text{obs}} -  \u \circledast u_i^{\text{obs}}  \right] \sum_{r} \delta(\x-\x^r) ,
\label{eq.cross-convolution_adj_source}
\end{align}
where we have introduced $w_r$ as a time window that allows us the possibility of using only part of the waveform. Eq. \eqref{eq.cross-convolution_adj_source} is not very useful when we do not have data to compare at hand and at the same time we look to gain some insight into the cross-convolution finite-frequency dependence of waves of interest. To overcome this limitation we can assume the following cross-convolution misfit function
\begin{align}
\E =  \mathtt{w} \sum_r \delta(\x-\x^r) .
\label{eq.crossconvolution_misfit1}
\end{align} 
Assuming that $\u^{\text{obs}} = \u + \delta \u$ \citep{marquering1999three,dahlen2000frechet}, allows us to write eq. \eqref{eq.cross-convolution_measure} as follows
\begin{align}
\begin{aligned}
\mathtt{w} = &  u_j^{\text{obs}}(t) \star u_i(\mathbf{m},t) - u_i^{\text{obs}}(t) \star u_j(\mathbf{m},t) , \\
= &  \int_{t_1}^{t_2} u_j(\tau) u_i(t-\tau) \dif \tau + \int_{t_1}^{t_2} \delta u_j(\tau) u_i(t-\tau) \dif \tau \\
& - \int_{t_1}^{t_2} u_i(\tau) u_j(t-\tau) \dif \tau  - \int_{t_1}^{t_2} \delta u_i(\tau) u_j(t-\tau) \dif \tau, \\
= & \gamma_{ji}(\tau) + \delta \gamma_{ji}(\tau) + \gamma_{ij}(\tau) + \delta \gamma_{ij}(\tau) ,
\end{aligned}
\end{align} 
with
\begin{align}
\gamma_{ji}(\tau) = \int_{t_1}^{t_2} u_j(t-\tau) u_i(\tau) \dif \tau , \qquad \delta \gamma_{ji} (\tau) = \int_{t_1}^{t_2}  \delta u_j(\tau) u_i(t-\tau) \dif \tau,
\end{align}
and with an equivalent definition by interchanging the sub-indexes $ij$. Using the Taylor series expansion we can write
\begin{align}
\begin{aligned}
\mathtt{w}(\tau) = & \gamma_{ji}(0) + \partial_{\tau} \gamma_{ji}(0) \delta \tau + \frac{1}{2}\partial_{\tau\tau} \gamma_{ji}(0) (\delta \tau)^2 + \delta \gamma_{ji}(0) + \partial_{\tau} \delta \gamma_{ji}(0) \delta \tau + \frac{1}{2}\partial_{\tau\tau} \delta \gamma_{ji}(0) (\delta \tau)^2 + ... \\
& + \gamma_{ij}(0) + \partial_{\tau} \gamma_{ij}(0) \delta \tau + \frac{1}{2}\partial_{\tau\tau} \gamma_{ij}(0) (\delta \tau)^2 + \delta \gamma_{ij}(0) + \partial_{\tau} \delta \gamma_{ij}(0) \delta \tau + \frac{1}{2}\partial_{\tau\tau} \delta \gamma_{ij}(0) (\delta \tau)^2 + ...
\end{aligned}
\end{align}
A critical point can be found using $\partial_{\tau} \mathtt{w}(\tau)=0$, i.e.,
\begin{align}
\begin{aligned}
\partial_{\tau} \mathtt{w}(\tau) = & \partial_{\tau} \gamma_{ji}(0) + \partial_{\tau\tau } \gamma_{ji}(0) \delta \tau + \partial_{\tau} \delta \gamma_{ji}(0) + \partial_{\tau \tau} \delta \gamma_{ji}(0) \delta \tau + ... \\
& + \partial_{\tau} \gamma_{ij}(0) + \partial_{\tau\tau } \gamma_{ij}(0) \delta \tau + \partial_{\tau} \delta \gamma_{ij}(0) + \partial_{\tau \tau} \delta \gamma_{ij}(0) \delta \tau + ... = 0,
\end{aligned}
\end{align}
rearranging we can write
\begin{align}
\delta \tau \approx - \frac{\partial_{\tau} \gamma_{ji} (0) + \partial_{\tau} \delta \gamma_{ji} (0) + \partial_{\tau} \gamma_{ij} (0) + \partial_{\tau} \delta \gamma_{ij} (0)}{\partial_{\tau \tau} \delta \gamma_{ji} (0) + \partial_{\tau \tau} \gamma_{ji}(0) +\partial_{\tau \tau} \delta \gamma_{ij} (0) + \partial_{\tau \tau} \gamma_{ij}(0)}.
\end{align}
Setting $\delta_{\tau} \gamma_{ij} (0)=\delta_{\tau} \gamma_{ji} (0)=0$ and assuming $\delta_{\tau \tau} \delta \gamma_{ij} (0)=\delta_{\tau \tau} \delta \gamma_{ji} (0)=0$, we thus write
\begin{align}
\delta \tau \approx - \frac{\partial_{\tau} \delta \gamma_{ji} (0) + \partial_{\tau} \delta \gamma_{ij} (0)}{\partial_{\tau \tau} \gamma_{ji}(0) + \partial_{\tau \tau} \gamma_{ij}(0)}.
\end{align}
Upon making the identification $\delta \tau = \delta T$ and assuming that $\delta u_i \to 0$ and $\delta u_j\to 0$ and that $\delta u_i = \hat{n}_i \delta\u$ and $ \delta u_j=\hat{n}_j \delta \u$, we can write 
\begin{align}
\delta T = - \left( \frac{\int_{t_1}^{t_2} \partial_t u_i  \dif t}{\int_{t_1}^{t_2}  2 \u \partial^2_t u_i  \dif t} +  \frac{\int_{t_1}^{t_2} \partial_t u_j \dif t}{\int_{t_1}^{t_2}  2\u \partial^2_t u_j   \dif t} \right) \delta \u .
\end{align}
We can identify the (data free) cross-convolution adjoint source as
\begin{align}
f^{*} = - \left( \frac{\int_{t_1}^{t_2} \partial_t u_i  \dif t}{\int_{t_1}^{t_2}  2 \u \partial^2_t u_i  \dif t} +  \frac{\int_{t_1}^{t_2} \partial_t u_j \dif t}{\int_{t_1}^{t_2}  2\u \partial^2_t u_j   \dif t} \right)  \sum_{r} \delta(\x-\x^r)  .
\label{eq.crossconvolution_adj_src_no_data}
\end{align}

In eq. \eqref{eq.crossconvolution_adj_src_no_data}, we have introduced a new way to compute cross-convolution sensitivity kernels without the need of operator theory and/or the Born approximation. Since we are running 2.5D SH simulations only, we leave numerical illustrations of Fr\'echet kernels using eq. \eqref{eq.crossconvolution_adj_src_no_data} for future contributions.

\subsection{Rotational Seismology: Can We Only Image Shallow Structures?}

Rotational seismology is a field in growth. New information obtained from portable broadband rotation sensor technology \citep[e.g.][]{bernauer2012measurements,bernauer2018blueseis3a,brokevsova2012rotaphone,jaroszewicz2012usefulness} combined with the information provided by conventional seismometers, opens a broad spectrum of applications for i) better resolving the Earth structure  \citep[e.g.][]{igel2021romy,fichtner2009sensitivity,bernauer2020dynamic,bernauer2012measurements,trifunac2006effects,bernauer2009inferring,bernauer2014reducing,reinwald2016improved,abreu2023deep}, ii) tilt corrections \citep[e.g.][]{bernauer2020dynamic,lindner2017seafloor},  (iii) earthquake source characterization \citep[e.g.][]{yuan2021seismic,donner2018retrieval,donner2016inversion,reinwald2016improved,cao2021simulation}, (iv) seismic exploration  \citep[e.g.][]{sollberger20186,li2017tutorial}, (v) volcano seismology \citep[e.g.][]{wassermann2020six}, (vi) structural engineering, microzonation planning \citep[e.g.][]{trifunac2006effects,schreiber2009application,zembaty2021rotation,murray2021characterization,gueguen2021torsional,simonelli2021monitoring} and (vii) seismic anisotropy \citep[e.g.][]{Noe2022,tang2023single}. Recent reviews can be found in e.g.  \cite{li2017tutorial,schmelzbach2018advances}. 

The adjoint method was first applied in rotational seismology by \cite{fichtner2009sensitivity} using operator theory. In the next, we present a new and (what we consider to be a) simplified version using the variational formalism.

\subsubsection{Transverse Waves}

A technique of rotational seismology states that the ratio between the transverse acceleration $a_T$ recorded by translational seismometers and the vertical rotation rate $\dot{\Omega}_z$ recorded by rotational seismometers is proportional to the apparent shear wave velocity as follows \citep{igel2005rotational}
\begin{align}
\beta_a = -\frac{1}{2} \frac{a_T}{\dot{\Omega}_z}  = \frac{1}{p},
\label{eq.plane_wave_ratio1}
\end{align}
where $\beta_a=\omega/k$ is the apparent shear wave velocity with $k$ the wave number and $\omega$ the angular frequency, and $p$ [s/km] is the horizontal slowness or ray parameter \citep{fichtner2009sensitivity,igel2005rotational,wassermann2016toward,schmelzbach2018advances}. 

We can define the rotational misfit function as the squared normalized difference between the observed $(\beta_a)_{\text{obs}}$ and synthetic $(\beta_a)_{\text{syn}}$ apparent shear wave velocities as follows
\begin{align}
\E = \frac{1}{2} \int_{\Omega} \int_{0}^{T} \left[\frac{(\beta_a)_{\text{obs}}-(\beta_a)_{\text{syn}}}{(\beta_a)_{\text{syn}}}\right]^2 \sum_r \delta(\x-\x^r) \dif t  \dif^3 \x ,
\label{eq.rotational_misfit}
\end{align}
where $(\beta_a)_{\text{obs}}$ are the ratio of amplitudes defined by eq. \eqref{eq.amplitude_definitions} and given by
\begin{align}
(\beta_a)_{\text{obs}} =  -\frac{1}{2} \frac{(\partial_t \u)_{\text{obs}}}{(\crl \u)_{\text{obs}}} = -\frac{1}{2} \frac{(\partial^2_t \u)_{\text{obs}}}{(\crl \partial_t \u)_{\text{obs}}}.
\label{eq.Observed_log_beta1}
\end{align}
The variation of the misfit function eq. \eqref{eq.rotational_misfit} can be written as follows 
\begin{align}
\begin{aligned}
\delta \E  = & \int_{0}^{T} \int_{\Omega} \left[\frac{(\beta_a)_{\text{obs}}}{(\beta_a)_{\text{syn}}}-1\right] \delta \ln (\beta_a)_{\text{syn}} \sum_r \delta(\x-\x^r) \dif^3 \x \dif t .
\end{aligned}
\label{eq.misfit_func_grad}
\end{align}
Using the definition of $\beta_a$ given by eq. \eqref{eq.Observed_log_beta1} and dropping the mnemonic subscript upon the synthetic displacement can write the following
\begin{align}
\begin{aligned}
\delta \ln \beta_a \sum_r \delta(\x-\x^r) & = \left[\delta \ln (\crl \u) - \delta \ln (\partial_t \u) \right] \sum_r \delta(\x-\x^r) , \\
& = \left[ \frac{\delta(\crl \u)}{(\crl \u)} - \frac{\delta(\partial_t \u)}{(\partial_t \u)} \right] \sum_r \delta(\x-\x^r), \quad (\text{using definitions in eq. } \eqref{eq.amplitude_definitions}) \\
& = \left[ \frac{\int_{t} \crl \u \, \crl( \delta \u) \dif t}{\int_{t} (\crl \u)^2 \dif t} - \frac{\int_{t} \partial_t \u \delta (\partial_t \u) \dif t}{\int_{t} (\partial_t \u)^2 \dif t} \right] \sum_r \delta(\x-\x^r) , \quad \text{(integration by parts)}\\
& = - \frac{\int_{t}(\epsilon \cdot \crl \u )\dif t}{2\int_{t} (\crl \u)^2 \dif t} \sum_r \nabla \delta(\x-\x^r) \delta \u  + \frac{\int_{t} (\partial^2_t \u) \dif t}{\int_{t} (\partial_t \u)^2 \dif t} \sum_r \delta(\x-\x^r) \delta \u , 
\end{aligned}
\label{eq.adjoint_sour}
\end{align}
where $\epsilon$ is the Levi-Civita symbol and we have used $\crl \u = \frac{1}{2} \epsilon \cdot \nabla \u$. Note that we can also write eq. \eqref{eq.adjoint_sour} is as follows
\begin{align}
\delta \ln \beta_a \delta(\x-\x^r) = \left[ \frac{\int_{t} \crl (\crl \u ) \dif t}{\int_{t} (\crl \u)^2 \dif t}  + \frac{\int_{t} (\partial_t^2 \u) \dif t}{\int_{t} (\partial_t \u)^2 \dif t} \right] \sum_r \delta(\x-\x^r) \delta \u . 
\label{eq.eq.adjoint_sour2}
\end{align} 
We can identify the adjoint source $(f^{*})$ given by the following expression
\begin{align}
f^{*} = \left[\frac{(\beta_a)_{\text{obs}}}{(\beta_a)_{\text{syn}}}-1\right] \left[-\frac{ \epsilon  \cdot  \crl \u }{2\int_{t} (\crl \u)^2 \dif t} \sum_r \nabla \delta (\x-\x^r) + \frac{\partial^2_t \u }{\int_{t}  (\partial_t \u)^2 \dif t} \sum_r \delta(\x-\x^r)\right] .
\label{eq.adjoint_sources}
\end{align}
It has been previously noted that the terms in the adjoint source eq. \eqref{eq.adjoint_sources} involving the curl are sources described by an asymmetric moment tensor M given by the following expression \citep{fichtner2009sensitivity,bernauer2012measurements}
\begin{align}
\begin{aligned}
\text{M}_{\theta \theta} &= \text{M}_{\phi \phi} = \text{M}_{zz} =0 , \quad 	\text{M}_{\phi \theta} = -\text{M}_{\theta \phi}, \quad \text{M}_{z \theta} = - \text{M}_{\theta z}, \quad \text{M}_{z \phi} = - \text{M}_{\phi z} , \\
\text{M}_{\theta \phi} & = \frac{(\crl \u)_z}{2\int ((\crl \u)_z)^2 \dif t}, \quad \text{M}_{\theta z} = \frac{-(\crl \u)_{\phi} }{2\int ((\crl \u)_{\phi})^2\dif t} , \quad \text{M}_{\phi z} = \frac{(\crl \u)_{\theta}}{2\int ((\crl \u)_{\theta})^2 \dif t} .	
\end{aligned}
\end{align}

\subsubsection{Longitudinal Waves}

In analogy to eq. \eqref{eq.plane_wave_ratio1}, the apparent $P$ wave velocity is defined as \citep{bernauer2012measurements}
\begin{align}
    \alpha_a =  \frac{\partial_t \u}{\tr \strain} ,
    \label{eq.apparent_P_wave}
\end{align}
where $\strain$ denotes the second-order strain tensor defined in eq. \eqref{eq.stress_and_strain}. We can define the rotational misfit function as the squared normalized difference between the observed $(\alpha_a)_{\text{obs}}$ and synthetic $(\alpha_a)_{\text{syn}}$ apparent longitudinal wave velocities as follows
\begin{align}
\E = \frac{1}{2} \int_{\Omega} \int_{0}^{T} \left[\frac{(\alpha_a)_{\text{obs}}-(\alpha_a)_{\text{syn}}}{(\alpha_a)_{\text{syn}}}\right]^2 \sum_r \delta(\x-\x^r) \dif t  \dif^3 \x ,
\label{eq.rotational_misfit_P}
\end{align}
where $(\alpha_a)_{\text{obs}}$ are the ratio of amplitudes defined by eq. \eqref{eq.amplitude_definitions} and given by
\begin{align}
(\alpha_a)_{\text{obs}} = \frac{(\partial_t \u)_{\text{obs}}}{(\tr  \strain)_{\text{obs}}} = \frac{(\partial^2_t \u)_{\text{obs}}}{(\tr \partial_t  \strain)_{\text{obs}}}.
\label{eq.Observed_log_alpha1}
\end{align}
The variation of the misfit function eq. \eqref{eq.rotational_misfit_P} can be written as follows 
\begin{align}
\begin{aligned}
\delta \E  = & \int_{0}^{T} \int_{\Omega} \left[\frac{(\alpha_a)_{\text{obs}}}{(\alpha_a)_{\text{syn}}}-1\right] \delta \ln (\alpha_a)_{\text{syn}} \sum_r \delta(\x-\x^r) \dif^3 \x \dif t .
\end{aligned}
\label{eq.misfit_func_grad_P}
\end{align}
Using the definition of $\alpha_a$ given by eq. \eqref{eq.Observed_log_alpha1} and dropping the mnemonic subscript upon the synthetic displacement can write the following
\begin{align}
\begin{aligned}
\delta \ln \alpha_a \sum_r \delta(\x-\x^r) & = \left[\delta \ln (\partial_t \u) - \delta \ln (\tr  \strain)\right] \sum_r \delta(\x-\x^r) , \\
& = \left[\frac{\delta(\partial_t \u)}{(\partial_t \u)} -  \frac{\delta(\tr  \strain)}{(\tr  \strain)}\right] \sum_r \delta(\x-\x^r), \quad (\text{using definitions in eq. } \eqref{eq.amplitude_definitions}) \\
& = \left[ \frac{\int_{t} \partial_t \u \delta (\partial_t \u) \dif t}{\int_{t} (\partial_t \u)^2 \dif t} - \frac{\int_{t} \tr  \strain \, \tr( \delta  \strain) \dif t}{\int_{t} (\tr  \strain)^2 \dif t}\right] \sum_r \delta(\x-\x^r) , \quad \text{(integration by parts)}\\
& = \frac{\int_{t}(\tr  \strain)\dif t}{\int_{t} (\tr  \strain)^2 \dif t} \sum_r \tr \nabla \delta(\x-\x^r) \delta \u  - \frac{\int_{t} (\partial^2_t \u) \dif t}{\int_{t} (\partial_t \u)^2 \dif t} \sum_r \delta(\x-\x^r) \delta \u , 
\end{aligned}
\label{eq.adjoint_sour_alpha}
\end{align}
where we can identify the adjoint source $(f^{*})$ given by the following expression
\begin{align}
f^{*} = \left[\frac{(\alpha_a)_{\text{obs}}}{(\alpha_a)_{\text{syn}}}-1\right] \left[\frac{\tr  \strain}{\int_{t} (\tr  \strain)^2 \dif t} \sum_r \tr \nabla \delta (\x-\x^r) - \frac{\partial^2_t \u }{\int_{t}  (\partial_t \u)^2 \dif t} \sum_r \delta(\x-\x^r)\right] .
\label{eq.adjoint_sources_rot_P}
\end{align}

It has been previously noted that the terms in the adjoint source eq. \eqref{eq.adjoint_sources_rot_P} involving the trace are sources described by an moment tensor M given by the following expression \citep{bernauer2012measurements}
\begin{align}
\begin{aligned}
	\text{M}_{\phi \theta} & = \text{M}_{\theta \phi} = \text{M}_{z \theta} =  \text{M}_{\theta z} = \text{M}_{z \phi} = \text{M}_{\phi z} = 0, \\
\text{M}_{\theta \theta} & =  \text{M}_{z z} =  \text{M}_{\phi \phi } = \frac{\tr  \strain}{\int (\tr  \strain)^2 \dif t}.	
\end{aligned}
\end{align}

\subsubsection{Imaging the Deep arth} 

If we consider a simplified earth model made of vertical layers with different velocities, the application of Snell's law yields the definition of the ray parameter $p$ which is constant along the ray and provides an estimate of the horizontal velocity as follows
\begin{align}
	p = \frac{\sin i}{v} = s \sin i,
	\label{eq.slowness_def}
\end{align} 
where $i$ is the incidence angle of the ray, $s$ is the slowness $(s=1/v)$ and $v$ the velocity of the layer \citep{shearer2019introduction}.  
\begin{figure}
	\begin{center}
		\includegraphics[width=1\textwidth]{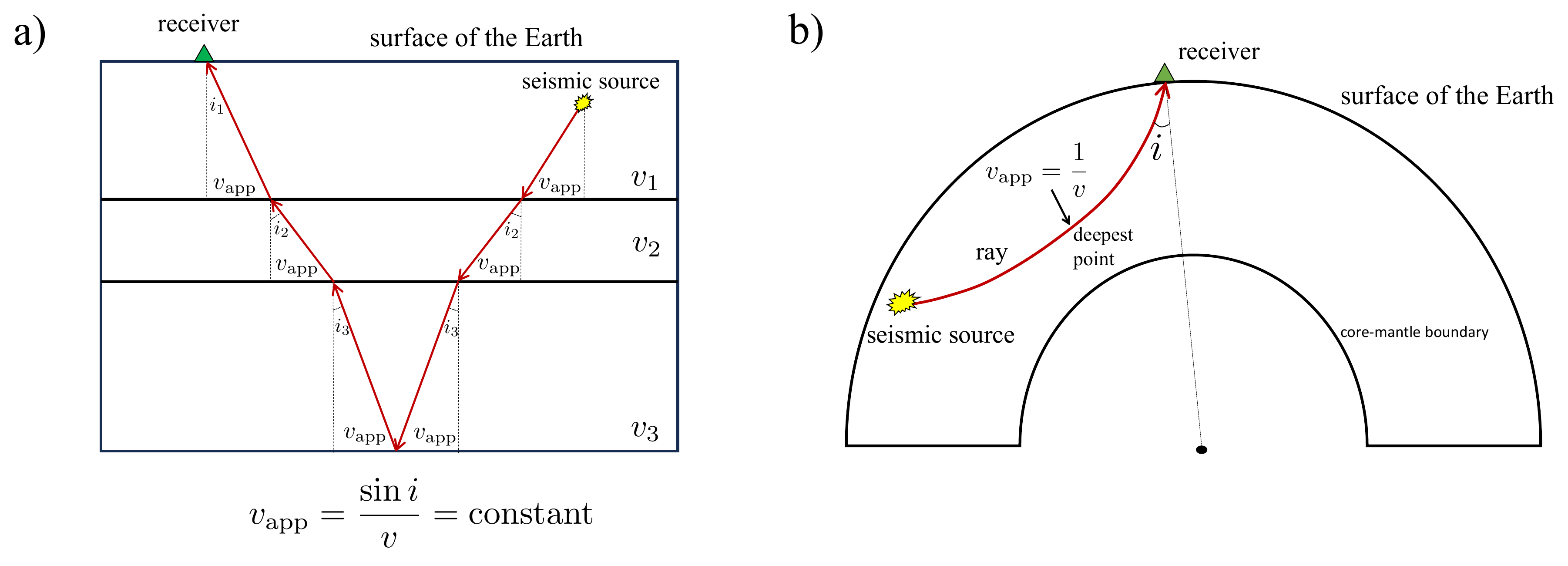}
		\caption{a) Diagram representing the constant value of the apparent velocity within a vertically layered Earth. b) Diagram representing the equivalence between the apparent and layer velocities at the deepest point of the wave.}
		\label{Fig.Snell_law}
	\end{center}
\end{figure}

From eq. \eqref{eq.slowness_def} we can observe that if we have information of the slowness (apparent velocity) and of the incidence angle of the ray, then we can obtain information of the velocity of the layer (see Fig. \ref{Fig.Snell_law}-a). If we consider a location at which $\sin i=1$, i.e., the location at which the wave travels horizontally (deepest and/or turning point), as a consequence, the ray parameter $p$ can be related to the velocity of the medium as follows \citep{stein2009introduction}
\begin{align}
	p = \frac{1}{v_d},
	\label{eq.ray_slowness_diff_depths1}
\end{align}
where the subscript $(d)$ refers to the turning (or deepest) point of the ray (see Fig. \ref{Fig.Snell_law}-b). Note that if the wave reflects at an interface (ScS wave for example), the deepest point of the ray will not travel horizontally ($i_d\neq90^{\circ})$, therefore the velocity of the medium at the deepest point of the ray will not be equal to the inverse of the slowness (see Fig. \ref{Fig.Snell_law}-a).

Using information of S waves and combining translational and rotational data, we can obtain local information of the velocity of the medium at the deepest point of the ray, as follows \citep{abreu2023deep}
\begin{align}
	v_d^S = \frac{1}{(p)^S} = - \frac{1}{2} \left(\frac{a_T}{\dot{\Omega}_z}\right)^S ,
	\label{eq.S_wave_slowness}
\end{align}
where $(p)^S$ is the slowness of the S wave using eq. \eqref{eq.plane_wave_ratio1}. We can write eq. \eqref{eq.S_wave_slowness} normalized with respect to a certain 1D earth model as follows
\begin{align}
	\left(\frac{\delta v}{v}\right)^{S} = \frac{v_d^{S(\text{obs})}-v_d^{S(\text{model})}}{v_d^{S(\text{model})}} = \frac{\left( p \right)^S_{\text{model}} }{\left(p\right)^{S}_{\text{obs}}} - 1   = - \frac{1}{2} \left(\frac{a_T}{\dot{\Omega}_z}\right)^S \left( p \right)^S_{\text{model}} - 1,
	\label{eq.normalized_vel_ratios}
\end{align} 
where $\left(p\right)_{\text{obs}}$ stands for the observed horizontal slowness (or ray parameter) and $\left( p \right)^S_{\text{model}} $ to the horizontal slowness obtained from any 1D earth model, e.g., PREM \citep{DZIEWONSKI1981297}, STW105 \citep{kustowski2008anisotropic}, AK135 \citep{kennett1995constraints}, IASP91 \citep{kennett1991traveltimes}. Using eq. \eqref{eq.apparent_P_wave}, for P waves we can write
\begin{align}
	 	\left(\frac{\delta v}{v}\right)^{P} = \frac{v_d^{P(\text{obs})}-v_d^{P(\text{model})}}{v_d^{P(\text{model})}} = \frac{\left( p \right)^P_{\text{model}} }{\left(p\right)^{P}_{\text{obs}}} - 1   = \left(\frac{\partial_t \u}{\tr \strain} \right)^P \left( p \right)^P_{\text{model}} - 1 . 
\end{align}

We can observe that determining the slowness values of seismic arrivals computed using either array techniques \citep[e.g.][]{rost2002array} and/or with the use of rotational and translational recordings, we can obtain local P and S velocity perturbations of the Earth's mantle \citep{abreu2023deep} at the deepest point of the P and S wave paths.

\subsection{Other Misfits Measures}

Although we have limited our presentation to certain misfit measures, several others can be constructed depending on specific requirements, e.g., noise cross-correlations \citep[e.g.][]{tromp2010noise,fichtner2015source,ermert2016cross,sager2018sensitivity,sager2020global,fichtner2020optimal}, phase correlation measures \citep{xu2013sensitivity}, receiver functions \citep{ammon1990nonuniqueness,de2022sensitivity}, Rayleigh wave ellipticity \citep{maupin20173}, using an unwrapped phase \citep{djebbi2014traveltime}, using local-similarity attributes \citep{zhang2019local}, double-difference measurements \citep{orsvuran2020double}, in the time/frequency domain \citep[e.g.][]{pratt1999seismic,gee1992generalized,kristekova2006misfit,fichtner2008theoretical,rickers2012imaging,choi2013frequency,alkhalifah2012taming}, in the $(\tau,p)$ and/or $(f,k)$ domains \citep[e.g.][]{pratt1999seismic,pratt1999seismic_part2,cao1990simultaneous,kormendi1991nonlinear,snieder1989retrieving}, 2D waveform inversions \citep[][]{shipp2002two,romdhane2011shallow,ernst2007application,borisov2020application}, among a considerable number of others. For each scientific study, an adequate misfit measure can/should be constructed that will allow to extract, as much as possible, information from the data.

\section{Discussion}

\subsection{Variational Formulations and Applications}

In this paper, we have used the variational formalism as the pillar of our theoretical developments. The variational formulation has remarkable advantages in several influential areas of mathematics. For instance, the variational description of non-standard (enhanced) continuum mechanics formulations is a powerful tool used today in (non-conventional) engineering applications such as the design of materials with acoustical and mechanical behaviors that are not found in nature, usually called seismic metamaterials \citep[e.g.][]{dell2020discrete,neff2014unifying,madeo2018relaxed,grekova2020reduced}. Such enhanced continuum mechanical formulations allow to model materials with microstructure as a continuum, i.e., a continuum model that has average properties compared to the micro-structured material. This becomes specially useful for the design of large engineering structures, where the numerical modeling of the material with exact microstructure becomes prohibited due to the large computational power required.   

Variational formulations offer high-quality processing capabilities for image processing. These techniques have been widely developed and applied in the last two decades, and bring the advantage of finding an optimal image by using a minimizer of an energy functional. This allows us to develop image processing methods that are more accurate and require less data \citep[e.g.][]{chan2005image,vese2016variational,hansen2006deblurring,sapiro2006geometric,sethian1999level,scherzer2009variational}.

The variational formulation of adjoint tomography was introduced by \cite{tromp2005seismic,Liu01122006} with further applications to poroelastic media \citep{morency2009finite} and noise cross-correlations \citep{tromp2010noise}. Today these formulations are used to build the accurate models of the deep Earth interior \citep{bozdaug2016global,lei2020global} in combination with the SEM \citep{komatitsch1998spectral,komatitsch1999introduction,chaljub2007spectral}.

\subsection{The Adjoint (Equation) Method: History}

The adjoint (equation) method dates back to early 1950's when it was introduced by \citet[p. 299]{Morse1953} as the authors called it \textit{a mathematical trick}. The original motivation was used in a variational problem to define the Lagrangian of non-conservative equations of motion, e.g., the heat equation \citep[e.g.][]{Morse1953,Gurtin1964,tonti1973variational}. Early applications to define the Lagrangian of the elastic wave equation including viscoelasticity can be found in \cite[e.g.][]{gurtin1963variational,leitman1966variational,kline1970variational}.

It seems that this motivation, and the connection to the community of mathematicians, has been lost in the Earth sciences, since more than a decade later, the adjoint method appears in the field of petroleum engineering \citep{chavent1975history} and later in seismic inversion  \citep{bamberger1979stability,bamberger1982inversion,tarantola1984inversion,Tarantola1988,gauthier1986two}, with no references to be found to the initial mathematical development and motivation of the method. Adjoint operators (matrix transpose) in geophysical inversion, however, can be found earlier in the literature \cite[e.g.][]{claerbout1976fundamentals}.

While the original motivation of the method was to find ways to define non-conservative Lagrangians, the main application in Earth sciences is to compute the Fr\'echet derivative of an objective functional with respect to model parameters. This allows us, among other applications, to perform full-waveform inversions and to extract, as much as possible, information from seismological data.

Despite that the adjoint method, as understood today by the Earth sciences community, brings many beneficial opportunities for scientific development, the lack of connection to the mathematical original motivations of the method prevents further scientific development. Specifically, this happens in the case of viscoelasticity. Viscoelastic effects of seismic wave propagation are not well understood today in the Earth sciences as well as its connections to Earth processes. Earth scientists include viscoelastic effects in the elastic wave equation in an ad hoc way with little connection to the mathematical description of the physical processes that causes the effects. One may claim that this happens not only the Earth sciences. This situation may be remediable by properly understanding the variational definition of the viscoelastic seismic equation. Concretely, understanding how to define the Lagrangian of the (viscoelastic) seismic wave equation using seismological data as an inverse problem. For doing this, the direct collaboration between mathematicians and Earth scientists is crucial.

The adjoint method requires the definition of a new variable that is commonly known is the literature as the adjoint variable. This new variable, unlike the original variable of the equation in study (e.g. displacement), has not well defined/understood physical meaning. Attempts to give a proper physical understanding of the adjoint variable date back to early 1960's in the area of reactor physics for transport phenomena \citep[e.g.][]{lewins1960derivation,lewins1960approximate,lewins1965importance}. It was \cite{tonti1973variational} that found that the adjoint variable can be simply defined as the original variable of the equation in study but evaluated at the instant $(T-t)$. This has been the main interpretation in the Earth sciences community. We have shown however that while this interpretation is correct, it is not the only that can be attributed to it. We have suggested that linear operator can be applied to the original time-reversed variable in study and the mathematics will still be consistent. This interpretation has allowed us to show the equivalence between waveform, travel-time and amplitude Fr\'echet kernels, which are commonly used in FWI. We reiterate that the physical interpretation and full understanding of the adjoint variable is, however, far from being a solved problem. 

\subsection{Adjoint Tomography: Viscoelastic Adjoint Equations}

We have shown, for the first time, that the adjoint equations have always the same mathematical structure as the initial equation under study. The presence of odd-order terms and/or viscoelastic terms (odd-order terms including time) in the equations of motion do not change their sign as has been previously claimed. This is a major finding since physical interpretation and numerical implementations are highly sensitive to changes in signs. For instance, while viscoelastic effects can be fairly easily numerically implemented and physically understood as loss of energy due to the presence of inhomogeneities in the material, the opposite is exactly the contrary. The numerical implementation of viscoelastic effects are very prone to numerical instabilities and their physical justification is nothing more than a mathematical tool since it is difficult to justify the gain of energy in the system when no time-dependent external forces are applied. As a consequence, adjoint viscoelastic wave equations are routinely interpreted as mathematical tools with little or no connection to reality.

Not all this changes with our mathematical findings. The adjoint equation is a mathematical tool that has the same mathematical structure of the initial wave equation (in the case of seismology); thus, viscoelastic effects continue to extract energy from the system and not to inject it into the system. This solves problems of numerical implementations and physical interpretations. The adjoint equation and its adjoint variable are still mathematical tools with little to no connection to reality. We, however, consider this to be an unsolved problem. Clearer understanding can be obtained when analyzing in detail variational formulations and adjoint equations in fields outside Earth sciences.

\subsection{Adjoint Tomography: Travel-Time Fr\'echet Kernels}

We have designed a new (envelope) travel-time adjoint source based on the numerical implementation of the Dirac delta distribution. This allowed us to show, for the first time, that travel-time Fr\'echet kernels have zero sensitivity along the ray path only when data and observations are similar enough. When there is no similarity between data and synthetics, (envelope) travel-time Fr\'echet kernels show an extremum (maximum or minimum) along the ray path (as one can physically expect). We claim that it is the assumption of the similarity between data and synthetics that allows us to perform travel-time cross correlation measurements and that it is the cause of the lack of sensitivity along the ray path that has been previously observed in numerous studies and has been a topic of debate for around two decades.

Having no sensitivity along the ray path but around it when data and synthetics are similar enough is physically sound since one can expect that the ray path has no influence anymore in the measurements and information around the ray path has to be taken/modified. This result represents a new perspective that explains, from a physical point of view, and validates previous results unifying different perspectives/opinions about travel-time sensitivity kernels.

\subsection{Adjoint Tomography with Cross-correlation or Envelope Travel-Time Measurements?}

Travel-time measurements can be computed under any desired assumption. There is no consensus on which is the best way to compute travel-time measurements from seismological data. We can find, however, large databases of body-waves travel-time measurements computed using cross-correlations \citep[e.g.][]{engdahl1998global,weston2018isc,lai2019global,engdahl2020isc}. The reason is simply because we can obtain long wavelength waveforms using 1D earth models that are relatively similar enough compared to data. Keeping this in mind, we can use these large databases \citep[e.g.][]{engdahl1998global,weston2018isc,lai2019global,engdahl2020isc}, as well as any others \citep[e.g.][]{li2023efficient}, to perform adjoint (envelope) travel-time inversions of body waves since, on one hand body wave travel-times are (nearly) frequency independent and on the other hand, the technique used to compute travel-times should not affect the tomographic model.

Performing adjoint (envelope) travel-time inversions gives a new way to obtain Fr\'echet sensitivity kernels that are travel-time dependent (see Fig. \ref{Fig.data_envelope_S_kernels}). This combined with source encoding techniques \citep[e.g.][]{krebs2009fast,castellanos2015fast,moghaddam2013new,ben2011efficient,schuster2011theory,tromp2019source}, will allow us to make use of all travel-time databases already available to improve tomographic images in a unique and pragmatic way.

\subsection{Adjoint Tomography: Numerical Implementation}

To develop new methodologies in adjoint inversion it is imperative to master the numerical implementation of the method. We have seen that the adjoint method requires access to the \textit{time reversed} wavefield, which can be reconstructed from the last snapshots of the forward simulation. This will depend on the time integration scheme that we use and whether the substitution $\Delta t \to - \Delta t$ leaves the integration scheme intact or not.  

Specifically, when solving the second-order (displacement) formulation of the (elastic/acoustic) wave equation and employing the leapfrog method for time integration, by a simple flip in the initial conditions we can reconstruct the time reversed wavefield, without the need to modify the original forward solver. When dealing with velocity-stress formulations of the (elastic/acoustic) wave equation and/or Newmark-beta \citep{newmark1959method} or Verlet \citep{verlet1967computer} time integration methods, the forward solver must be modified according to the substitution $\Delta t \to - \Delta t$ and no need to flip initial conditions is required. Finally, to obtain correct sensitivity kernels it is imperative to run the adjoint simulation to a total time equal to the total forward simulation time $T$ minus the time delay $(t_s)$ of the initial source time function $(T-t_s)$.

\subsection{Adjoint Tomography: Emerging Fields}

In the ideal scenario, one should be able to use all information contained in the seismograms and to obtain an ensemble of earth models that match those observations equally well. We have discussed that due to computational and human limitations (to process large amount of seismological data) this is not feasible. Emerging fields in FWI methodologies thus aim to include statistical techniques to overcome these limitations. 

There is a large amount of statistical techniques that are being included in FWI and our aim is not to mention all of them or to state which one is better but to mention that, among all of these techniques, we can find variational methods like, i) variational inference formulations recently emerging \citep[e.g.][]{blei2017variational,zhang2018advances,Valentine2021,lopez2021deep,zhang2020seismic,nawaz2020variational,zhang2020variational,ZHANG202173,Zhao2024}, where the advantage is that it allows to approximate a high-dimensional Bayesian posterior distribution by solving an optimization problem, thus reducing computational costs associated with the inversions. ii) Hamilton's variational principle \citep{fichtner2021autotuning,gebraad2020bayesian,fichtner2019hamiltonian,fichtner2019hamiltonian2,fichtner2018hamiltoniansources}, used to obtain a family of models that satisfy seismological observations.

Emerging fields in adjoint tomography also aim to include new data to build more constrained earth models such as Distributed Acoustic Sensing (DAS) \citep{daley2013field,parker2014distributed,eaid2020multiparameter,fichtner2022introduction} and rotational data \citep{bernauer2018blueseis3a,igel2021romy,fichtner2009sensitivity}. We have presented a new and simple variational formulation of the adjoint method applied to rotational seismology that does not require the use of complex mathematical concepts.

\section{Summary of Major Contributions of this Work}

We have presented a simple variational formulation of the adjoint method that does not require the use of advanced mathematical concepts like Green's functions, the Born approximation and/or the representation theorem. We have clarified some mathematical inconsistencies found in seminal works related to the treatment of viscoelastic effects and in general to odd-order derivatives in the equations of motion. Specifically, we have shown that the adjoint equations result in the original equations even in the presence of viscoelastic effects and/or odd-order derivatives. We have shown that the understanding of the adjoint method completely depends on the arbitrarily chosen definition of the inner product and norm in the model and data spaces. We have defined a new envelope travel-time misfit function that allows us to show that travel-time Fr\'echet kernels are insensitive to the theoretical ray path only when it is assumed that data and synthetics are similar enough, which allows cross correlation travel-time measurements to be performed. The advantages of the introduced envelope misfit function iare that, unlike cross-correlation, it does not require processing task each iteration of the adjoint inversion. This gives us the opportunity to re-utilize old and new travel-time databases \citep[e.g.][]{engdahl1998global,weston2018isc,lai2019global,engdahl2020isc}, where waveform information has not been kept, to perform adjoint inversions. The main advantages of the adjoint (envelope) travel-time inversion over any other conventional theory are that arbitrary crustal corrections are completely avoided and highly heterogeneous media (where ray theory and normal modes fail) can now be inverted relying on purely numerical simulations, therefore bridging a gap between ray theory and waveform inversions. In a future contribution we will discuss in depth the implications of the presented methodology in viscoelastic media.

\section{Acknowledgements}

Dedicated to the memory of Prof. Enzo Tonti. R.A. acknowledges the deep and constructive reviews of Barbara Romanowicz, Jean-Paul Montagner, Jeffrey Gu, Michael J. Rycroft and an anonymous reviewer that allowed us to considerably improve the clarity and quality of the presented work. R.A. acknowledges computational resources kindly provided by Fabian Bonilla. R.A. was partially supported by the DFG project EARLY AB887/1-1. 

\footnotesize
\bibliographystyle{apalike}
\bibliography{Biblio}

\end{document}